\documentclass[iop]{emulateapj}
\usepackage{amsmath}
\usepackage{txfonts}
\usepackage{natbib}
\usepackage{graphicx}
\usepackage{color} % zum Hinweise markieren

\begin{document}

\title{Self-sustained asymmetry of lepton-number emission: A new phenomenon
during the supernova shock-accretion phase in three dimensions}

\shorttitle{Self-sustained asymmetry of lepton-number emission in supernovae}

\shortauthors{Tamborra et al.}

% authors
\author{Irene Tamborra\altaffilmark{1~*},
  Florian Hanke\altaffilmark{2,3},
  Hans-Thomas Janka\altaffilmark{2},
  Bernhard M\"uller\altaffilmark{2~**},
  Georg G.~Raffelt\altaffilmark{1}, and
  Andreas Marek\altaffilmark{4}
  }

\affil{\altaffilmark{1} Max-Planck-Institut f\"ur Physik
                        (Werner-Heisenberg-Institut), F\"ohringer Ring 6,
                        80805 M\"unchen, Germany\\
       \altaffilmark{2} Max-Planck-Institut f\"ur Astrophysik,
                        Karl-Schwarzschild-Str.~1, 85748 Garching, Germany\\
       \altaffilmark{3} Physik Department, Technische Universit\"at M\"unchen,
                        James-Franck-Str.~1, 85748 Garching, Germany\\
       \altaffilmark{4} Rechenzentrum der Max-Planck-Gesellschaft,
                        Boltzmannstr.~2, 85741 Garching, Germany}

\altaffiltext{*}{Present address: GRAPPA Institute, University of
Amsterdam,
Science Park 904, 1098 XH Amsterdam, The Netherlands}

\altaffiltext{**}{Present address: Monash Center for Astrophysics, School of
  Mathematical Sciences, Building 28, Monash University, Victoria
  3800, Australia}

\begin{abstract}
  During the stalled-shock phase of our three-dimensional, hydrodynamical
  core-collapse simulations with energy-dependent, three-flavor neutrino
  transport, the lepton-number flux ($\nu_e$ minus $\bar\nu_e$) emerges
  predominantly in one hemisphere. This novel, spherical-symmetry breaking
  neutrino-hydrodynamical instability is termed LESA for ``Lepton-number
  Emission Self-sustained Asymmetry.'' While the individual $\nu_e$ and
  $\bar\nu_e$ fluxes show a pronounced dipole pattern, the heavy-flavor
  neutrino fluxes and the overall luminosity are almost spherically
  symmetric. Initially, LESA seems to develop stochastically from
  convective fluctuations. It exists for hundreds of milliseconds or more
  and persists during violent shock sloshing associated with the standing
  accretion shock instability. The $\nu_e$ minus $\bar\nu_e$ flux asymmetry
  originates predominantly below the neutrinosphere in a region of
  pronounced proto-neutron star (PNS) convection, which is stronger in the
  hemisphere of enhanced lepton-number flux. On this side of the PNS, the
  mass-accretion rate of lepton-rich matter is larger, amplifying the
  lepton-emission asymmetry, because the spherical stellar infall deflects
  on a dipolar deformation of the stalled shock. The
  increased shock radius in the hemisphere of less mass accretion and
  minimal lepton-number flux ($\bar\nu_e$ flux maximum) is sustained by
  stronger convection on this side, which is boosted by stronger neutrino
  heating due to \hbox{$\langle\epsilon_{\bar\nu_e}\rangle >
  \langle\epsilon_{\nu_e}\rangle$}. Asymmetric heating thus supports
  the global deformation despite extremely nonstationary convective
  overturn behind the shock. While these different elements of the
  LESA phenomenon form a consistent picture, a full understanding remains
  elusive at present. There may be important implications for
  neutrino-flavor oscillations, the neutron-to-proton ratio in the
  neutrino-heated supernova ejecta, and neutron-star kicks, which remain to
  be explored.
\end{abstract}


%%%%%%%%%%%%%%%%%%%%%%%%%%%%%%%%%%%%%%%%%%%%%%%%%%%%%%%%%%%%%%%%%%%%%%%%%%%%%%%
\section{Introduction}
%%%%%%%%%%%%%%%%%%%%%%%%%%%%%%%%%%%%%%%%%%%%%%%%%%%%%%%%%%%%%%%%%%%%%%%%%%%%%%%

Nonradial hydrodynamic instabilities play an important role in the
postbounce dynamics of collapsing stellar cores on their way to the
onset of supernova (SN) explosions. They also accompany the formation
and cooling of the proto-neutron star (PNS). These phenomena include
convection in the PNS, large-scale convective overturn below the
stalled shock wave during the accretion-shock phase, and the standing
accretion shock instability (SASI).  We presently add a new phenomenon
to this list which we call LESA for ``Lepton-number Emission
Self-sustained Asymmetry.'' Its most conspicuous manifestation is
lepton-number flux ($\nu_e$ minus $\bar\nu_e$) emission primarily in
one hemisphere, but it also involves dipolar asymmetry of
convection inside the PNS and different strengths of large-scale 
convective overturn below the stalled shock in both hemispheres.

We first recall these traditional spherical-symmetry breaking instabilities
and begin with convection. Prompt postshock convection leads to the decay of
the negative entropy and electron-number gradients left behind by the
weakening bounce shock and the shock-breakout burst of electron neutrinos,
respectively \citep{Burrows_1992,Janka_1993,Mueller_1994}. It fosters
shock expansion and acts as a source of gravitational-wave emission for a
period of some ten milliseconds after core bounce
\citep{Mueller_1997,Mueller_2013}.
Inside the nascent neutron star (NS), i.e.\ below
(and possibly also around) the neutrinosphere, Ledoux convection was expected
to occur because of the negative lepton number gradient produced by the
inward progression of the deleptonization wave associated with neutrino
losses \citep{Epstein_1979,Burrows_1988}. In two-dimensional (2D)
simulations (axial symmetry), PNS convection was first studied by
\citet{Keil_1996} and later again by \citet{buras_06b} and
\citet{Dessart_2006}.

%------------------------------------------------------------------------------
\begin{figure*}[t!]
\begin{center}
  \includegraphics[width=\textwidth]{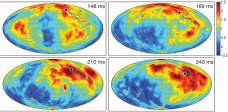}
  \caption{Lepton-number flux ($\nu_e$ minus $\bar\nu_e$) for our
  $11.2\,M_\odot$ model as a function of direction for the indicated times
  post bounce.
  The latitudes and longitudes, indicated by dotted lines, correspond
  to the angular coordinates of the polar grid of the numerical simulation.
  The flux in each panel is normalized to its average, i.e.,
  the quantity   $(F_{\nu_e}-F_{\bar\nu_e})/\langle
  F_{\nu_e}-F_{\bar\nu_e}\rangle$ is color coded.
  The lepton-number emission
  asymmetry is a large-scale feature, which at later times has clear
  dipole character. The black dots indicate the
  positive dipole direction of the flux distribution, the black crosses mark
  the negative dipole direction. The dipole track between 70 and 340\,ms is
  shown as a dark-gray line.  Once the dipole is strongly developed, its
  direction remains essentially
  stable and shows no correlation with the $x$-, $y$-,
  and $z$-axes of the numerical grid. The dipole direction is also independent
  of polar hot spots, which are persistent, local features of moderate
  amplitude and an artifact connected with numerical peculiarities near
  the z-axis as coordinate singularity of the polar grid.}
   \label{fig:leptonskyplots}
\end{center}
\end{figure*}
%------------------------------------------------------------------------------

Large-scale convective overturn below the stalled shock was predicted by
\citet{bethe_90} and confirmed by the first 2D hydrodynamical simulations
\citep{Herant_1992,Herant_1994,Burrows_1995,Miller_1993,Janka_1995,Janka_1996,Mezzacappa_1998}
as well as in 3D \citep{Fryer_2002,Fryer_2004}.
The driving force is a negative entropy gradient that
develops a few tens of milliseconds after core bounce in the neutrino-heating
region between the gain radius (where neutrino heating begins to exceed
neutrino cooling) and the stalled shock. Convective flows stretch the
dwelling time of matter in the gain layer and thus increase the energy
deposition by neutrinos. This effect can provide crucial support to the
delayed neutrino-heating mechanism: multi-D simulations can yield explosions
even when spherically symmetric models fail (e.g.,
\citealt{Janka_1996,Murphy_2008,Nordhaus_2010,Hanke_2012,Dolence_2013,Couch_2013,Couch_2013a}).

The delayed-explosion mechanism is also aided by SASI, which leads to violent
shock sloshing motions. This effect expands the shock, increases the gain
layer and, again, can enhance the efficiency of neutrino-energy deposition
\citep{marek_09} even when convection is weak or its growth is suppressed
because of a small shock-stagnation radius and correspondingly fast infall
velocities in the gain layer \citep{Foglizzo_2006,Scheck_2008}. This nonradial
instability was first observed in 2D simulations with a full 180$^\circ$ grid
\citep{Janka_1996,Mezzacappa_1998,Janka_2003,Janka_2004},
but not immediately recognized as a new effect beyond
large-scale convection. It was unambiguously identified in 2D hydrodynamical
simulations of idealized, adiabatic (and thus non-convective) postshock
accretion flows \citep{Blondin_2003}. SASI was found to possess the highest
growth rates for the lowest-order (dipole and quadrupole) spherical harmonics
\citep{Blondin_2006,Foglizzo_2007,Iwakami_2008} and to give rise to
spiral-mode mass motions in 3D simulations
\citep{Blondin_2007,Iwakami_2009,Fernandez_2010,Hanke_2013} or in 2D setups
without the constraint of axisymmetry
\citep{Blondin_2007,Yamasaki_2008,Foglizzo_2012}.
The instability can be explained by an advective-acoustic cycle of
amplifying entropy and vorticity perturbations in the cavity between
accretion shock and PNS surface
\citep{Foglizzo_2002,Foglizzo_2007,Scheck_2008,Guilet_2012}
and has important consequences for NS kicks
\citep{Scheck_2004,Scheck_2006,Nordhaus_2010b,Nordhaus_2012,Wongwathanarat_2010,Wongwathanarat_2013}
and spins \citep{Blondin_2007,Rantsiou_2011,Guilet_2013},
quasi-periodic neutrino emission modulations
\citep{Marek_2009,Lund_2010,Tamborra_2013},
and SN gravitational-wave signals \citep{Marek_2009,Murphy_2009,Mueller_2013}.

Here we report the discovery of a new type of low-mode nonradial instability,
LESA, which we have observed in 3D hydrodynamical simulations with detailed,
energy-dependent, three-flavor neutrino transport using the
\textsc{Prometheus-Vertex} code.  Our current portfolio of simulated 3D
models includes an $11.2\,M_\odot$ model that shows violent large-scale
convection, but no obvious signs of SASI activity during the simulated period
of postbounce evolution, and two models (20$\,M_\odot$ and 27\,$M_\odot$)
in which episodes of SASI alternate with
phases of dominant large-scale convection
\citep{Hanke_2013,Tamborra_2013,Tamborra_2014}.
While all models exhibit LESA, with different orientations of the
emission dipole, the clearest case is the $11.2\,M_\odot$ model, because the
new effect is not overlaid with SASI activity.

To provide a first impression of our new and intriguing phenomenon we show in
Fig.~\ref{fig:leptonskyplots} the distribution of lepton-number emission
($\nu_e$ minus $\bar\nu_e$) for the $11.2\,M_\odot$ model over the stellar
surface at postbounce (p.b.) times of 148, 169, 210, and 240\,ms. In each
panel, the lepton-number flux is normalized to the instantaneous average and
the color scale covers the range from $-$0.5 to 2.5 of this relative measure.
We indicate the positive dipole direction with a black dot, the negative
direction with a cross. We also show the track of the positive dipole
direction as a dark-gray line, ranging from 70\,ms p.b., where the dipole
begins forming, to the end of the simulation at 340\,ms. While at 148\,ms the
dipole pattern is not yet strong---a quadrupole component is clearly visible
and the dipole is still building up as we will see later---the subsequent
snapshots reveal a strong dipole pattern with large amplitude: In the
negative-dipole direction, the lepton-number flux is around zero, and even
negative in some small regions, whereas in the positive direction it is
roughly twice the average and even larger in some small regions.

LESA is a large and conspicuous effect concerning the deleptonization flux.
At the same time, it is associated with more subtle dipole deformations of
other quantities. In particular, the dipolar lepton-number emission is linked
to anisotropic PNS convection, which leads to an aspherical electron
distribution in the PNS mantle layer. The emission dipole is additionally fed
by a hemispheric mass-accretion asymmetry, which might cause the one-sided
enhancement of PNS convection. This accretion asymmetry in turn is a
consequence of a dipolar shock deformation that deflects the accretion flow
preferentially to one hemisphere.  Despite vigorous and highly time-variable,
nonstationary convective overturn stirring the postshock region, the shock
deformation and mass-accretion asymmetry are maintained for hundreds of
milliseconds by a neutrino-heating asymmetry that is tightly linked to the
neutrino-emission anisotropy: Because $\bar\nu_e$ have somewhat harder
spectra than $\nu_e$, neutrino heating on the side of a relatively higher
$\bar\nu_e$ flux (the side with lowest lepton-number flux and lowest mass
accretion rate) is stronger and sustains the dipolar shock deformation that
produces the hemispheric asymmetry of the postshock accretion flow.

This preliminary interpretation suggests that LESA is not a purely
hydrodynamical phenomenon, in contrast to the traditional instabilities, but
depends on a complex interaction of hydrodynamic mass flow and neutrino
emission and heating.

Our new effect is predominantly a hemispheric asymmetry in these quantities
and as such is not a generic 3D effect, but it has not been previously
reported in the context of 2D simulations. Actually, indications of LESA may
be present in the O-Ne-Mg core explosions of \citet{Wanajo_2011} and in a
$15\,M_\odot$ explosion model of \citet{mueller_12}. However,
it seems difficult to make a strong case for this
neutrino-hydrodynamical instability based on 2D simulations. The constraint
of axisymmetry resticts any dipole asymmetry to the direction of
the polar axis of the grid. This symmetry axis, where
reflecting boundary conditions are imposed, defines a preferred direction and
has various problematic consequences. It tends to
artificially create hemispheric differences by deflecting the converging
flows either inward or outward, and the grid axis also directs
shock-sloshing motions. Strong bipolar motions of the postshock accretion
layer in most 2D simulations, where shock expansion and contraction
alternate violently between the poles, could interfere with the LESA
phenomenon or even create asymmetries of different nature.

In the following, we first describe briefly, in Sect.~\ref{sec:numerics}, the
numerical setup of our 3D simulations and the overall properties of our three
progenitor models. In Sect.~\ref{sec:results} we study various manifestations
of our new phenomenon, ranging from dipole deformations of
neutrino-emission properties to asymmetric PNS convection. Next we turn in
Sect.~\ref{sec:feedback-loop} to more subtle manifestations in the form of
asymmetric accretion and neutrino heating, which however form a feedback loop
and as such are the driving engine of the overall effect. In
Sect.~\ref{sec:explanation} we string the different elements together and
provide an overall scenario that involves the outer feedback mechanism
consisting of asymmetric mass accretion and neutrino heating and the inner
mechanism of asymmetric electron-density distribution and PNS convection. We
conclude in Sect.~\ref{sec:conclusions} with a summary and a discussion of
possible implications.

%------------------------------------------------------------------------------
\begin{figure*}[t!]
\begin{center}
\hbox to\textwidth{\includegraphics[width=.33\textwidth]{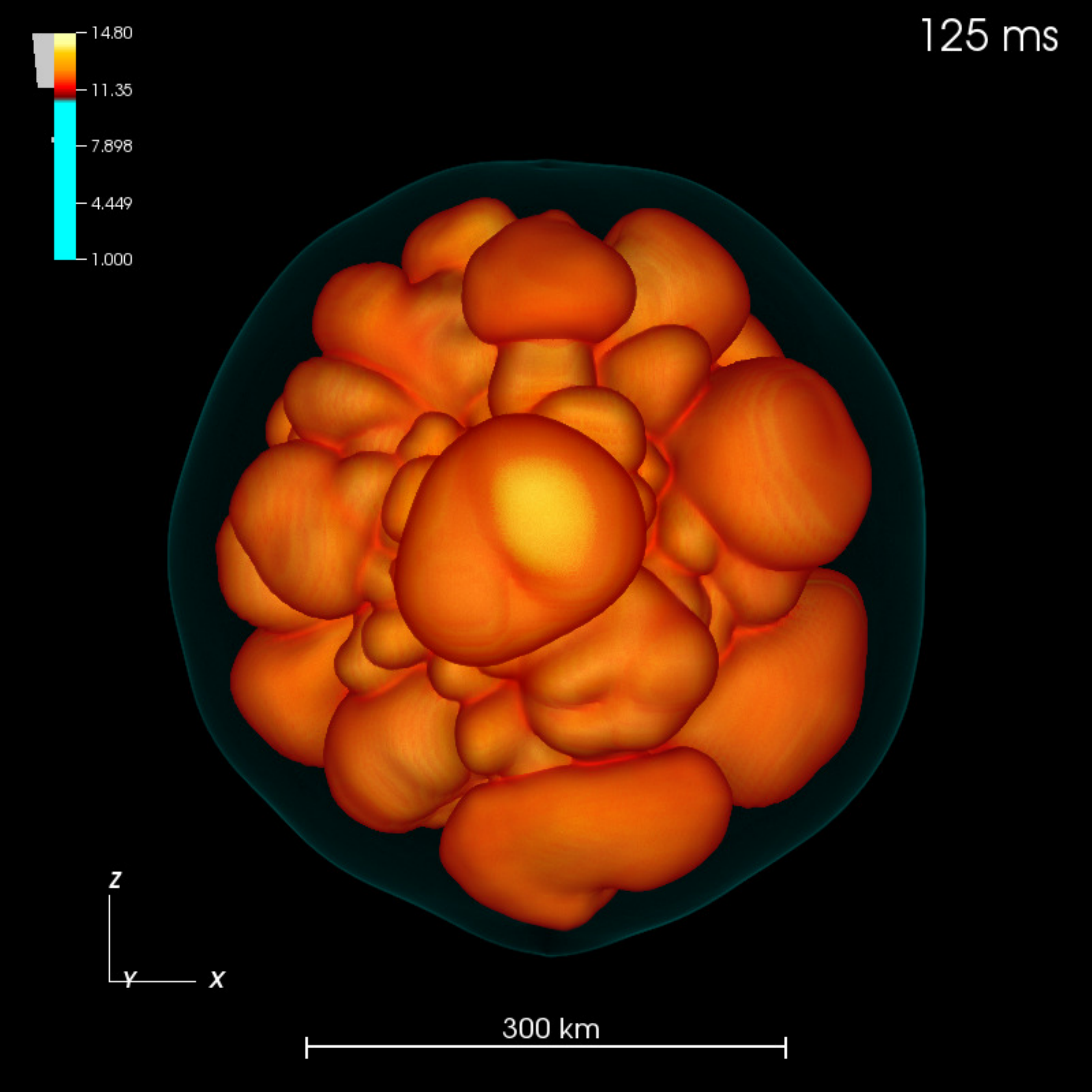}\hfil
\includegraphics[width=.33\textwidth]{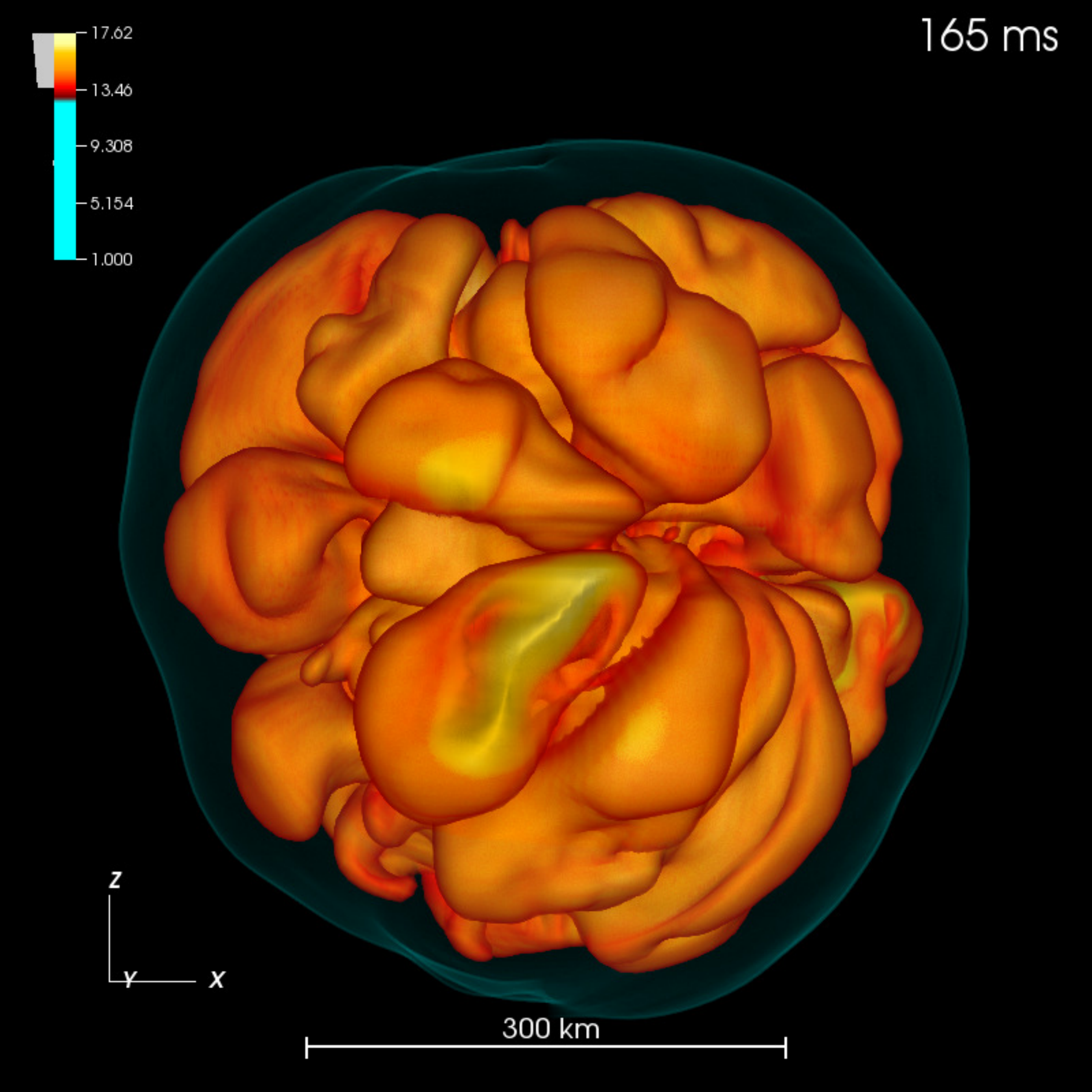}\hfil
\includegraphics[width=.33\textwidth]{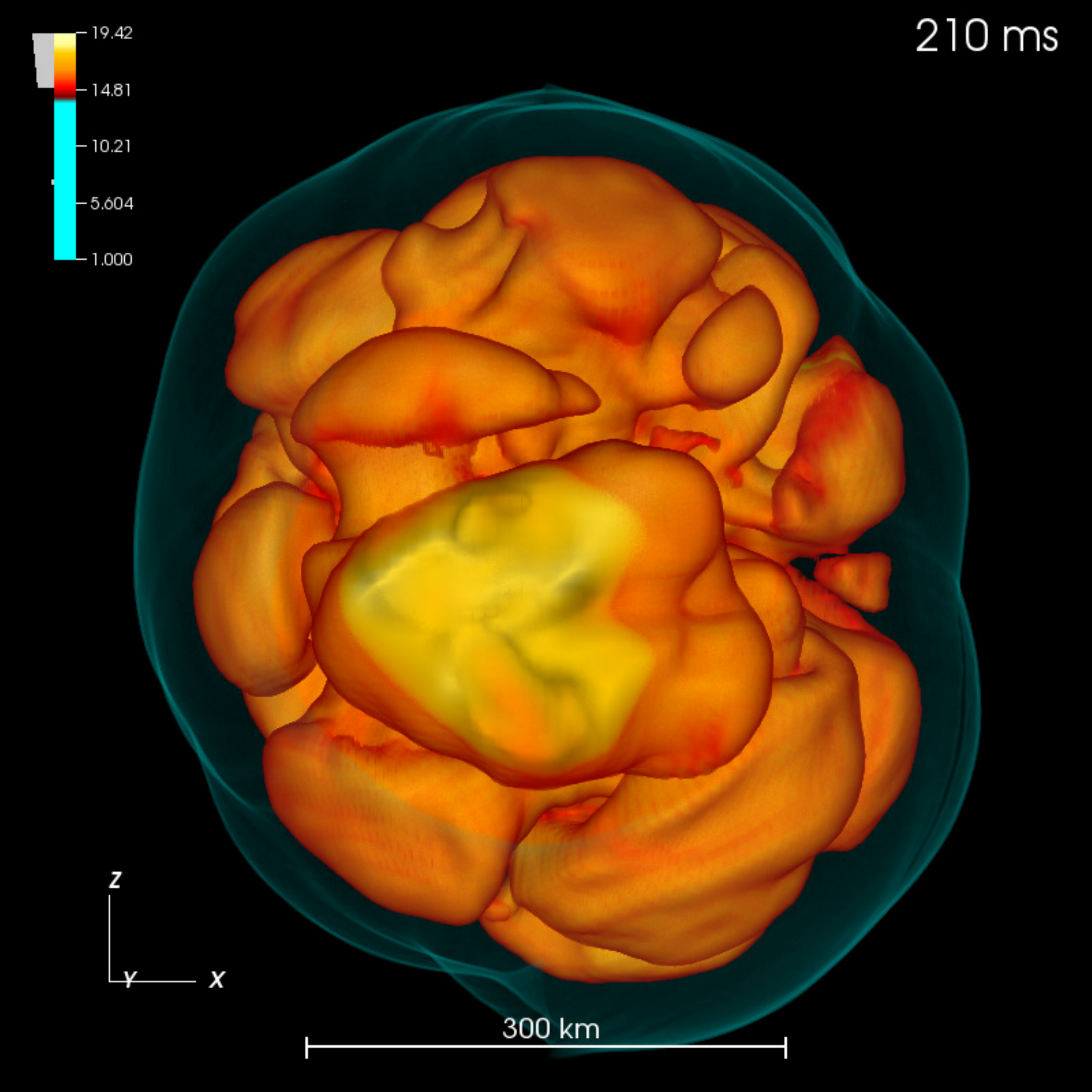}}
\vskip1.5pt
\hbox to\textwidth{\includegraphics[width=.33\textwidth]{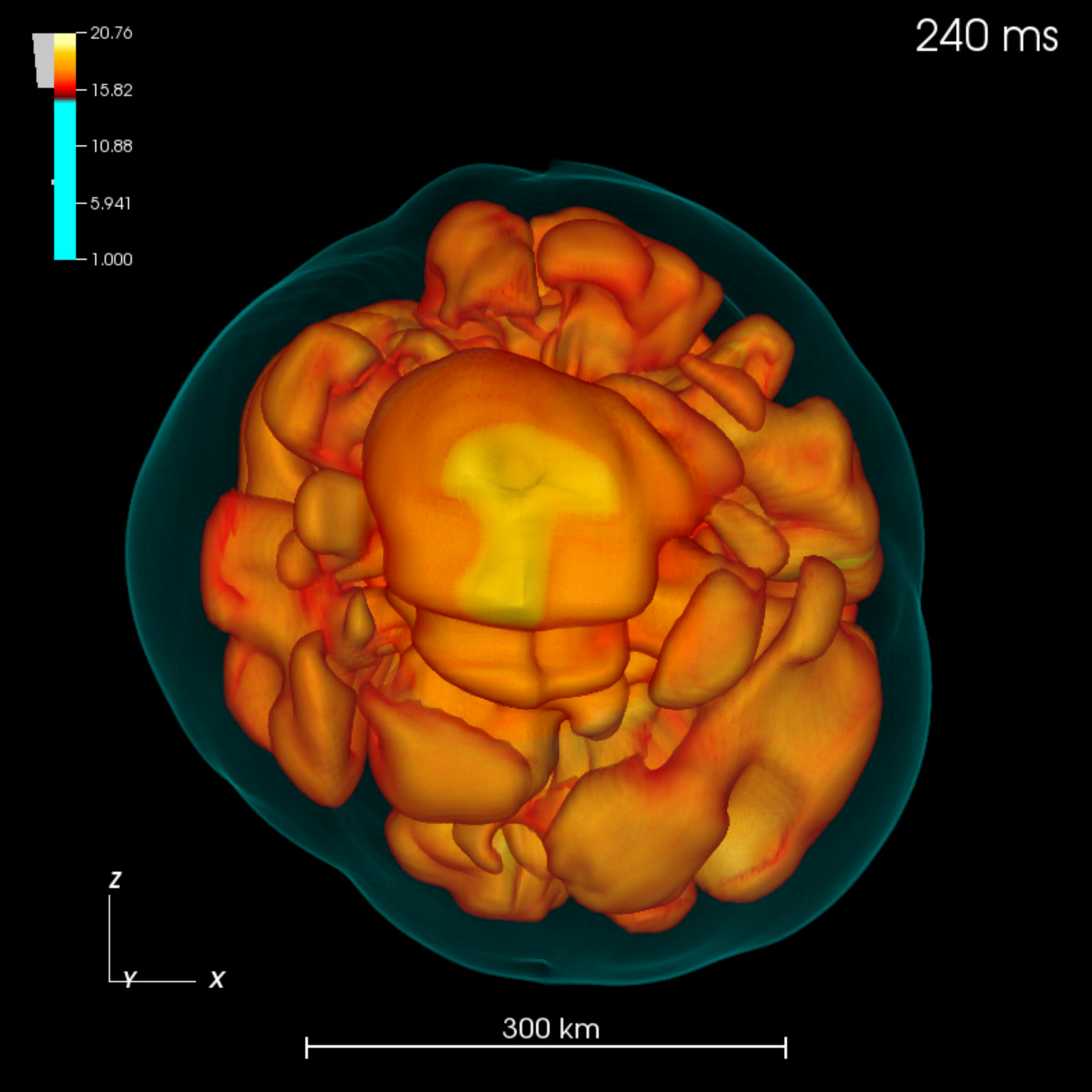}\hfil
\includegraphics[width=.33\textwidth]{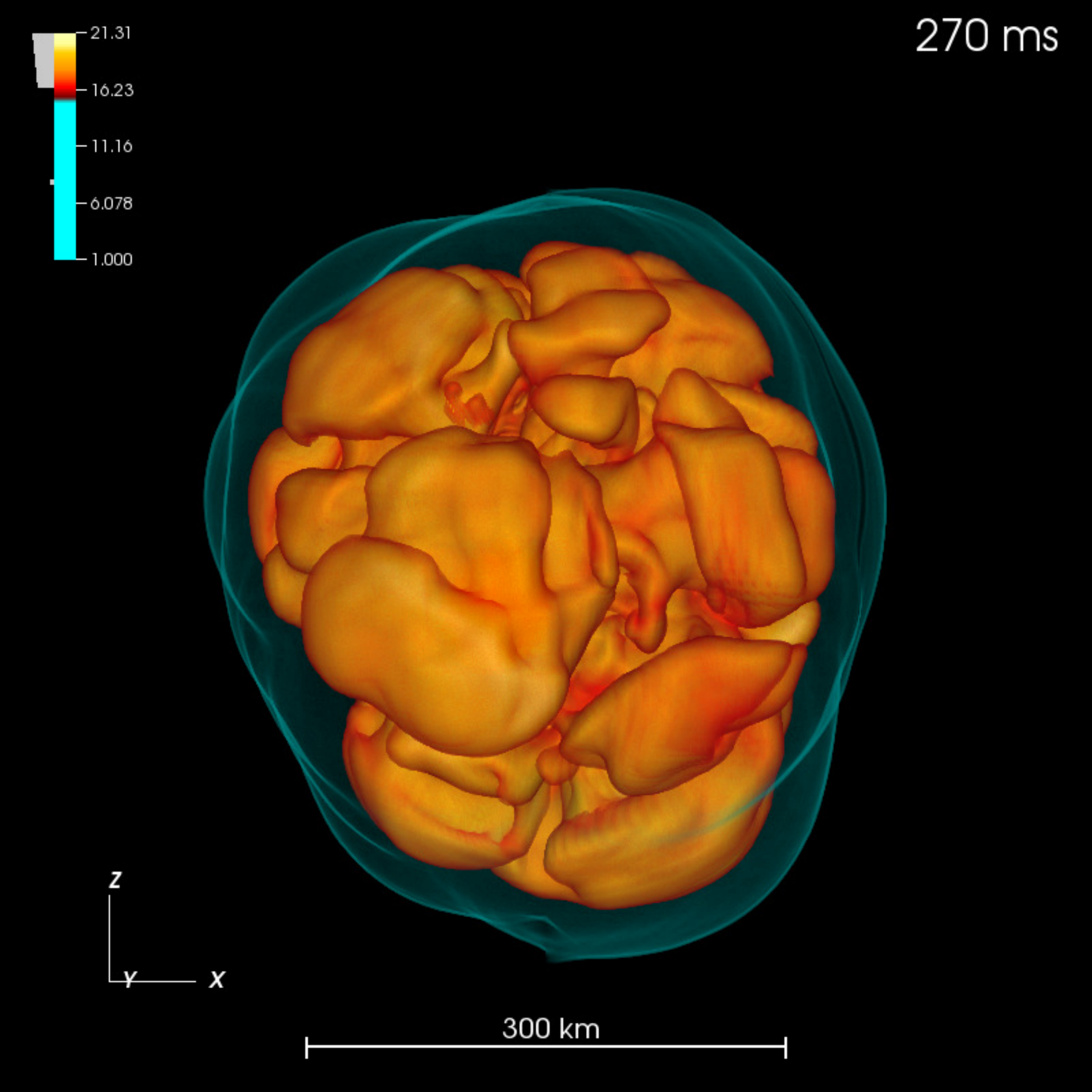}\hfil
\includegraphics[width=.33\textwidth]{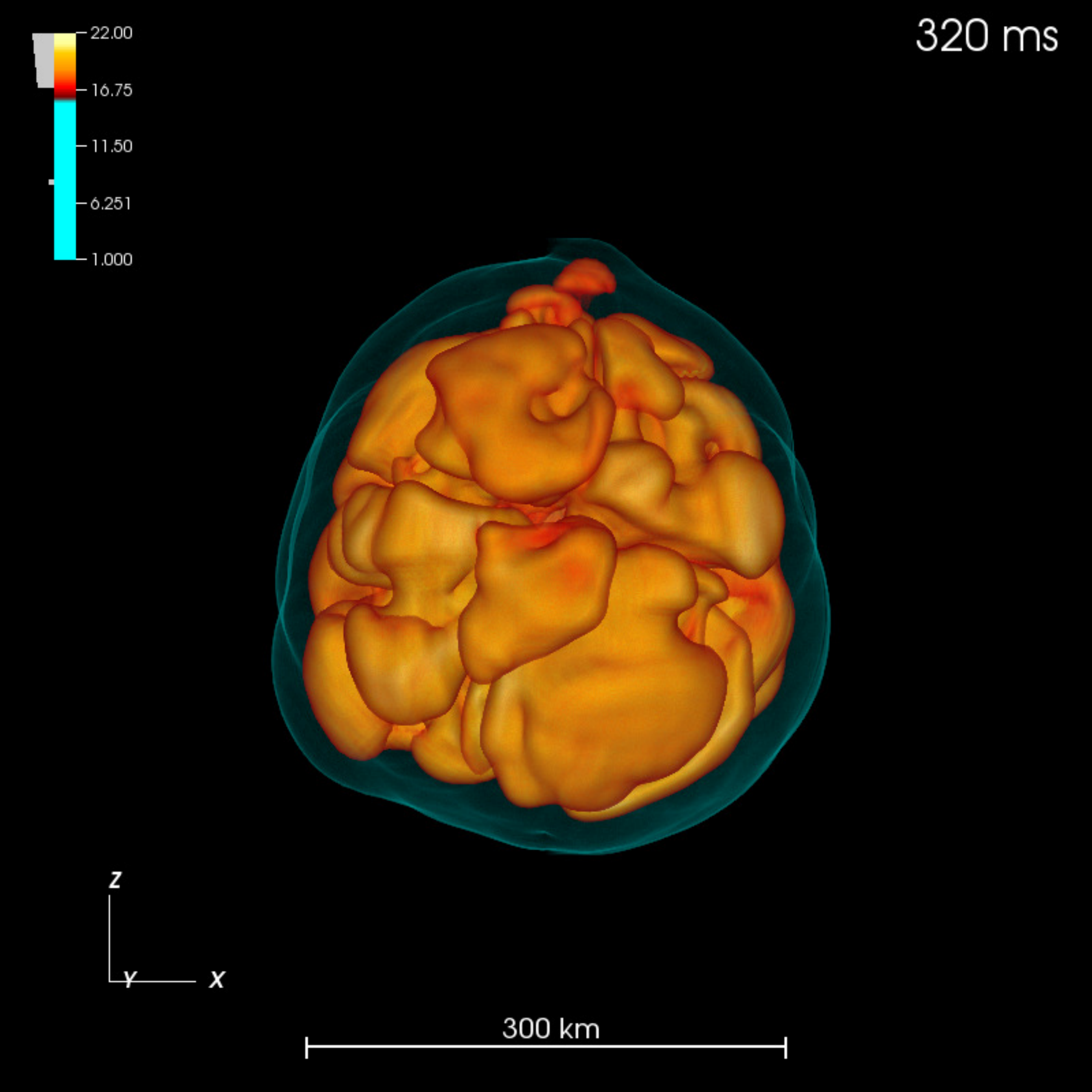}}
\caption{Snapshots of the convective overturn activity during the accretion
phase of the 11.2\,$M_\odot$ model at the indicated p.b.\ times.
Visualized are surfaces of constant entropy: The bluish, semi-transparent
envelope is the SN shock, the red-orange-yellow surfaces are entropy structures
in the postshock region. Neutrino-heated high-entropy matter  expands in
typical mushroom-like, buoyant plumes. These features are  highly time
dependent with bubbles continuously appearing, expanding and  rising, and
disappearing to give way to a new generation of convective  plumes. The
direction of strongest lepton-number emission, i.e., the largest excess of the
radiation of $\nu_e$ compared to $\bar\nu_e$ (see text), points  towards the
observer, above the $x$-$y$-plane at roughly 45$^\circ$  inclination to the
$z$-axis. A corresponding global, persistent dipolar deformation of the
accretion shock is present but
can hardly be recognized without a detailed analysis.} \label{fig:3dconv11}
\end{center}
\end{figure*}
%------------------------------------------------------------------------------

%%%%%%%%%%%%%%%%%%%%%%%%%%%%%%%%%%%%%%%%%%%%%%%%%%%%%%%%%%%%%%%%%%%%%%%%%%%%%%%
\section{Numerical 3D Models}
\label{sec:numerics}
%%%%%%%%%%%%%%%%%%%%%%%%%%%%%%%%%%%%%%%%%%%%%%%%%%%%%%%%%%%%%%%%%%%%%%%%%%%%%%%

The calculations of our 3D models were performed with the elaborate
neutrino-hydrodynamics code \textsc{Prometheus-Vertex}. This SN simulation
tool combines the hydrodynamics solver \textsc{Prometheus}
\citep{fryxell_89}, which is a dimensionally-split implementation of the
piecewise parabolic method (PPM) of \citet{colella_84}, with the neutrino
transport module \textsc{Vertex} \citep{rampp_02}. \textsc{Vertex} solves the
fully energy-dependent moment equations for the neutrino energy and momentum
densities, with ${\cal O}(v/c)$ velocity dependence, for spherically symmetric
transport problems defined to be associated with every angular bin of the
polar grid (``radial rays'') used for the multi-dimensional simulations. The
moment equations are closed by a variable Eddington factor relation that is
provided by the formal solution of a model Boltzmann equation. An up-to-date
set of neutrino interaction rates is applied in \textsc{Vertex} (see, e.g.,
\citealt{mueller_12}). In the multi-dimensional case, our ``ray-by-ray-plus''
approach \citep{buras_06a} includes non-radial neutrino advection and
pressure terms in addition to the radial transport solves. The ray-by-ray
approximation implies that we assume the neutrino radiation field to be
axially symmetric around the radial direction and thus ignore nonradial
components of the neutrino flux. In the simulations presented here, we
adopted monopolar gravity but included general relativistic corrections by
means of an effective gravitational potential \citep{marek_06}.

We have performed 3D simulations for the evolution of the 11.2\,$M_\odot$ and
27\,$M_\odot$ progenitors of \citet{woosley_02} and the 20\,$M_\odot$ model
of \citet{Woosley_2007}, using the high-density equation of state (EoS) of
\citet{lattimer_91} with a nuclear incompressibility of $K=220$\,MeV. The
11.2\,$M_\odot$ and 27\,$M_\odot$ stars had been previously investigated in
2D by \citet{buras_06b}, \citet{marek_09}, \citet{mueller_12}, and
\citet{mueller_12b}. Our 3D models were computed on a spherical polar grid
with an initial resolution of $n_r \times n_\theta \times n_\phi = 400 \times
88 \times 176$ zones. Later, refinements of the radial grid ensured adequate
resolution in the PNS surface region. The innermost 10\,km were treated in
spherical symmetry to avoid excessive time-step limitations near the polar
grid axis. Doing so we took special care to ensure that the convectively
unstable layer below the neutrinosphere and the corresponding undershooting
region were fully covered by the 3D grid during the entire simulations. Seed
perturbations for aspherical instabilities were imposed by hand 10\,ms after
core bounce by introducing random perturbations of 0.1\% in density on the
entire computational grid. None of these models led to successful explosions
during the simulation periods of 350\,ms for the 11.2\,$M_\odot$ model,
420\,ms for the $20\,M_\odot$ progenitor, and 575\,ms for the $27\,M_\odot$
case.

The postbounce hydrodynamics of the $27\,M_\odot$ model, in particular the
prominent presence of SASI sloshing and spiral modes, was described in detail
in a previous paper \citep{Hanke_2013}. Basic properties of the neutrino
signal and its detection were subject of a recent paper by
\citet{Tamborra_2013},
highlighting the large-amplitude, quasi-periodic modulations of
the radiated luminosities and mean energies associated with the SASI
activity. In addition, information about the expected neutrino signal from
the $20\,M_\odot$ and $11.2\,M_\odot$ runs was shown. More details on
the neutrino emission of all three 3D simulations and implications for the
direction dependent detectability of SN neutrino signals will be discussed in
\citet{Tamborra_2014}.

The 27\,$M_\odot$ and the 20\,$M_\odot$ models both show periods of strong
SASI activity. In the former case, which was simulated until 575\,ms
p.b., a first phase of violent SASI occurs between $\sim$170\,ms
and $\sim$260\,ms \citep{Hanke_2013}, and a second SASI episode sets in
at $\sim$420\,ms after an intermediate period of essentially pure
convective overturn. In the 20\,$M_\odot$ case strong SASI mass motions
take place from $\sim$170\,ms until $\sim$305\,ms, and SASI
reappears after 400\,ms, shortly before the simulation run was stopped.
On the other hand, the 11.2\,$M_\odot$ model does not
exhibit any clear evidence of SASI motions but develops the typical
signatures of postshock convective overturn in the neutrino-heating layer as
evident from our Fig.~\ref{fig:3dconv11}, to be compared with the
27\,$M_\odot$ case in Fig.~1 of \citet{Hanke_2013}. In the 11.2\,$M_\odot$
model, the short-timescale neutrino-emission variations are distinctly
smaller than in the SASI-active models \citep{Tamborra_2013,Tamborra_2014}.

In the 11.2\,$M_\odot$ simulation, first indications of postshock convection
become visible at about 80\,ms p.b., shortly after a gain region below the
stalled shock has developed.  Buoyant, mushroom-like plumes appear, which are
initially small and then successively replaced by larger ones. At 100\,ms
p.b., first evidence of shock deformation occurs, and after about 140\,ms,
convective overturn has attained its full strength with a ratio of maximum to
minimum shock radius of up to $R_\mathrm{s,max}/R_\mathrm{s,min} \sim 1.35$.
The expansion of the accretion shock continues until about 210\,ms p.b., when
the average shock radius reaches a maximum of $\sim$260\,km (see
Fig.~\ref{fig:accretionasymmetry} below). It is followed by a slow but monotonic
recession of the average shock radius to only 150\,km at the end of the
simulation at 350\,ms. No explosion has set in until this stage, in contrast
to the corresponding 2D calculation with the same microphysics, same
numerical treatment, and in particular the same radial and angular grid
resolution. In the 2D case, the shock continues to expand, supported by
large-amplitude shock-sloshing motions along the symmetry axis. More and more
favorable conditions for an explosion develop until finally, at roughly
350\,ms p.b., the shock accelerates and triggers an outgoing blast wave,
whereas the 3D case at this time shows little promise of an explosion.

Apparently, the 3D setup with the chosen angular resolution (limited by the
requirements of computational resources, which are prohibitive for our
sophisticated treatment of neutrino transport) is less beneficial for the
possibility of a SN explosion by the neutrino-driven mechanism. This finding
is in line with recent investigations based on cruder treatments of neutrino
physics, namely a neutrino-light bulb description with simple heating and
cooling terms \citep{Hanke_2012,Couch_2013}, ray-by-ray neutrino trapping
with a parametrized heating strength \citep{Couch_2013a}, and a
ray-by-ray implementation of the isotropic diffusion source approximation
\citep{Takiwaki_2013}. However, the difference between 2D and 3D models is
not subject of our present discussion and we next turn to the new phenomenon
of asymmetric lepton-number emission.

%------------------------------------------------------------------------------
\begin{figure*}[t!]
\begin{center}
  \includegraphics[width=\textwidth]{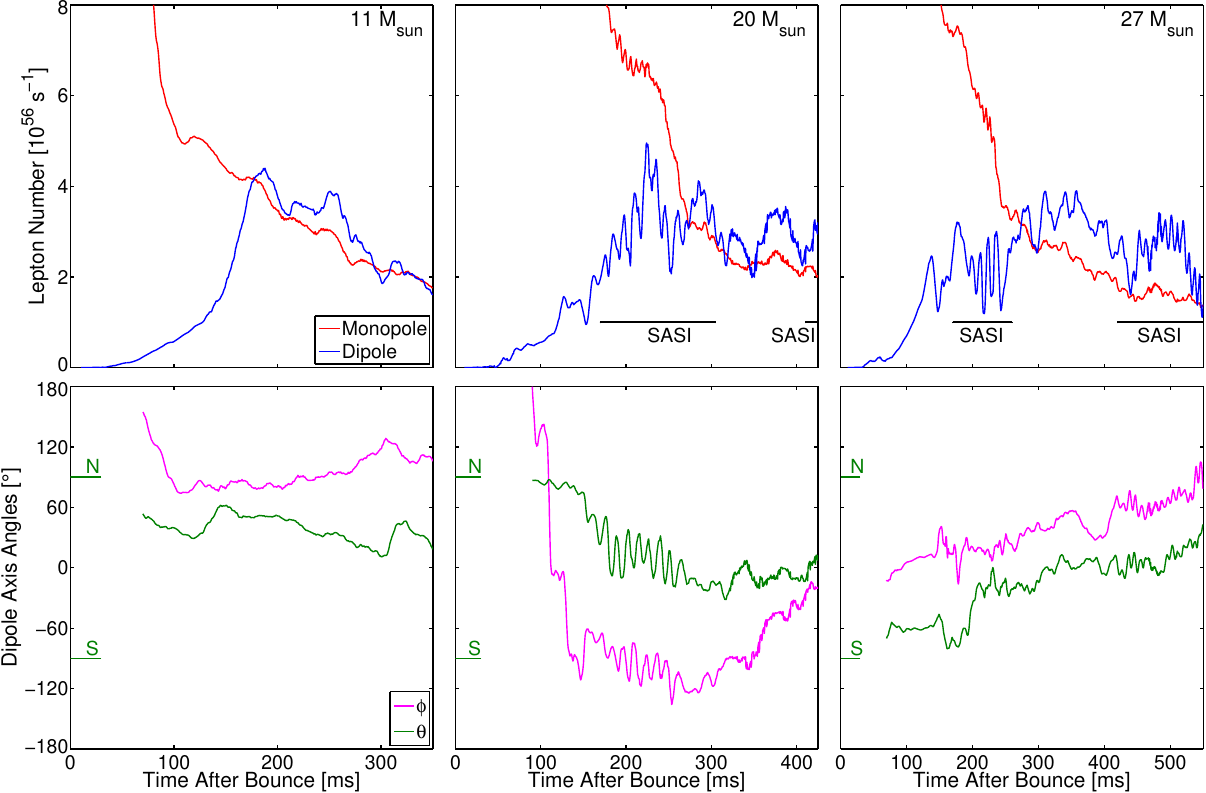}
\caption{Time evolution of the
  lepton-number emission ($\nu_e$ minus $\bar\nu_e$) for the
  11.2, 20 and $27\,M_\odot$ models as labelled. For each model,
  the upper panels show the overall lepton number flux
  (monopole of the angular distribution; red curve) and its dipole component
  (blue curve). Episodes of dominant SASI activity are marked by horizontal
  bars. The lower panels display the zenith angle $\theta$
  (green line) and the azimuth angle $\phi$ (magenta line)
  of the dipole direction, which describes the
  track shown for the $11.2\,M_\odot$ case in Fig.~\ref{fig:leptonskyplots}.
  For the zenith angle we indicate
  the north- and south-polar grid directions at $\pm90^\circ$ on the vertical
  axis. The monopole evolution depends strongly on the accretion rate and
  varies between the models,
  whereas the maximum dipole amplitude is similar in all
  cases and shows a similar initial growth phase.
  The dipole persists (and can even grow) during
  the indicated phases of pronounced SASI activity. The dipole directions are
  different in all cases, bear no correlation to the numerical grid, and they
  drift only slowly even during SASI phases.}
\label{fig:dipole11}
\end{center}
\end{figure*}
%------------------------------------------------------------------------------

%\newpage

%%%%%%%%%%%%%%%%%%%%%%%%%%%%%%%%%%%%%%%%%%%%%%%%%%%%%%%%%%%%%%%%%%%%%%%%%%%%%%%%
\section{Persistent Dipole Asymmetries of 3D Models}
\label{sec:results}
%%%%%%%%%%%%%%%%%%%%%%%%%%%%%%%%%%%%%%%%%%%%%%%%%%%%%%%%%%%%%%%%%%%%%%%%%%%%%%%%

\subsection{Evolution of lepton-number emission dipole}
\label{sec:dipole-time}

We have recently explored the flavor-dependent neutrino emission of our 3D
simulations to forecast possible signatures of hydrodynamical instabilities
in large-scale neutrino detectors \citep{Tamborra_2013} and as a
prerequisite for flavor oscillation studies. A systematic analysis has
revealed a long-lasting, slowly evolving dipole asymmetry of the
lepton-number ($\nu_e$ minus $\bar\nu_e$) emission from the newly formed NS.
In Fig.~\ref{fig:leptonskyplots} we have shown typical directional
distributions of the lepton-number flux for our $11.2\,M_\odot$ model. This
pronounced asymmetry builds up in parallel to the development of large-scale
convective overturn behind the stalled shock and shows a fairly stable
direction, which has no particular correlation with the numerical coordinate
grid\footnote{The orientation of the coordinate system in our sky-plots of
Figs.~\ref{fig:leptonskyplots}, \ref{fig:skymaps11}, \ref{fig:skymaps27},
and \ref{fig:lesasasisky}
is such that the north-south direction corresponds to the $z$-axis of the
numerical grid, the center of the plot is the $-x$ direction, and the left
and right extreme points correspond to the $+x$ direction. The half-way
points on the equator belong to the $+y$ (left) and $-y$ directions.}.

Before attempting a physical interpretation of this puzzling phenomenon, we
first collect a number of conspicuous phenomenological manifestations. A
natural first question is to see when and how this effect builds up in the
course of postbounce core-collapse evolution and if it is correlated with
other symmetry-breaking hydrodynamical instabilities.

To quantify the time evolution of our new effect we consider the lowest-order
multipole components of the lepton-number flux as a function of emission
direction. To clarify our normalization of the dipole component we note that
if the lepton-number flux distribution contains only a monopole and dipole
term, then the distribution is \hbox{$A_{\rm Monopole}+A_{\rm Dipole}
\cos\vartheta$} in coordinates aligned with the dipole direction. When the
ratio of these amplitudes is unity, the distribution is proportional to
$1+\cos\vartheta$ and the lepton-number flux vanishes in the direction of
minimal flux and is twice the average in the direction of maximal flux,
corresponding roughly to what we see in Fig.~\ref{fig:leptonskyplots}.
$A_{\rm Monopole}$ is nothing but the total rate of lepton number emitted by
the evolving PNS, whereas $A_{\rm Dipole}$ is 3 times the projection of the
total lepton-number flux onto the dipole direction.

Figure~\ref{fig:dipole11} shows the evolution of $A_{\rm Monopole}$ and
$A_{\rm Dipole}$ and the dipole direction for our three progenitor models.
The total lepton-number emission is at first off-scale, corresponding to the
usual prompt $\nu_e$ burst, and then decreases monotonically with small
modulations caused by large-scale convection and concomitant variations of
the postshock accretion flow. The overall lepton-number emission is fed by the
mass-accretion flow so that it is not surprising that the monopole strength
depends considerably on the progenitor model.

In all models, a dipole component becomes first discernible at about 50\,ms
p.b., grows for 100--150\,ms, and later begins to decrease, more or less in
parallel with the overall decline of the lepton-number emission. The
dipole decay is not monotonic and has an overall gradient that is different
in the different progenitors (steepest in the 11.2\,$M_\odot$ model).
In this later
phase, the dipole amplitude sometimes exceeds the monopole, meaning that in
the negative dipole direction, the lepton-number flux is somewhat negative
(excess of $\bar\nu_e$ over $\nu_e$ emission). The overall dipole strength is
similar in all three progenitor models, at the peak reaching a value of
$\sim$3--$4\times10^{56}\,{\rm s}^{-1}$.

While in all three models the dipole growth sets in and reaches large
amplitudes during an early, convection-dominated postbounce phase, before
SASI activity starts in the postshock layer, it is remarkable that SASI seems
to have different consequences for the LESA evolution in the 20\,$M_\odot$
and 27\,$M_\odot$ models. In the former case the lepton-number
emission dipole continues to grow even after the onset of SASI and reaches
its full strength during the SASI-dominated phase. In contrast,
in the 27\,$M_\odot$ model the onset of SASI seems to prevent a further
growth of the dipole amplitude, which instead performs quasi-periodic
oscillations around a stable level somewhat below the maximum dipole
amplitudes in the 11.2 and 20\,$M_\odot$ models.
The dipole continues to develop to its
peak amplitude only after the SASI episode has ended and convective
overturn becomes the dominant hydrodynamic instability in the postshock
accretion layer again.
We will come back to this interesting difference later in
our discussion of the physical elements and underlying physical
mechanism of LESA (see Sect.~\ref{sec:modelswithsasi}).

The dipole orientation on the computational polar grid is expressed by
latitudinal and azimuthal polar angles $\theta$ and $\phi$, respectively.
The zenith angle, $\theta$, varies from $\pi/2$ (north pole of
numerical grid) to $-\pi/2$ (south pole), whereas $-\pi\leq\phi\leq\pi$ is
measured relative to the $x$-axis of the grid.
The dipole direction slowly drifts, but remains fairly stable once the dipole
is well developed. This conclusion also follows from the dipole's sky-track
shown as a dark-gray line in Fig.~\ref{fig:leptonskyplots} for the
$11.2\,M_\odot$ case. It is remarkable that this behavior is quite similar in
the higher-mass (20 and 27\,$M_\odot$)
progenitors, where phases with dominant large-scale convection alternate
with phases of pronounced SASI activity, which are indicated in
Fig.~\ref{fig:dipole11} by horizontal bars. The SASI action is clearly
visible in the form of small, periodic modulations of the overall
deleptonization flow and much larger modulations of the dipole strength
and direction with typical oscillation periods of $\sim$10--20\,ms.
Yet, even these modulations are only relatively small variations of the
dipole strength and its orientation in space, not changing the
hemispheric location of the whole phenomenon. The initial, large
movement of the azimuthal angle from $+$180$^\circ$ to about $-$90$^\circ$
that happens between about 90 and 110\,ms in the case of the 20\,$M_\odot$
model (Fig.~\ref{fig:dipole11}) occurs in a phase when the dipole is
still building up and its amplitude is correspondingly small.
Therefore the dipole direction is not particularly meaningful and
the step-like jumps of its azimuth angle simply reflect stochastical
fluctuations. At later times the
dipole orientations in all three models exhibit only very slow drifts.

The lepton-emission dipole is a long-lasting, persistent, and only
slowly evolving phenomenon. Its dipole direction remains in the
same hemisphere for periods of hundreds of milliseconds, i.e.,
for nearly the whole computed postbounce evolution. A slow overall
drift of the dipole direction (mainly in the 20 and 27\,$M_\odot$ models)
happens on timescales much longer than the typical dynamical timescales
in the convective layers inside the PNS and between the gain radius and
stalled shock. With
convective velocities of up to a few $10^8$\,cm\,s$^{-1}$ in the former
case and up to more than $10^9$\,cm\,s$^{-1}$ in the latter, convective
overturn timescales are typically milliseconds in the PNS and around
$\sim$10\,ms in the neutrino-heating region. These timescales depend on
the radial diameter of the PNS convection zone (roughly 10\,km,
which also determines the angular scale of $\sim$20--40$^\circ$ of
convective cells there) and on the shock stagnation radius,
respectively. The dynamical timescales of the postshock region
are reflected by the SASI-imprinted modulations of dipole strength
and spatial orientation visible in Fig.~\ref{fig:dipole11}.
Convective activity in both regions is highly time dependent and
nonstationary, and individual convective cells and buoyant bubbles have
life times that are usually not much longer than one or two overturns
(for PNS convection this was discussed by \citealt{Keil_1996,Dessart_2006}).
It is therefore astonishing that despite such an extreme time variability
of the environment a hemispheric emission asymmetry can survive over many
tens of dynamical periods. Only
over long, secular timescales SASI or convectively induced changes
of the system seem to lead to a gradual, slow drift of the dipole
orientation as seen in all models but especially in those with
episodes of vigorous SASI activity (Fig.~\ref{fig:dipole11},
panels in the middle and right columns).

These simple observations already suggest that the LESA phenomenon must
physically depend on a complicated interplay of different effects. The
initial growth over 100--150\,ms parallels the growth of large-scale
convection in the gain region, suggesting gain-region convection as the
primary engine. On the other hand, the persistence throughout SASI episodes
and the near-universal dipole strength suggest that LESA must also be
anchored to deeper regions. We will see that indeed it originates in the PNS
convection region deep below the neutrinospheres.

\subsection{Overall neutrino emission properties}
\label{sec:neutrino-emission-properties}

\begin{figure}
\begin{center}
  \includegraphics[width=.85\columnwidth]{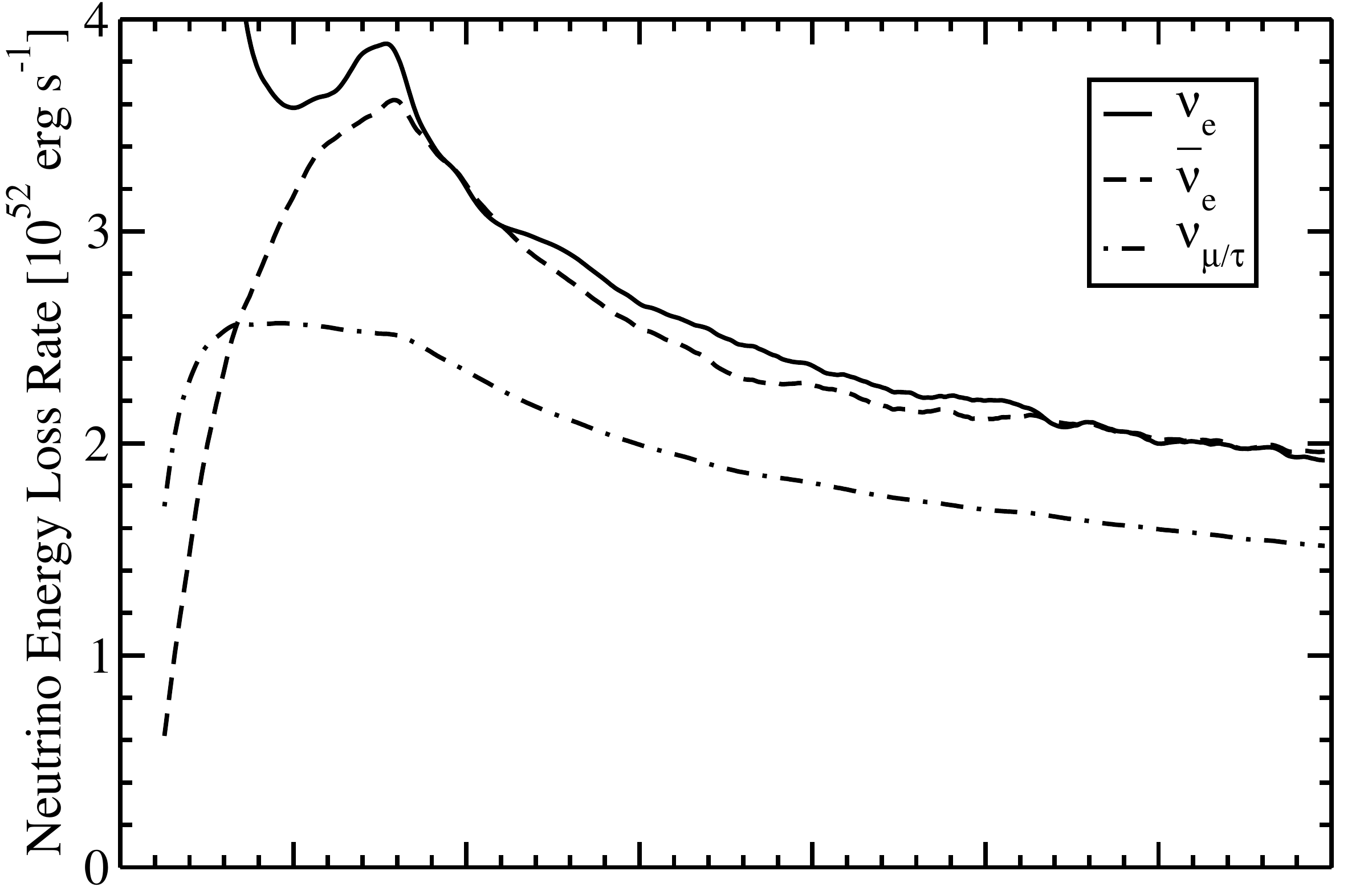}
  \includegraphics[width=.85\columnwidth]{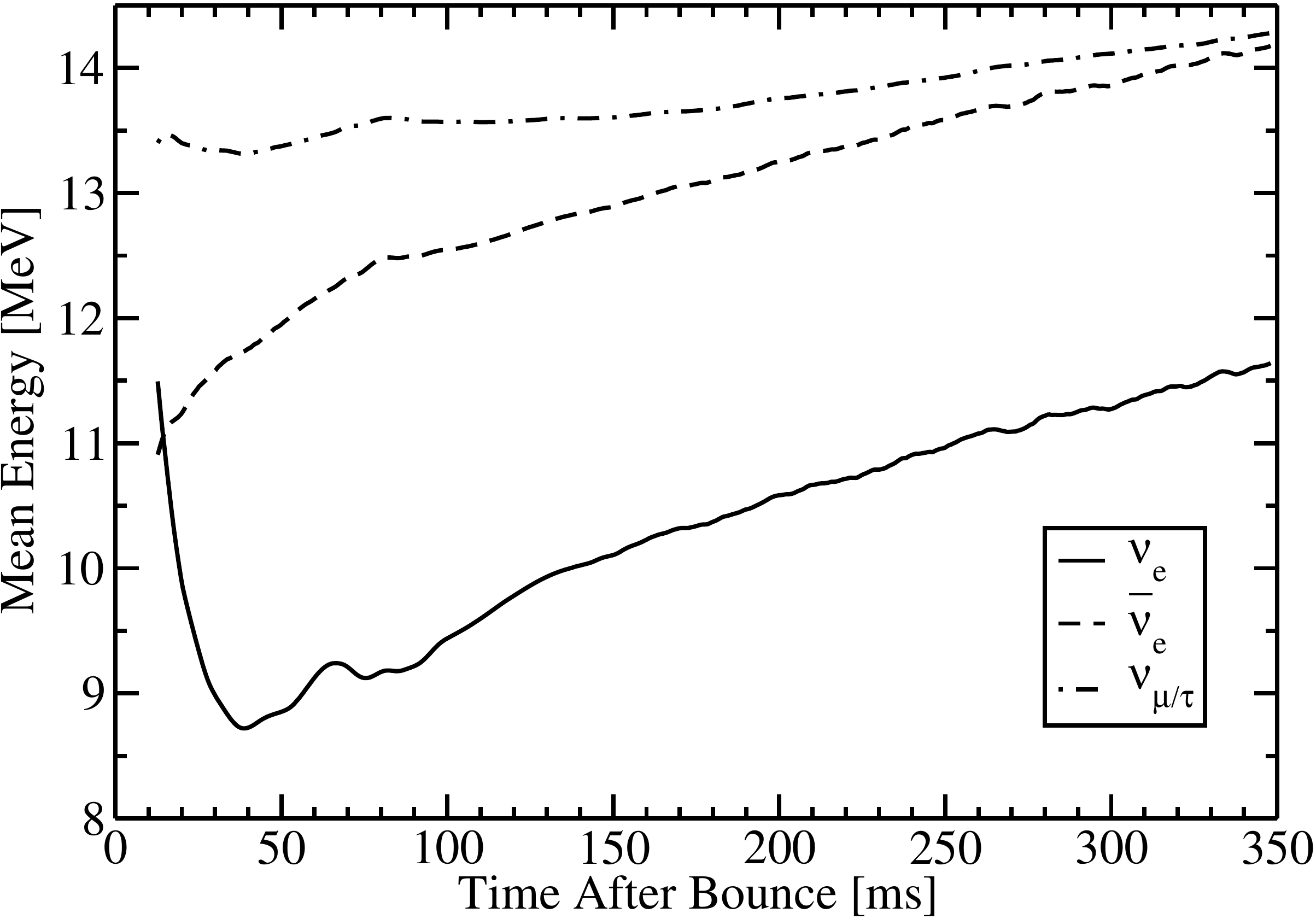}
\caption{Time evolution of spherically averaged
neutrino emission properties (laboratory frame)
for the 11.2\,$M_\odot$
model. {\em Top:} Energy loss rates, integrated over
all directions, for $\nu_e$, $\bar\nu_e$, and (one kind of)
$\nu_x$. {\em Bottom:} Mean energies, averaged
over all directions.
These results do not show any peculiarities and
are similar to comparable 1D and 2D simulations.
\label{fig:neutrinoemission}}
\end{center}
\end{figure}

Before turning to physical interpretations, however, we first continue with
our description of phenomenological observations in our numerical models. In
particular, one may wonder if the overall neutrino emission parameters
exhibit any peculiarities, but this is not the case. We specifically show in
Fig.~\ref{fig:neutrinoemission} the evolution of the overall energy-loss rate
for the $11.2\,M_\odot$ model in the three species $\nu_e$, $\bar\nu_e$, and
$\nu_x$, i.e., we display the $4\pi$-integrated energy fluxes,
\begin{equation}
\dot E_\nu(t)\equiv\int_{4\pi}\mathrm{d}\Omega\,R^2
F_\mathrm{e}(\textit{\textbf{R}},t)\,,
\label{eq:edot}
\end{equation}
where $F_\mathrm{e}(\textit{\textbf{R}},t)$ is the ray-by-ray computed
energy-flux density at a point $\textit{\textbf{R}}$ of a chosen sphere
with radius $R = |\textit{\textbf{R}}|$. The quantity $\dot E_\nu(t)$
is usually called ``luminosity'' but it is not measurable in
the multi-dimensional case by observers at any location. The bottom panel
of Fig.~\ref{fig:neutrinoemission}
shows the corresponding $4\pi$-averaged mean energies,
$\langle\epsilon_{\nu_i}\rangle$, which are defined as
the ratios of energy-loss rates to number-loss rates.

This figure reveals the usual behavior during the postbounce accretion phase
(compare, e.g., with results by \citealt{marek_09,Marek_2009,Janka_2012}).
$\dot E_{\nu_e}$ after the shock-breakout burst is very close to $\dot
E_{\bar\nu_e}$ or even slightly larger, while
$\langle\epsilon_{\bar\nu_e}\rangle$ exceeds $\langle\epsilon_{\nu_e}\rangle$
by a few MeV. Heavy-lepton neutrinos have significantly lower individual
luminosities because their production in the accretion layer of the PNS is
less efficient due to the lack of charged-current processes, and
$\langle\epsilon_{\nu_x}\rangle$ is only slightly larger than
$\langle\epsilon_{\bar\nu_e}\rangle$ because energy transfers in
neutrino-nucleon scatterings reduce the high-energy spectrum of $\nu_x$
diffusing outward from their deeper production layers
\citep{Raffelt_2001,Keil_2003}.

Here as well as in the following discussion we mostly focus on the
11.2\,$M_\odot$ star. This has two reasons. On the one hand,
the 11.2\,$M_\odot$ model does not possess the violent SASI episodes
which massively affect the neutrino-emission properties in the 20 and
27\,$M_\odot$ cases \citep{Tamborra_2013,Tamborra_2014,Hanke_2013},
where they lead
to time-dependent variations of the neutrino transport and radiation,
superimposed on the hemispheric asymmetry of the lepton-number emission dipole
(Fig.~\ref{fig:dipole11}). Such short-time fluctuations can hamper
the easy visibility of the LESA-specific features.
On the other hand, outside of the SASI episodes
diagnostic quantitities that we evaluate for the lepton-emission
dipole in the 20 and 27\,$M_\odot$ models look, qualitatively and
quantitatively, very similar to those that we present in more detail
for the 11.2\,$M_\odot$ case. This will be shown in
Fig.~\ref{fig:accretionasymmetry}.

%-------------------------------------------------------------------------------
%-------------------------------------------------------------------------------

\newpage

%-------------------------------------------------------------------------------
\begin{figure}
\begin{center}
  \includegraphics[width=0.85\columnwidth]{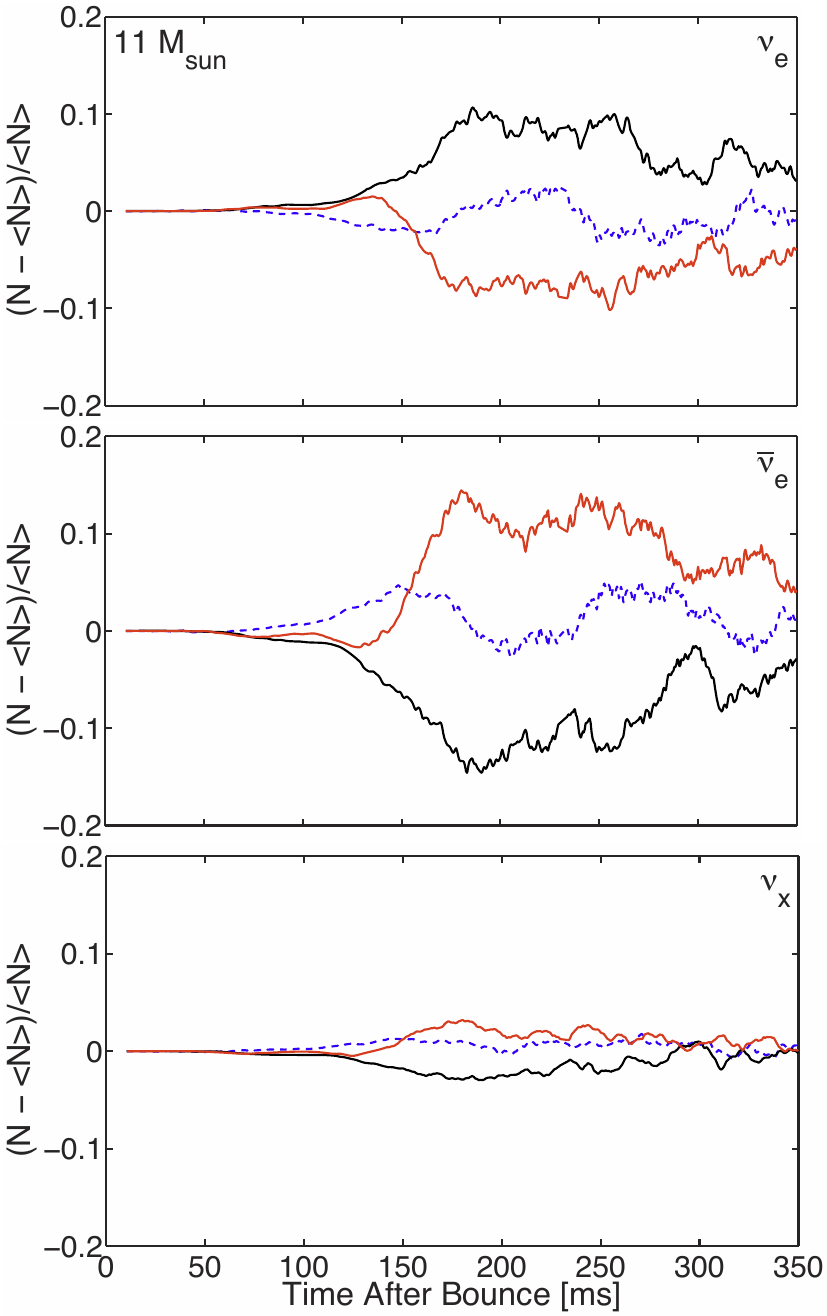}
  \caption{Time evolution of the $\nu_e$, $\bar\nu_e$, and heavy-lepton
  neutrino ($\nu_x$) number fluxes (top to bottom) relative to their
  directional averages for the 11.2\,$M_\odot$ model. We show hemispheric
  averages (accounting for projection effects; see text) as seen by
  distant observers who are
  located approximately in the direction of maximal lepton-number
  emission (black), in the opposite direction (red)
  and in a representative transverse direction (dashed blue).
  The $\nu_e$ and $\bar\nu_e$ fluxes each exhibit a strong dipolar asymmetry,
  (anti-)aligned with the lepton-number flux dipole, whereas the $\nu_x$ flux
  is nearly isotropic except for a small enhancement in the hemisphere of
  smallest lepton-number flux (red line).
\label{fig:fluxvariations}}
\end{center}
\end{figure}
%-------------------------------------------------------------------------------

%------------------------------------------------------------------------------------------
\begin{figure*}
\begin{center}
  \includegraphics[width=1\textwidth]{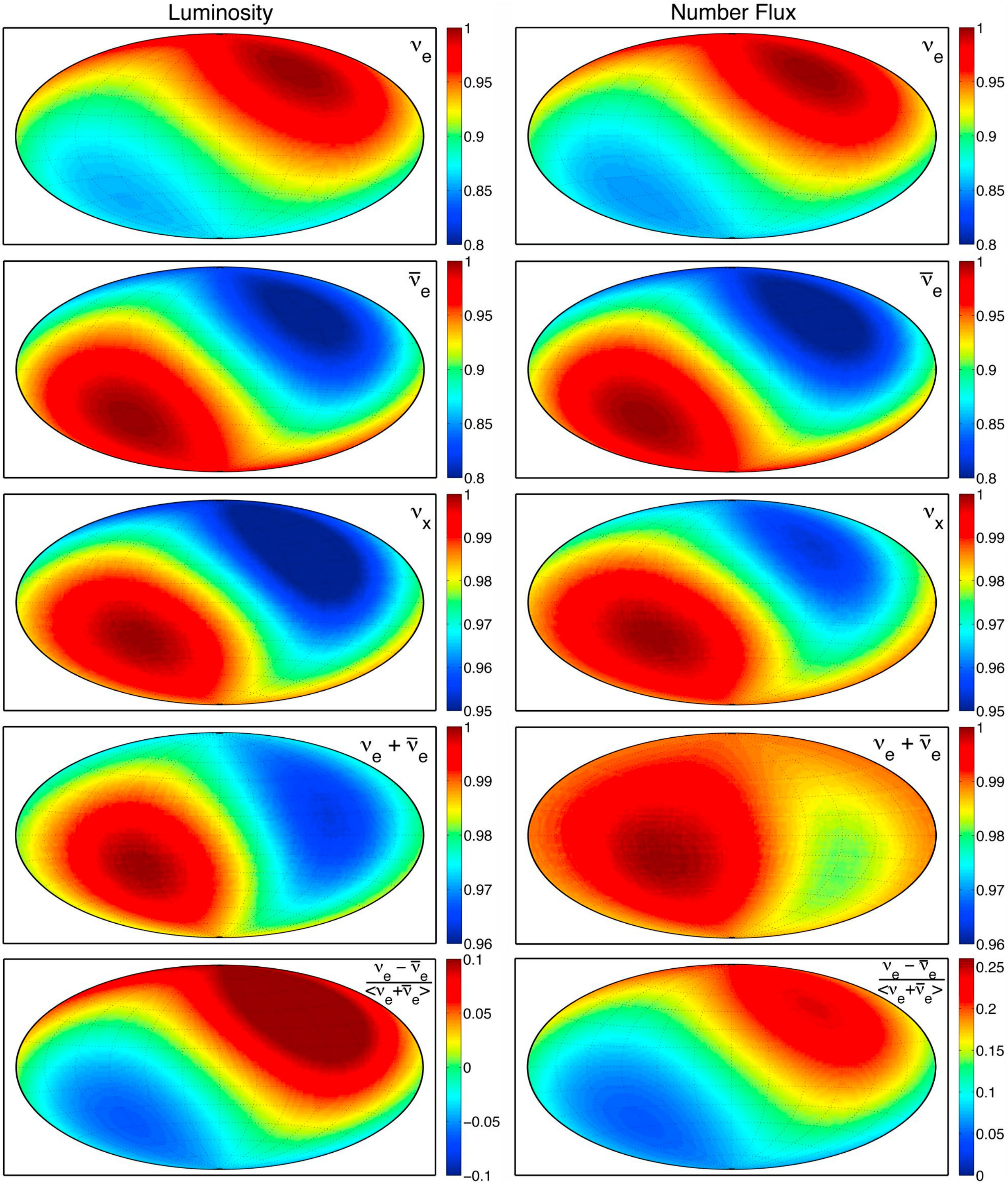}
\caption{Energy luminosity, $L$ ({\em left}), and number flux,
$N$ ({\em right}), for the
11.2\,$M_\odot$ simulation as functions of viewing direction for a distant
observer. The latitudes and longitudes indicated by dotted lines follow
the angular coordinate directions of the computational grid.
The quantities are hemispheric averages (including projection effects as
described in the text) and time integrated over 150--250\,ms post bounce. The first
three rows show the $\nu_e$, $\bar\nu_e$, and heavy-lepton neutrino ($\nu_x$)
fluxes, while the fourth row shows $L_{\nu_e}+L_{\bar\nu_e}$ and
$N_{\nu_e}+N_{\bar\nu_e}$. The plotted quantities are normalized to their
maximum. In each row the color scale of both images is the same, but the
ranges are different in different rows. The bottom row shows the relative
excess of $\nu_e$ over $\bar\nu_e$ emission, i.e.,
$(L_{\nu_e}-L_{\bar\nu_e})/\left\langle L_{\nu_e}+L_{\bar\nu_e}\right\rangle$
(left) and $(N_{\nu_e}-N_{\bar\nu_e})/\left\langle
N_{\nu_e}+N_{\bar\nu_e}\right\rangle$ (right). The denominators are averages
over all observer directions. In one hemisphere, the lepton number-emission
($\nu_e$ minus $\bar\nu_e$) is significantly smaller than the average, while
in this hemisphere the energy luminosity of $\bar\nu_e$ exceeds that of
$\nu_e$. In contrast, the number and energy fluxes of $\nu_e$ plus
$\bar\nu_e$ as well as those of $\nu_x$ deviate from isotropy only on
the few-percent level.}\label{fig:skymaps11}
\end{center}
\end{figure*}
%------------------------------------------------------------------------------------------

%------------------------------------------------------------------------------------------
\begin{figure}
\begin{center}
  \includegraphics[width=\columnwidth]{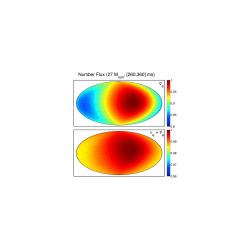}
  \caption{Sky-maps for the $27\,M_\odot$ model analogous to
  second and fourth panels in the right column of Fig.~\ref{fig:skymaps11}.
  The time interval of integration, 260--360\,ms p.b., is between two episodes
  of strong SASI activity. The behavior of the heavier-mass models
  is qualitatively similar to the $11.2\,M_\odot$ case but the
  spatial orientation of the lepton-emission dipole is different in
  each case.
  \label{fig:skymaps27}}
\end{center}
\end{figure}
%------------------------------------------------------------------------------------------

%------------------------------------------------------------------------------------------
\begin{figure*}
\begin{center}
  \includegraphics[width=.475\textwidth]{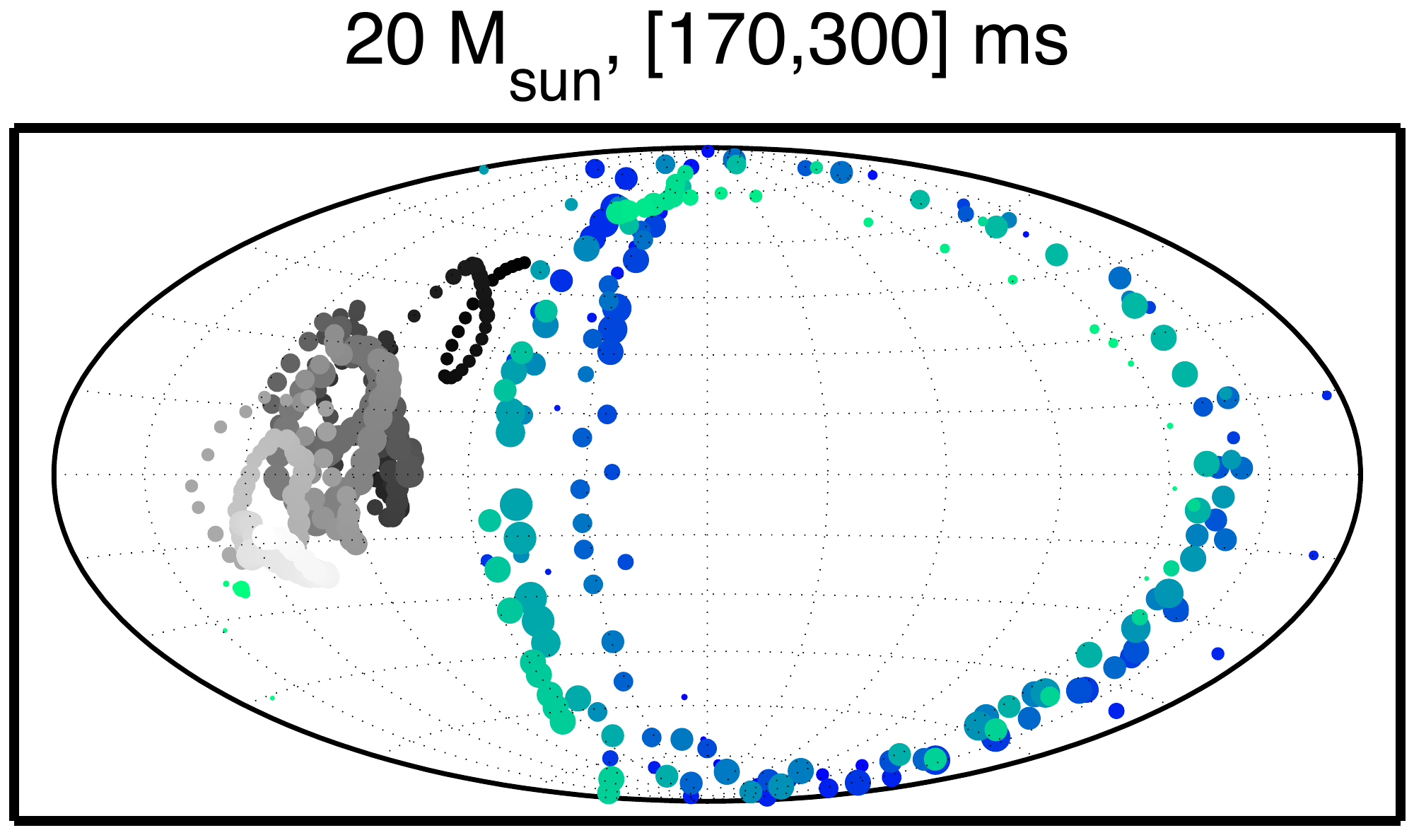}\hspace{2pt}
  \includegraphics[width=.475\textwidth]{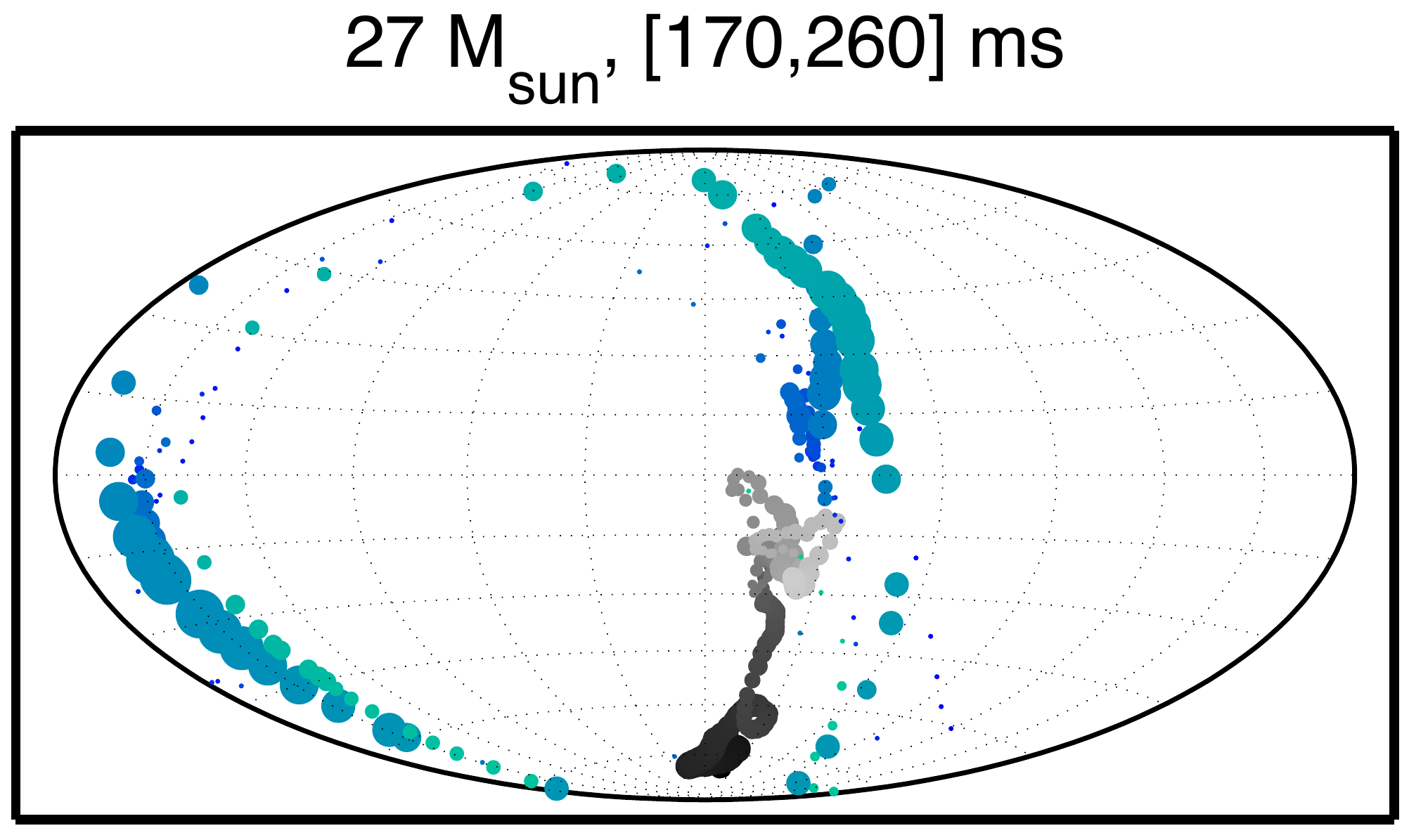}\\[2pt]
  \includegraphics[width=.475\textwidth]{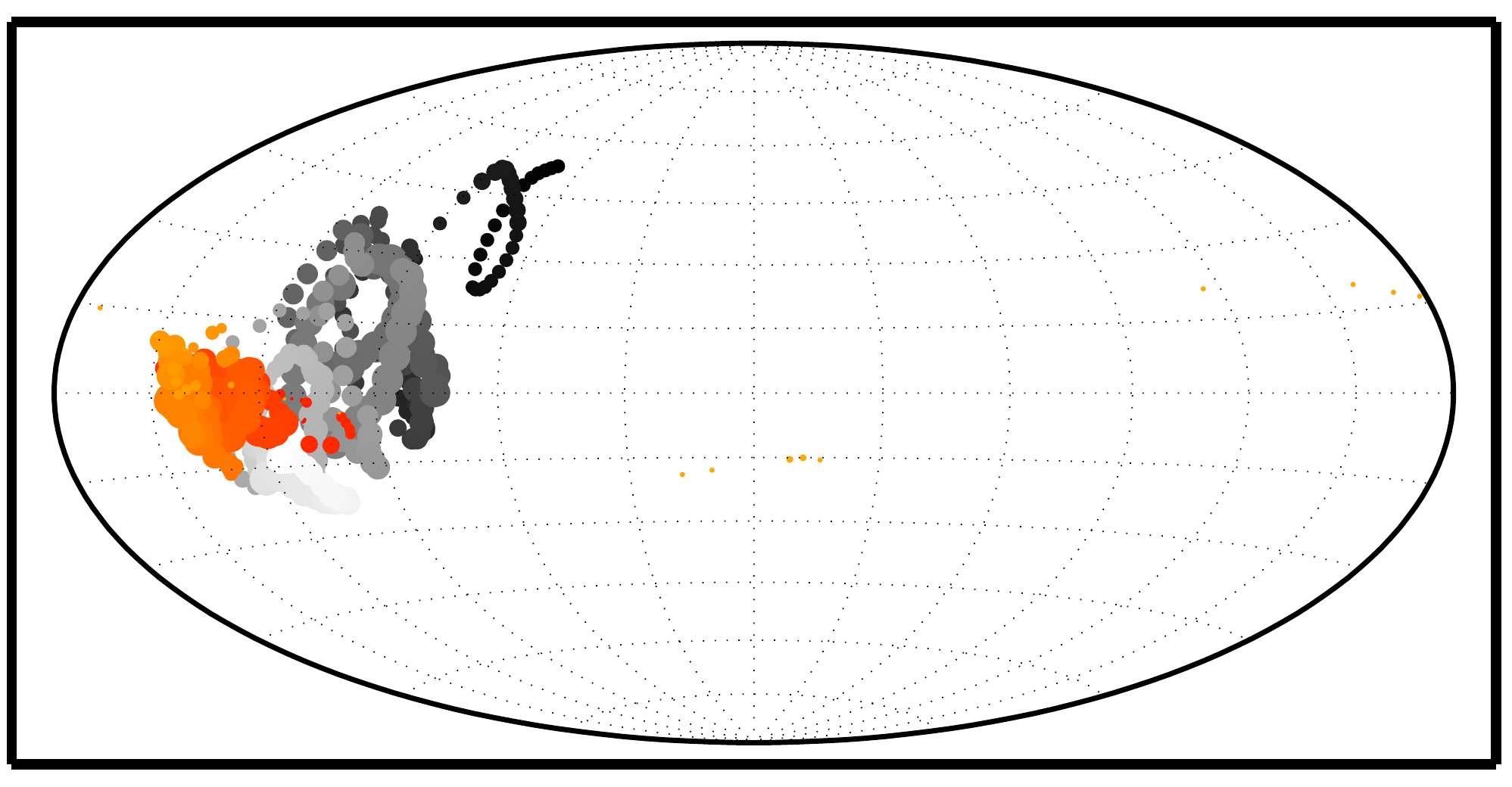}\hspace{2pt}
  \includegraphics[width=.475\textwidth]{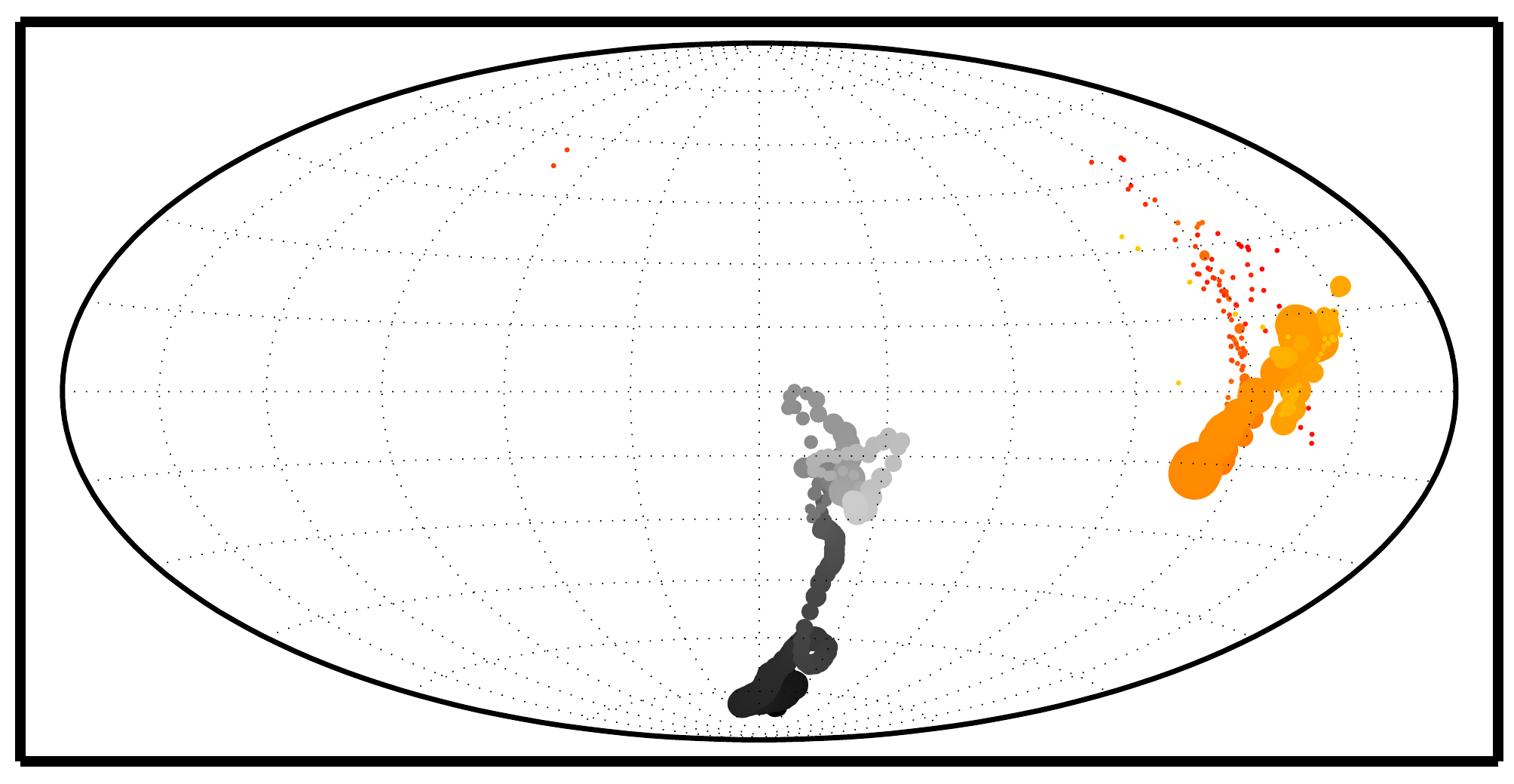}
\caption{Evolution of LESA and SASI directions during the first
SASI-dominated episodes of the 20\,$M_\odot$ model ({\em left})
and 27\,$M_\odot$ model ({\em right}).
The corresponding time intervals are
given above the columns. Black and gray dots mark the path of
the positive LESA dipole direction (associated with maximal
excess of the $\nu_e$ emission). Colored dots show the path of
the SASI shock-deformation vector, which describes the plane of
the SASI spiral motion ({\em upper panels}), and the path of the
instantaneous normal vector, which represents the direction
orthogonal to the SASI plane ({\em bottom panels}). The
size of the dots scales with the vector lengths and thus increases
with the LESA and SASI amplitudes, and the color hues become
lighter as time progresses. While the LESA dipole lies close
to the SASI plane in the 27\,$M_\odot$ model, both are nearly
perpendicular to each other in the 20\,$M_\odot$ case.
\label{fig:lesasasisky}}
\end{center}
\end{figure*}
%------------------------------------------------------------------------------------------

\subsection{Other flux asymmetries}
\label{sec:fluxasymmetries}

The LESA phenomenon is a conspicuous order-unity effect of the directional
lepton-number flux variation, but also shows up in other quantities, notably
in the directional variation of the individual $\nu_e$ and $\bar\nu_e$
fluxes. To illustrate this point we show in Fig.~\ref{fig:fluxvariations}
these number fluxes as they would be seen by a distant observer relative
to their directional averages. We use three
viewing directions oriented relative to the lepton-number dipole axis, i.e.,
an observer located in the direction of maximal lepton-number flux (black
lines), the opposite direction of minimal lepton-number flux (red lines), and
a typical direction transverse to the dipole axis (dashed blue lines).

A distant observer measures the integrated intensity over a hemisphere of the
radiating NS surface, projected on the viewing direction. The corresponding
``averaging'' over the visible hemisphere eliminates small-scale variations.
To evaluate the observational quantities we follow the procedure described in
Sect.~3.1 of \citet{Mueller_2012} and in Appendix~A of
\citet{Tamborra_2014} and calculate the observable flux, here
the number-flux $N$, from the ray-by-ray-computed number-flux densities,
$F_\mathrm{n}(\textit{\textbf{R}})$, at points $\textit{\textbf{R}}$ on the
radiating surface by an integration over the visible hemisphere, cf.\ Eq.~(7)
in \citet{Mueller_2012}:
\begin{equation}
  N(t) = 2\, \int_\mathrm{vis.hem.} {\mathrm d} A\,
           \cos\vartheta\, F_\mathrm{n}(\textit{\textbf{R}},t)\,
           \left (1 + \frac{3}{2} \cos\vartheta \right) \,.
\label{eq:applum-n}
\end{equation}
Here $\vartheta$ is the zenith angle at point $\textit{\textbf{R}}$ on the
radiating sphere, i.e., the angle between the normal vector of the surface
element ${\mathrm d} A$ and the observer direction. An analogous
relation pertains to the energy luminosity, $L$, as a function of the
energy-flux densities, $F_\mathrm{e}(\textit{\textbf{R}})$,
\begin{equation}
  L(t) = 2\, \int_\mathrm{vis.hem.} {\mathrm d} A\,
           \cos\vartheta\, F_\mathrm{e}(\textit{\textbf{R}},t)\,
           \left (1 + \frac{3}{2} \cos\vartheta \right)\,,
\label{eq:applum-e}
\end{equation}
cf.\ Eq.~(5) in \citet{Mueller_2012}. The integrands of
Eqs.~(\ref{eq:applum-n}) and~(\ref{eq:applum-e}) account for projection
effects of the radiating surface elements and limb darkening. In the
free-streaming limit, the flux density $F(\textit{\textbf{R}})$ declines like
$R^{-2}$ with distance $R = |\textit{\textbf{R}}|$ from the source center,
while the surface element ${\mathrm d} A = R^2{\mathrm d}\Omega$ (${\mathrm
d}\Omega$ being the solid angle) increases with $R^2$. Therefore, the product
${\mathrm d} A\,F(\textit{\textbf{R}})$ becomes constant and the integral
value is independent of the chosen surface of integration.

Figure~\ref{fig:fluxvariations} shows that in the early phase of postshock
convection the directional emission asymmetry remains small. At about
150\,ms, however, a stable dipolar pattern emerges and reaches maximum
amplitudes of around 10\% for $\nu_e$ and 15\% for $\bar\nu_e$ at roughly
180\,ms after bounce. A positive amplitude for $\nu_e$ is correlated with a
negative one for $\bar\nu_e$ and vice versa, and local maxima (minima) of the
$\nu_e$ emission generally coincide in time with minima (maxima) of the
$\bar\nu_e$ emission. After $t\sim 180$\,ms a slow, overall trend of decay of
the dipolar emission asymmetry begins, in agreement with our earlier finding
in the lepton-number flux, although the relative strength of the
lepton-number asymmetry remains large. Note also that we show here the
evolution as observed from a fixed direction so that overall trends of the
fluxes can be partly caused by a slight drift of the dipole direction.

The dipole asymmetry is large in the $\nu_e$ and $\bar\nu_e$ fluxes, whereas
heavy-lepton neutrinos, $\nu_x$, exhibit at most a few-percent effect. The
$\nu_x$ emission is slightly enhanced in the direction of small lepton-number
(high $\bar\nu_e$) flux.

Figure~\ref{fig:skymaps11} provides similar information for the
11.2\,$M_\odot$ model in the form of sky maps for all viewing directions of a
distant observer. The temporal stability of the dipole pattern justifies time
averaging instead of individual snapshots. In particular, we average over
150--250\,ms p.b., when the dipole effect is particularly strong. The left
column provides the luminosities, $L$, for $\nu_e$, $\bar\nu_e$, $\nu_x$, as
well as $\nu_e$ plus $\bar\nu_e$, and $\nu_e$ minus $\bar\nu_e$, all
normalized to their directional maxima except for the difference plot, which
is normalized to the all-sky average of $L_{\nu_e}+L_{\bar\nu_e}$. The right
column shows analogous information for the corresponding number fluxes, $N$.

Both luminosities and number fluxes clearly show the emission dipole
(anti-)aligned with the lepton-number dipole axis. While the maximum
variation amplitudes of $\nu_e$ and $\bar\nu_e$ fluxes are approximately
$\pm10\%$ in the two dipole directions, heavy-lepton neutrinos show just
about $\pm$2\% variations. In contrast to the individual luminosities and
number fluxes, the sums $L_{\nu_e}+L_{\bar\nu_e}$ and
$N_{\nu_e}+N_{\bar\nu_e}$, vary only on the few-percent level. We also note
that the relative variation of the energy-flux difference,
$L_{\nu_e}-L_{\bar\nu_e}$, is a bit smaller than the hemispheric
variation of the number-flux difference $N_{\nu_e}-N_{\bar\nu_e}$.
Most importantly, while the former can be positive or negative, the
latter is positive in all directions. This shift of the asymmetry
variation in the luminosity difference corresponds to considerably
larger $\bar\nu_e$ energies relative to $\nu_e$.

In contrast, the 27\,$M_\odot$ run with its episodes of strong SASI shock
sloshing and spiralling motions (see \citealt{Hanke_2013}) exhibits
large-amplitude, quasiperiodic neutrino emission modulations with dipolar
asymmetry in all flavors \citep{Tamborra_2013,Tamborra_2014}.
In addition, however, the 27\,$M_\odot$
model also shows a
steady lepton-number emission dipole, i.e., a long-lasting and
non-oscillating dipole in the lepton-number flux ($\nu_e$ minus $\bar\nu_e$).
We present a sky-map of the $\bar\nu_e$ number flux as well as the total $\nu_e$
plus $\bar\nu_e$ flux in Fig.~\ref{fig:skymaps27}. To avoid any confusion
with SASI activity, we show a time-averaged signal here as seen by a distant
observer taken between the two episodes of SASI activity, i.e., integrated
over the p.b.\ interval of 260--360\,ms. In qualitative agreement with the
$11.2\,M_\odot$ case, there is a clear dipole feature in the $\bar\nu_e$
flux, whereas in the sum flux the dipole variation is weak---the individual
$\nu_e$ and $\bar\nu_e$ fluxes are again anti-correlated.

The LESA dipole directions of our three stellar models do not show any
correlation with each other
(cf.\ Fig.~\ref{fig:dipole11}), and they are uncorrelated
with the numerical grid. Moreover, the LESA dipole direction has no clear
correlation with the main direction of SASI sloshing or with the plane
of SASI spiralling motions. Figure~\ref{fig:lesasasisky} displays the
evolution of the SASI and LESA directions during the first SASI-dominated
phases of the 20$\,M_\odot$ and 27$\,M_\odot$ models (170--300\,ms and
170--260\,ms, respectively). In the former case, the LESA dipole direction
is clearly far outside of the plane defined by the movements of the
SASI shock-deformation vector (implying that the
LESA dipole vector and the normal vector of the SASI plane are nearly
aligned), while in the latter case the LESA dipole happens to be close
to the SASI plane. This suggests that the relative orientations of SASI
and of the lepton-number emission dipole are chosen randomly and that both
effects are independent phenomena. Nevertheless, both seem to be able to
influence each other as we will discuss later
(Sect.~\ref{sec:modelswithsasi}).

%------------------------------------------------------------------------------
\begin{figure}
\begin{center}
  \includegraphics[width=.90\columnwidth]{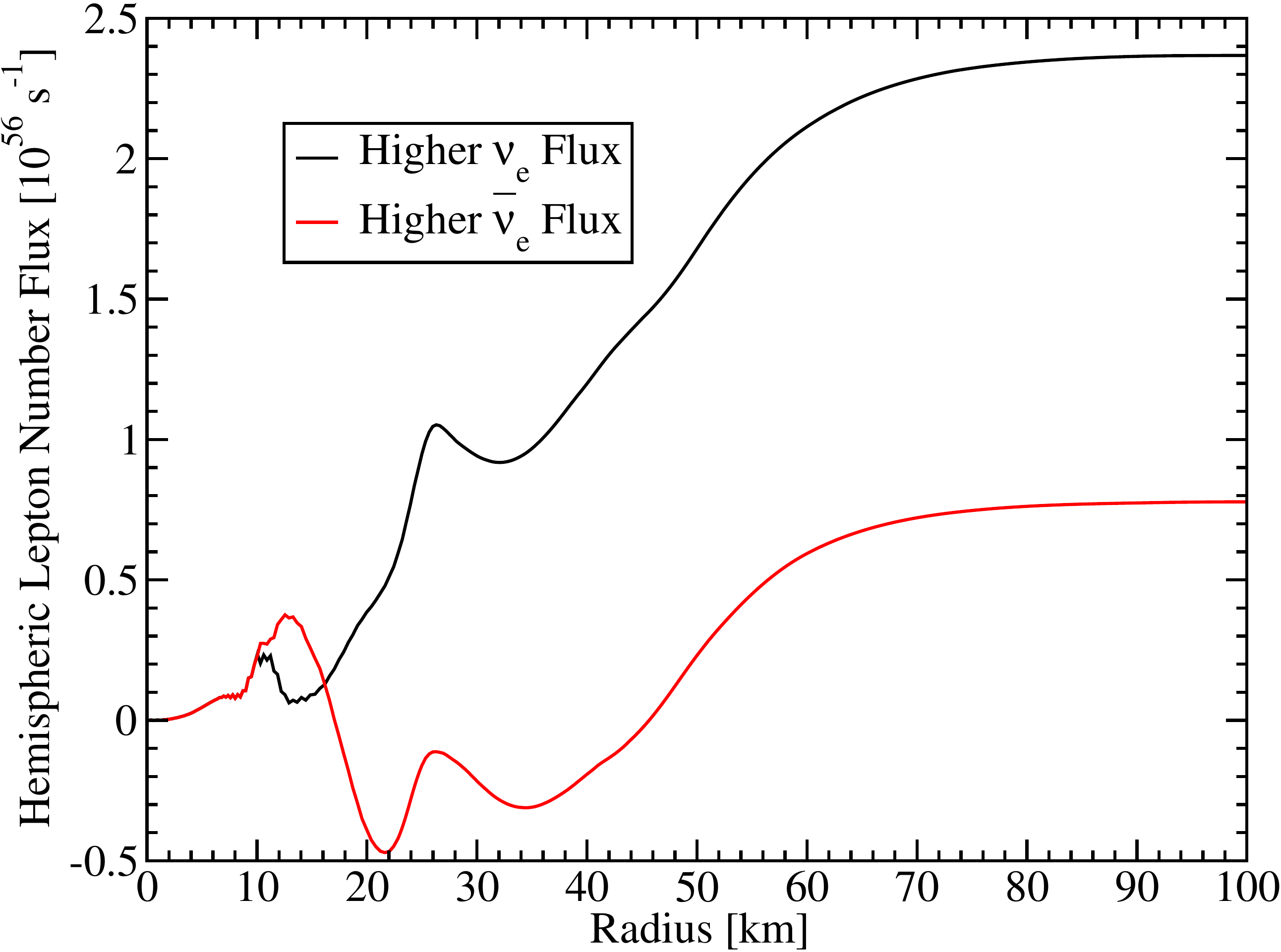}
  \caption{Radial evolution of the lepton-number emission in the hemisphere
  where the lepton flux is maximal (black) and minimal (red) for the
  11.2\,$M_\odot$ simulation at 210\,ms p.b. The fluxes are integrated
  over the hemispheres without projection effects so that their sum is
  the total lepton-number flux traversing a spherical surface of given
  radius. The lepton-number flux asymmetry originates mostly from
  deep inside the PNS, i.e., from the hot PNS mantle below the
  neutrinosphere that is located at approximately 35~km, whereas
  a more spherically symmetric component of the lepton-number
  flux develops in the surrounding, semi-transparent cooling layer
  and is fed by the accretion of lepton-rich material.
\label{fig:neutrinoasymmetry1}}
\end{center}
\end{figure}
%------------------------------------------------------------------------------

%------------------------------------------------------------------------------------------
\begin{figure*}
\begin{center}
  \includegraphics[width=.40\textwidth]{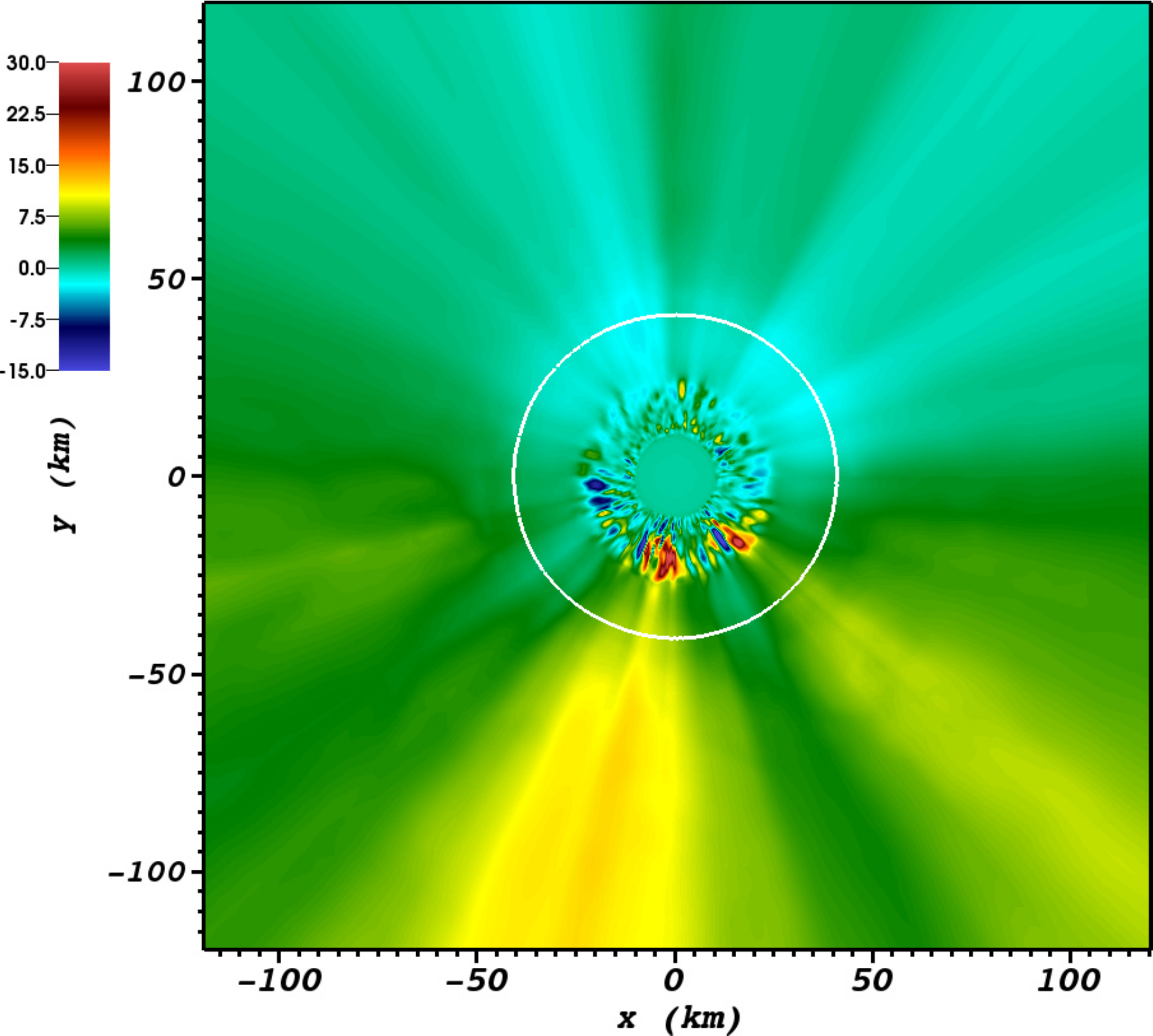}\hspace{8pt}
  \includegraphics[width=.40\textwidth]{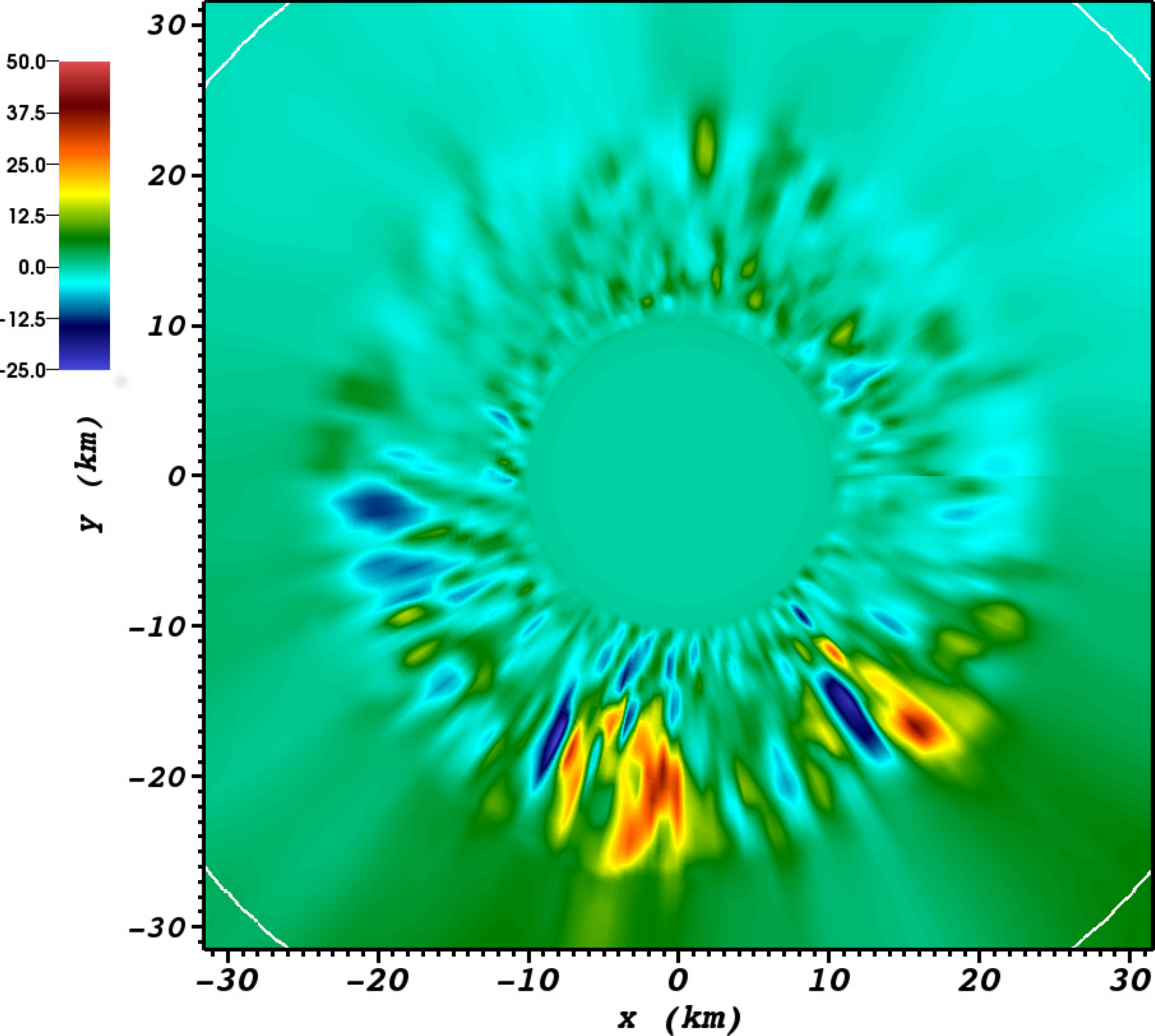}
\caption{Radial evolution of the lepton-number flux in the $11.2\,M_\odot$
model at 210\,ms p.b.\ (same moment as in Fig.~\ref{fig:neutrinoasymmetry1}).
Shown is the color-coded
isotropic equivalent of the lepton number flux, i.e., $4\pi\,r^2\,(F_{\nu_e}
- F_{\bar\nu_e})$ in $10^{56}$\,s$^{-1}$, along angular ``rays'' of the
transport simulation. The cut plane includes the direction of maximal lepton
emission (bottom of panels) and the opposite direction of minimal lepton
emission (top of cut). The average neutrinosphere is at about 35\,km (white
circle). The right panel is a zoom of the left one. PNS convection is
clearly visible, with stronger activity in the hemisphere of maximal
lepton-number flux (bottom direction).\label{fig:neutrinoasymmetry3}}
\end{center}
\end{figure*}
%------------------------------------------------------------------------------------------
\vskip4pt
%------------------------------------------------------------------------------------------
\begin{figure*}
\begin{center}
  \includegraphics[width=.40\textwidth]{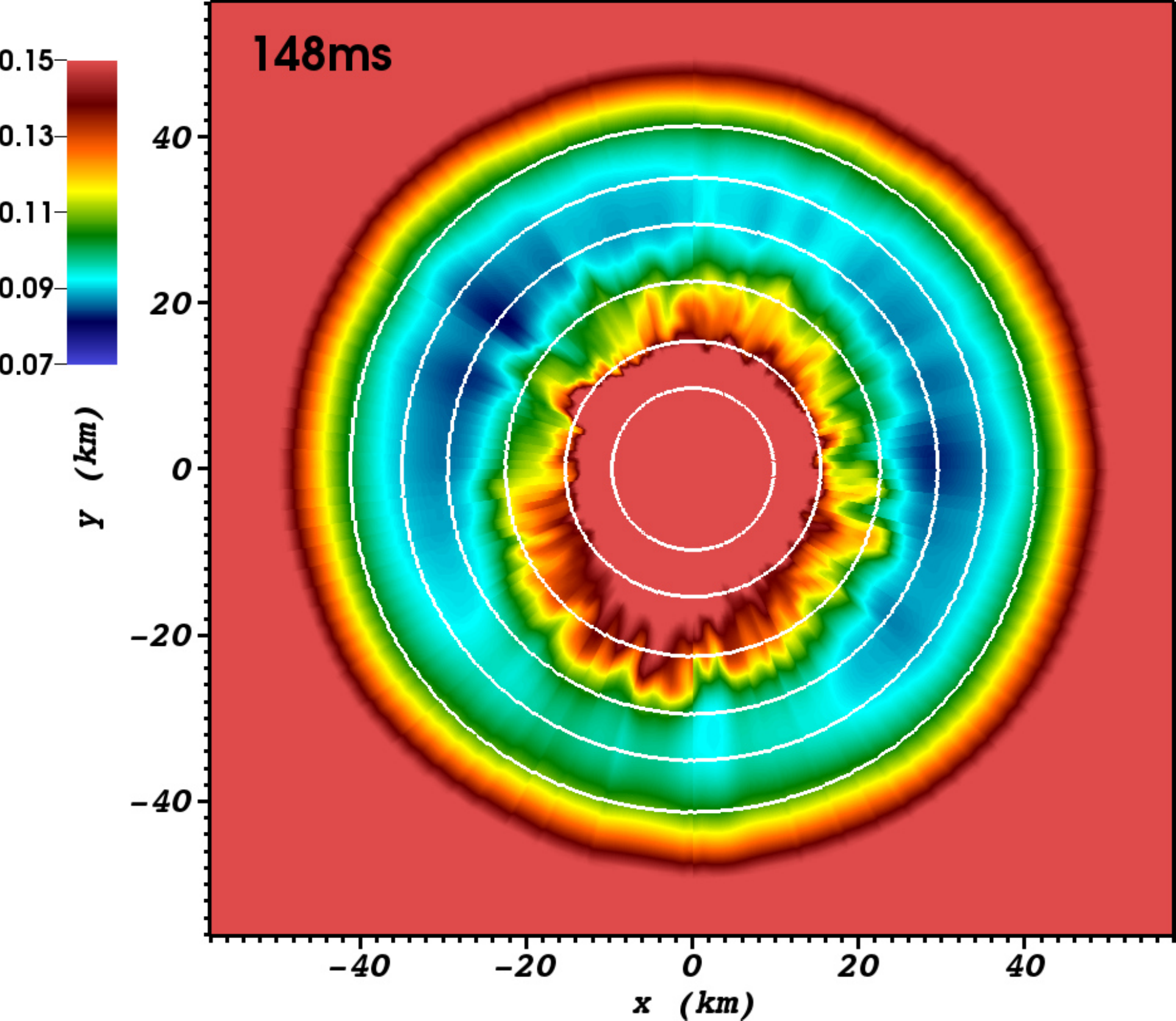}\hspace{8pt}
  \includegraphics[width=.40\textwidth]{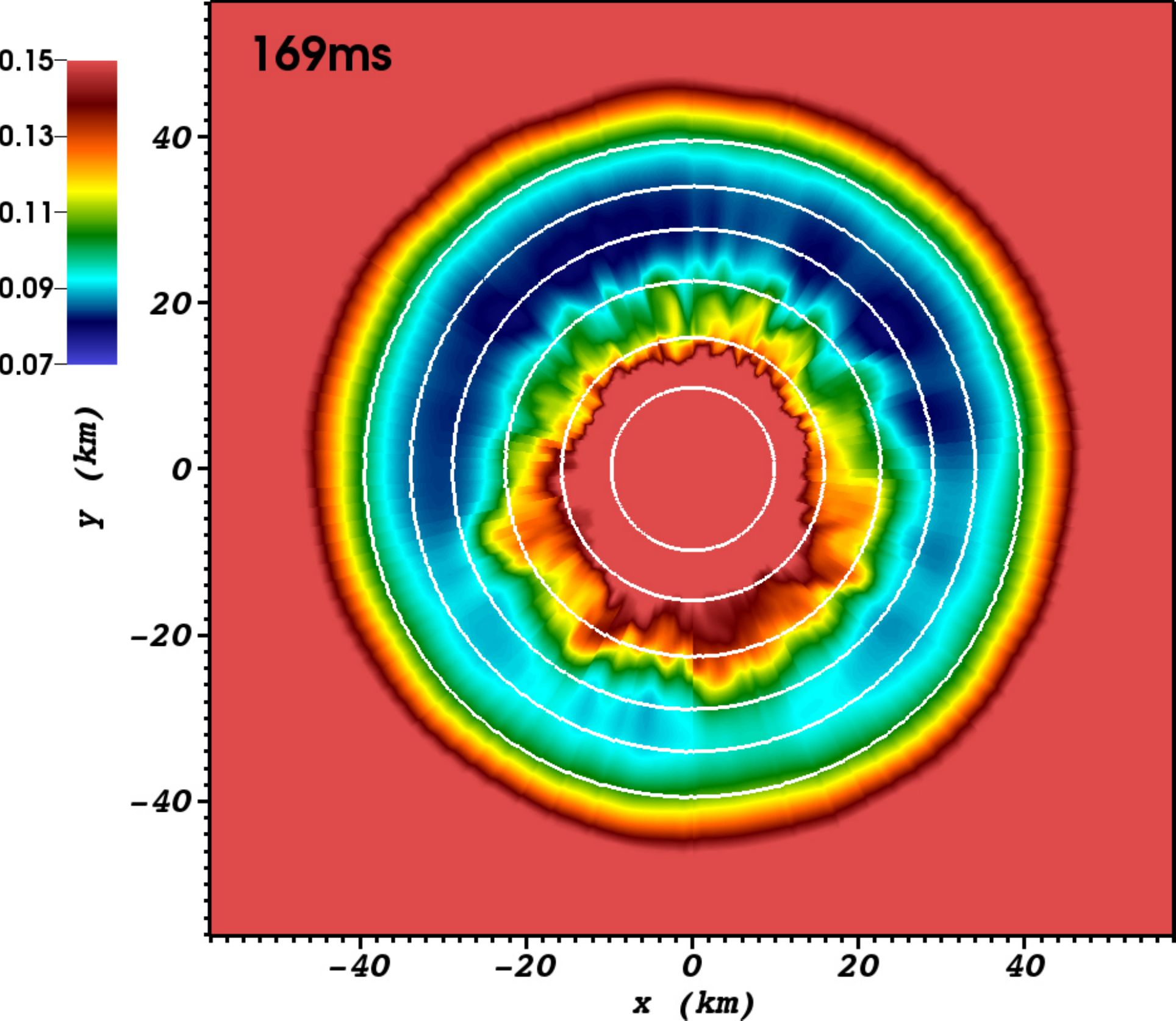}\\[3pt]
  \includegraphics[width=.40\textwidth]{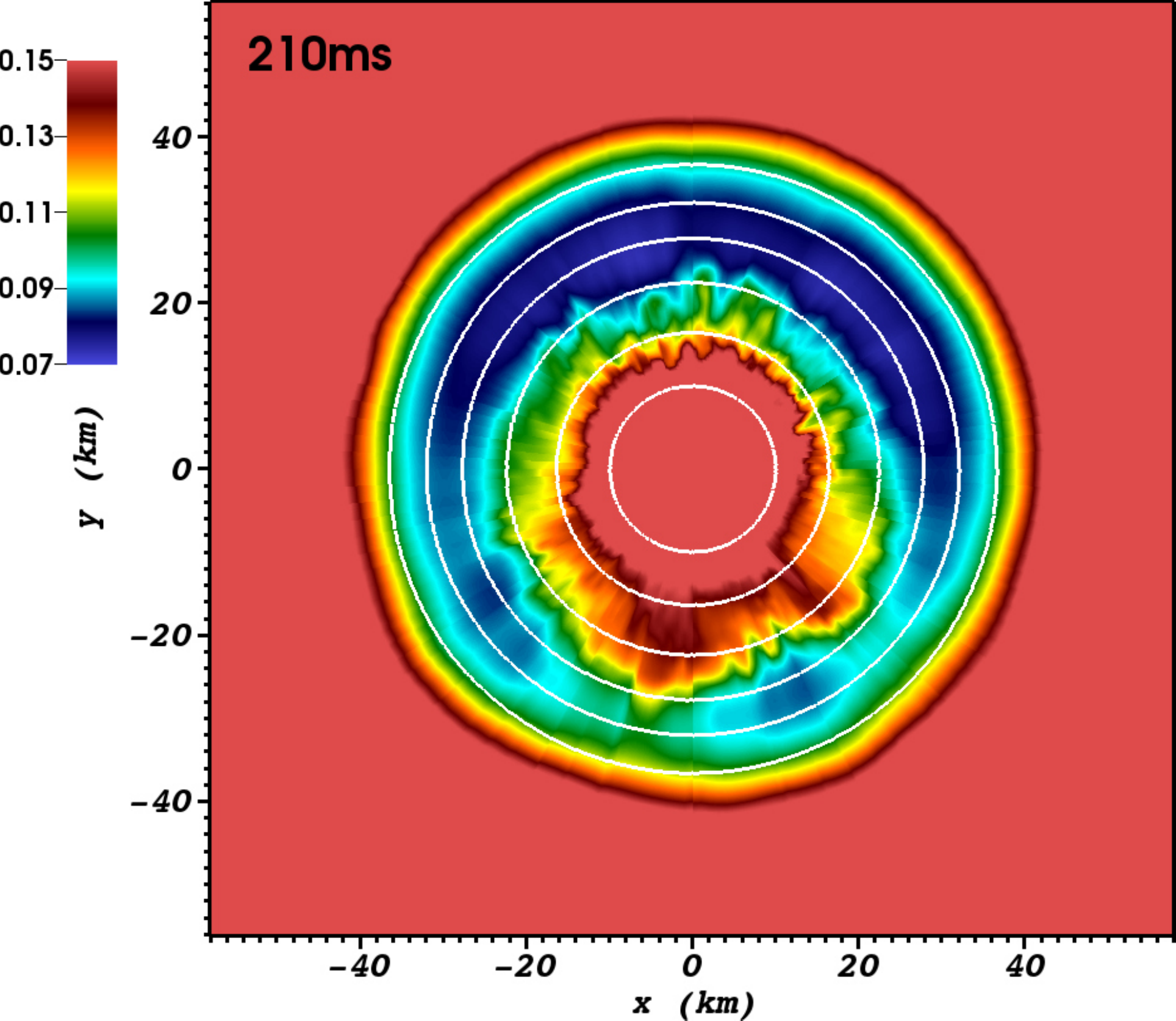}\hspace{8pt}
  \includegraphics[width=.40\textwidth]{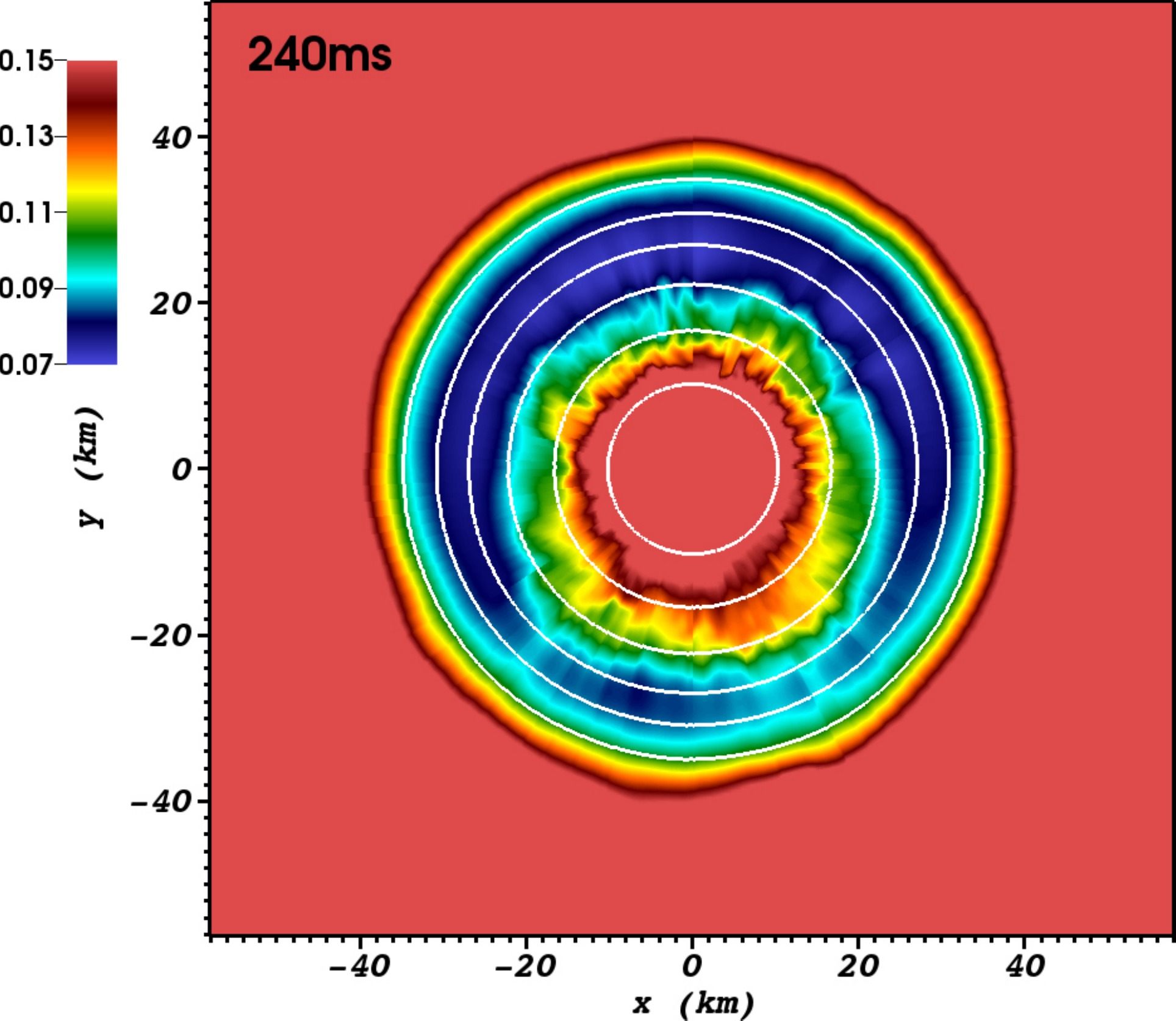}
\caption{Distribution of the electron fraction, $Y_e$, in the PNS and its
immediate surroundings for the 11.2\,$M_\odot$ model at the indicated p.b.\
times. The cut plane is the same as in Fig.~\ref{fig:neutrinoasymmetry3},
i.e., it contains the dipole axis with the direction of maximal lepton-number
emission being downward in these panels. The color scale saturates when $Y_e
> 0.15$ and was chosen to highlight the $Y_e$ variations in the PNS mantle
region around the central, lepton-rich core and below the neutrinosphere
(which roughly coincides with the outermost white circle). The white circles
are isodensity contours at the levels of $3{\times}10^{11}$, $10^{12}$,
$3{\times}10^{12}$, $10^{13}$, $3{\times}10^{13}$, and
$10^{14}$\,g\,cm$^{-3}$. Notice the development of a more strongly
deleptonized shell in the upper hemisphere (direction of minimal
lepton-number flux), while in the bottom hemisphere the lepton-number
fraction is larger. In this hemisphere, the mass accretion rate is larger,
supplying a larger amount of fresh lepton number. \label{fig:yecuts}}
\end{center}
\end{figure*}
%------------------------------------------------------------------------------------------

\subsection{Radial evolution of the emission dipole}
\label{sec:dipole-radius}

We next investigate the spatial origin of the lepton-number flux
asymmetry. To this end we consider the evolutionary stage at 210\,ms p.b.\ of
the $11.2\,M_\odot$ model when the lepton-number dipole has reached a large
value. Figure~\ref{fig:neutrinoasymmetry1} shows the radial evolution of the
lepton-number ($\nu_e$ minus $\bar\nu_e$) flux for the two hemispheres where
it is maximal and minimal, respectively. The integration avoids any
projection or observer effects---the sum of the two hemispheric values yields
the total lepton-number flux traversing a spherical shell of given radius. It
is intriguing that most of the hemispheric difference builds up in the PNS
mantle layer {\em below} the (average) neutrinosphere, which is located here
at around 35\,km. At this radius, the lepton-flux difference has nearly
reached its asymptotic value, whereas only about 20--25\% (or 3--$4\times
10^{55}\,{\rm s}^{-1}$) arise at larger radii and are therefore more directly
associated with the hemispheric asymmetry of the accretion flow (cf.\
Sect.~\ref{sec:mass-accretion}).

A different way of visualizing the radial evolution of the lepton-number flux
is to study it along individual radial ``rays''
of our transport scheme. To this end we
have performed a cut of our $11.2\,M_\odot$ model in a plane containing the
dipole direction at 210\,ms p.b. Figure~\ref{fig:neutrinoasymmetry3} shows
the result with a color coding corresponding to the radial lepton-number flux
as a function of location in this cut plane. The downward direction in
the plots is the direction of maximum lepton-number flux. It is apparent that
this hemisphere shows stronger convection inside the PNS than the other
hemisphere. The flux asymmetry arises far below the average neutrinosphere,
here indicated by a white circle.

Most of the overall lepton-number emission (the monopole of the emission
distribution) builds up in the envelope (i.e., the neutrino-cooling) region
above the NS and is fed by the accretion downflows of lepton-rich material,
whereas most of the dipole builds up around the PNS convection zone deep
inside the NS and below the neutrinosphere. While the accretion flow also
shows a dipole asymmetry as we will see, it is not responsible for the main
effect of the asymmetry of the lepton-number emission.

\subsection{Asymmetry of electron density distribution}
\label{sec:Ye}

Most of the lepton number stored in the PNS and its accretion layer is in the
form of electrons, whereas it is emitted in the form of a $\nu_e$-$\bar\nu_e$
number-flux difference. Therefore, it is instructive to inspect the electron
density distribution in those regions of the PNS where the lepton-flux
dipolar asymmetry originates. Figure~\ref{fig:yecuts} shows color-coded $Y_e$
distributions of the 11.2\,$M_\odot$ model
in cut planes containing the dipole axis in analogy to
Fig.~\ref{fig:neutrinoasymmetry3} and with the same orientation, i.e., bottom
is the hemisphere of largest lepton-number emission. We also show iso-density
contours as white circles---the density stratification is perfectly spherical
and concentric around the center of mass (which essentially coincides with
the coordinate origin) because of the extreme strength of the gravity field
of the PNS. The outermost contour, corresponding to $3\times
10^{11}$\,g\,cm$^{-3}$, is somewhat interior to the average neutrinosphere.

The four different postbounce moments correspond to the ones shown in
Fig.~\ref{fig:leptonskyplots} and span the time when the lepton-emission
dipole begins to form (148\,ms p.b.) all the way to a time when it is fully
developed, but still before any noticeable decay takes place (240\,ms). We
see the development of a more electron-depleted region in the upper hemisphere,
where a smaller lepton-number flux originates, while the bottom hemisphere,
where a larger lepton-number flux originates, exhibits more electron-rich
material. The growth of the hemispheric asymmetry of the lepton distribution
in the PNS mantle region below the neutrinosphere is clearly visible as the
compact remnant deleptonizes and contracts between $t \sim 150$\,ms (top
left) and $t = 210$\,ms (bottom left). At around this later time the most
extreme hemispheric difference is reached with an electron fraction
difference of up to $\Delta Y_e\sim 0.03$--0.06 on some density levels. As
time and lepton emission progress, the hemispheric differences tend to
decrease (bottom right).

The asymmetry of the $Y_e$ distribution not only explains the emission
dipoles of $\nu_e$ and $\bar\nu_e$, it also explains why the number flux of
heavy-lepton neutrinos, $\nu_x$, is somewhat amplified (on the percent
level) in the direction of the smaller
lepton-number flux, which is the direction of
stronger $\bar\nu_e$ emission (cf.\ Figs.~\ref{fig:fluxvariations} and
\ref{fig:skymaps11}). Because the annihilation of $e^+e^-$ and
$\nu_e\bar\nu_e$ pairs yields important contributions to the $\nu_x$
number flux, in particular at lower densities (whereas at high densities the
production by nucleon-nucleon bremsstrahlung dominates; see
\citealt{Raffelt_2001,Buras_2003,Keil_2003} for details),
the larger positron and
$\bar\nu_e$ abundances on this side of the PNS also foster the emission of
heavy-lepton neutrinos.

\section{Driving Mechanism of LESA}
\label{sec:feedback-loop}

\subsection{Asymmetry of mass-accretion flow}
\label{sec:mass-accretion}

The lepton-flux asymmetry originates deep inside the PNS, below the
neutrinosphere, and several phenomenological observations form a consistent
picture, e.g., the lepton-emission asymmetry, the PNS convection asymmetry,
the asymmetric $Y_e$ distribution, and the small $\nu_x$ emission asymmetry.
However, these manifestations do not yet provide a hydrodynamical explanation
of how these effects first arise and then stabilize themselves in a
long-lasting, only slowly evolving pattern.
Moreover, the initial growth of the dipole
distribution over 100--150\,ms is parallel to the growth of convective
overturn in the gain region below the stalled shock wave.

\begin{figure*}
\begin{center}
  \includegraphics[width=0.33\textwidth]{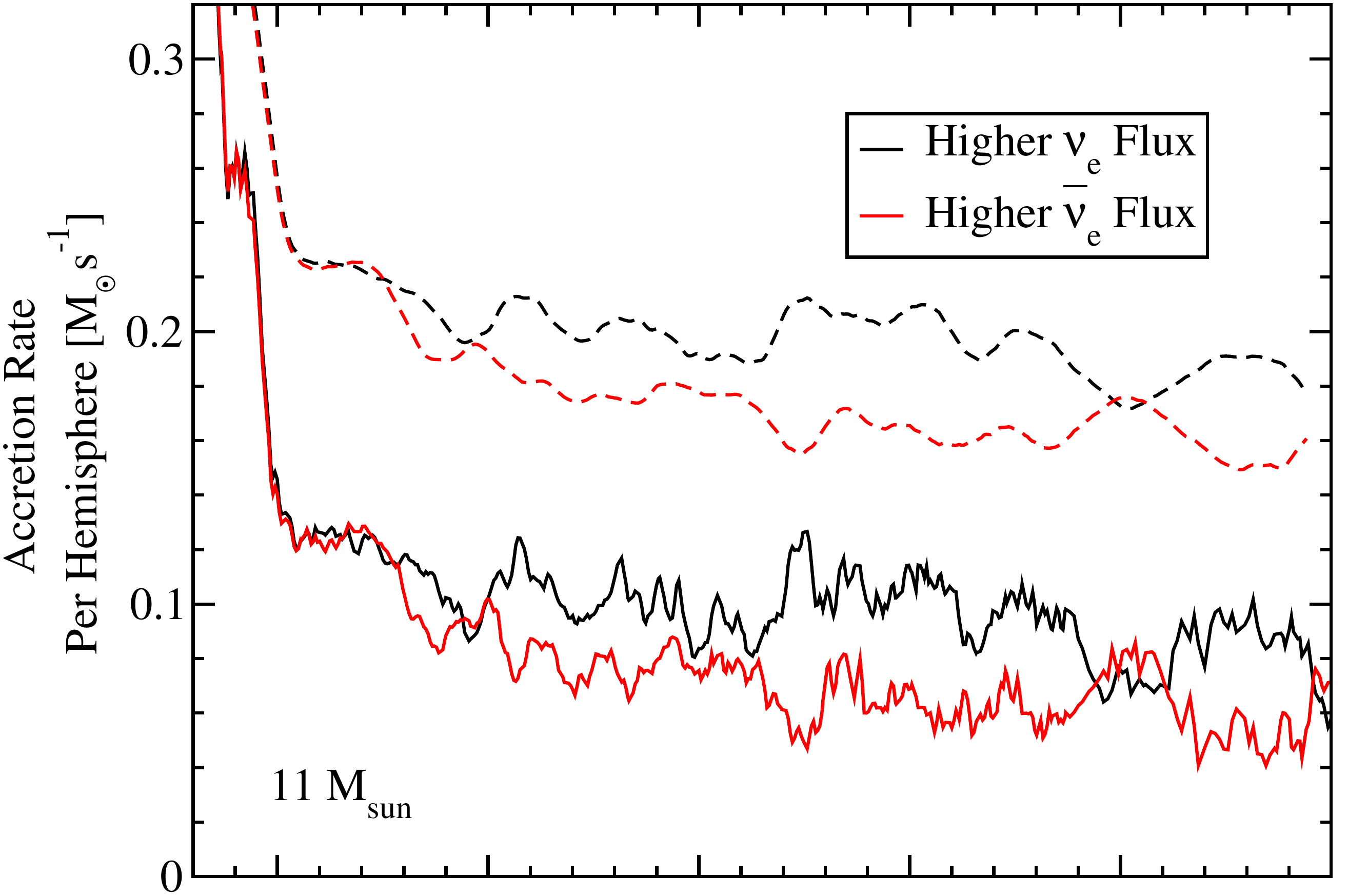}
  \includegraphics[width=0.33\textwidth]{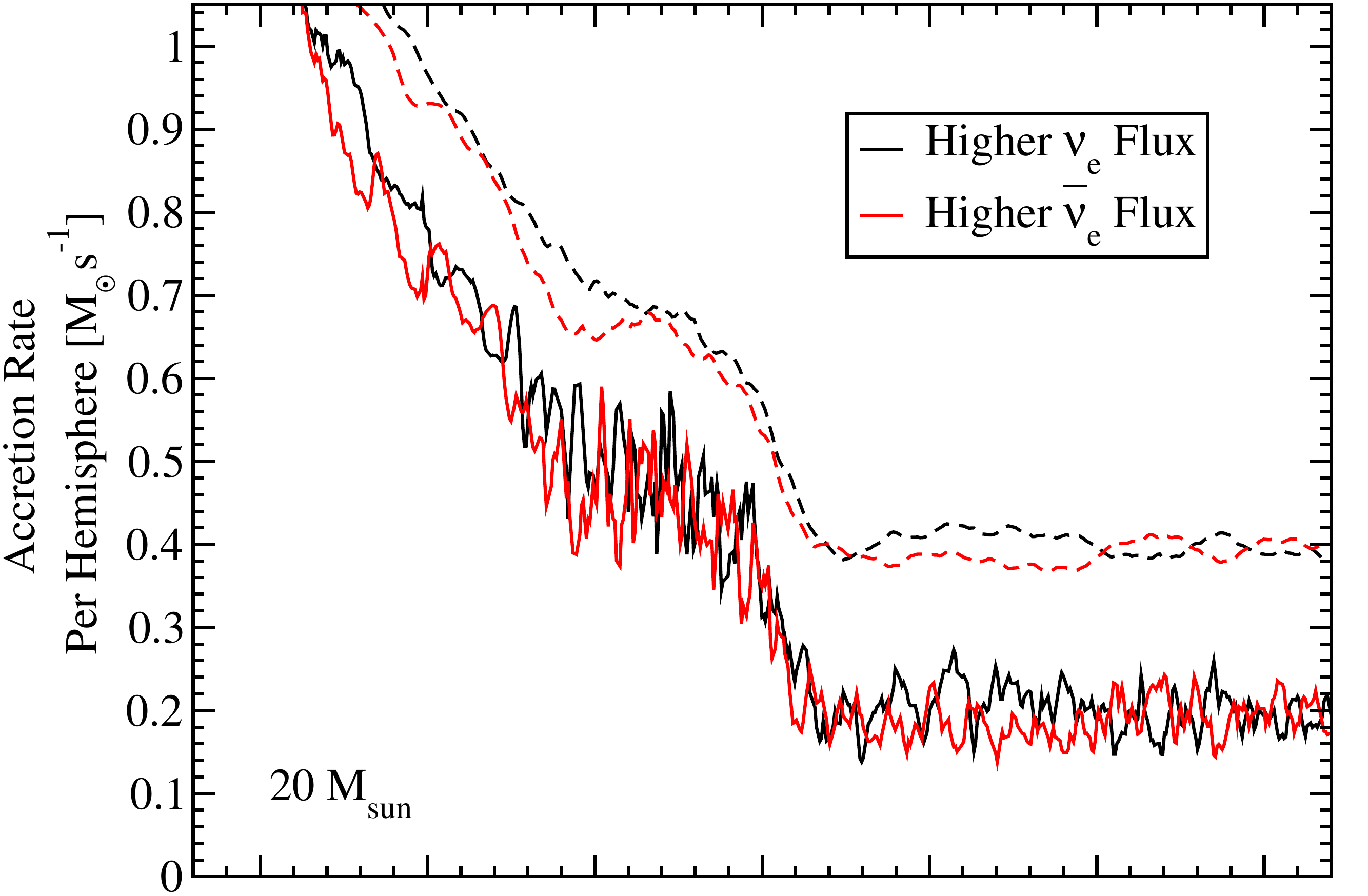}
  \includegraphics[width=0.33\textwidth]{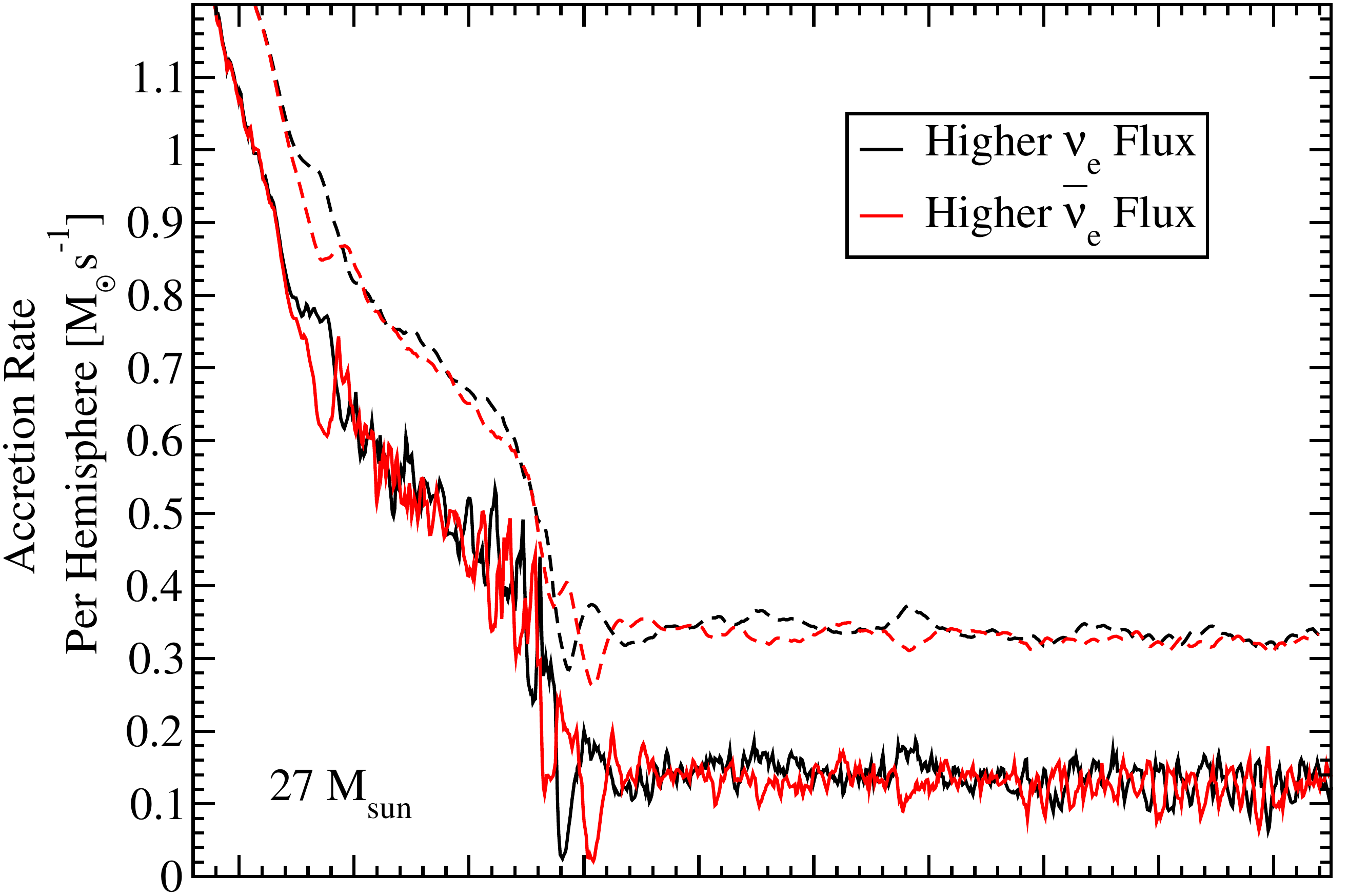}\\
  \includegraphics[width=0.33\textwidth]{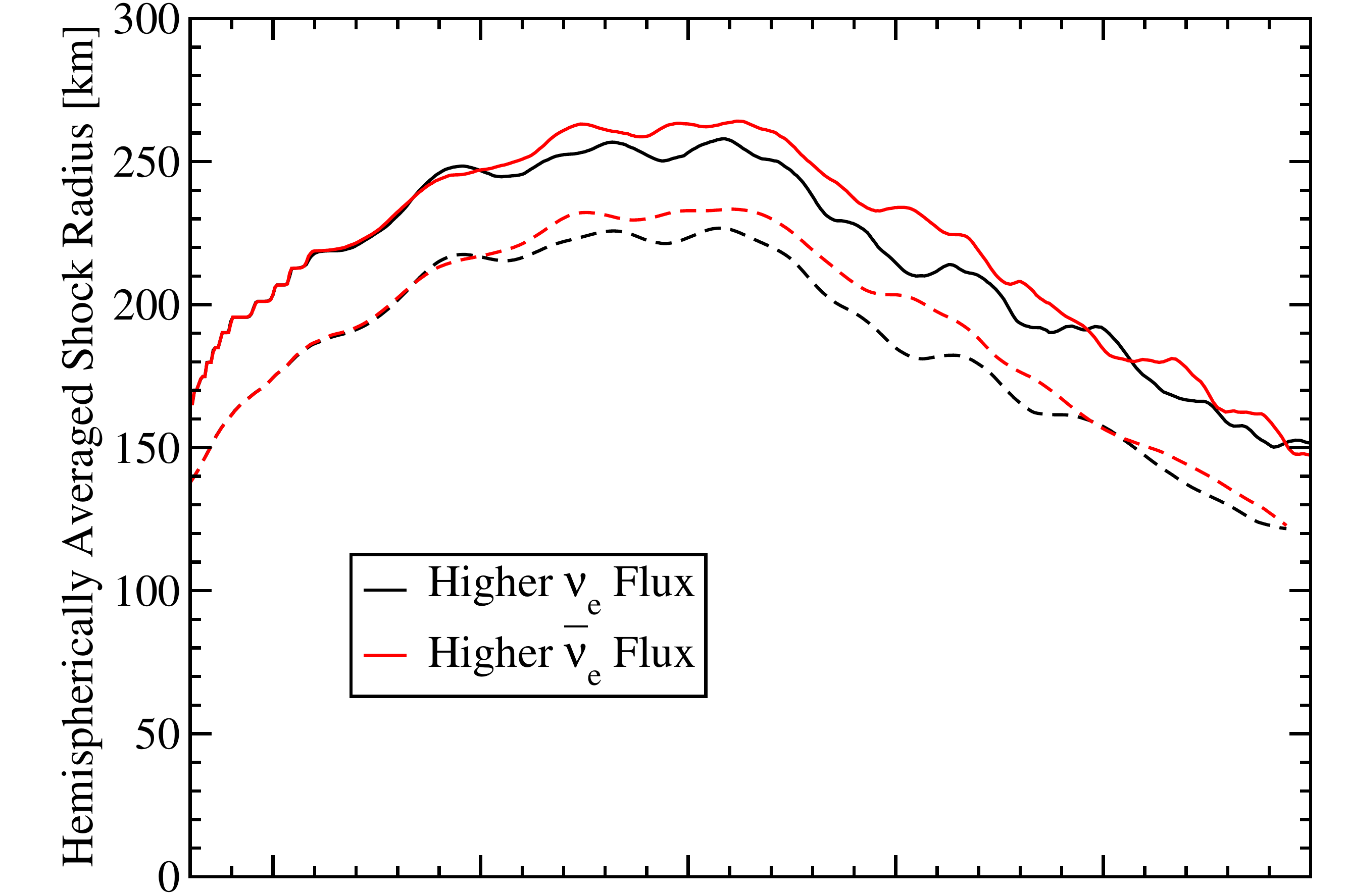}
  \includegraphics[width=0.33\textwidth]{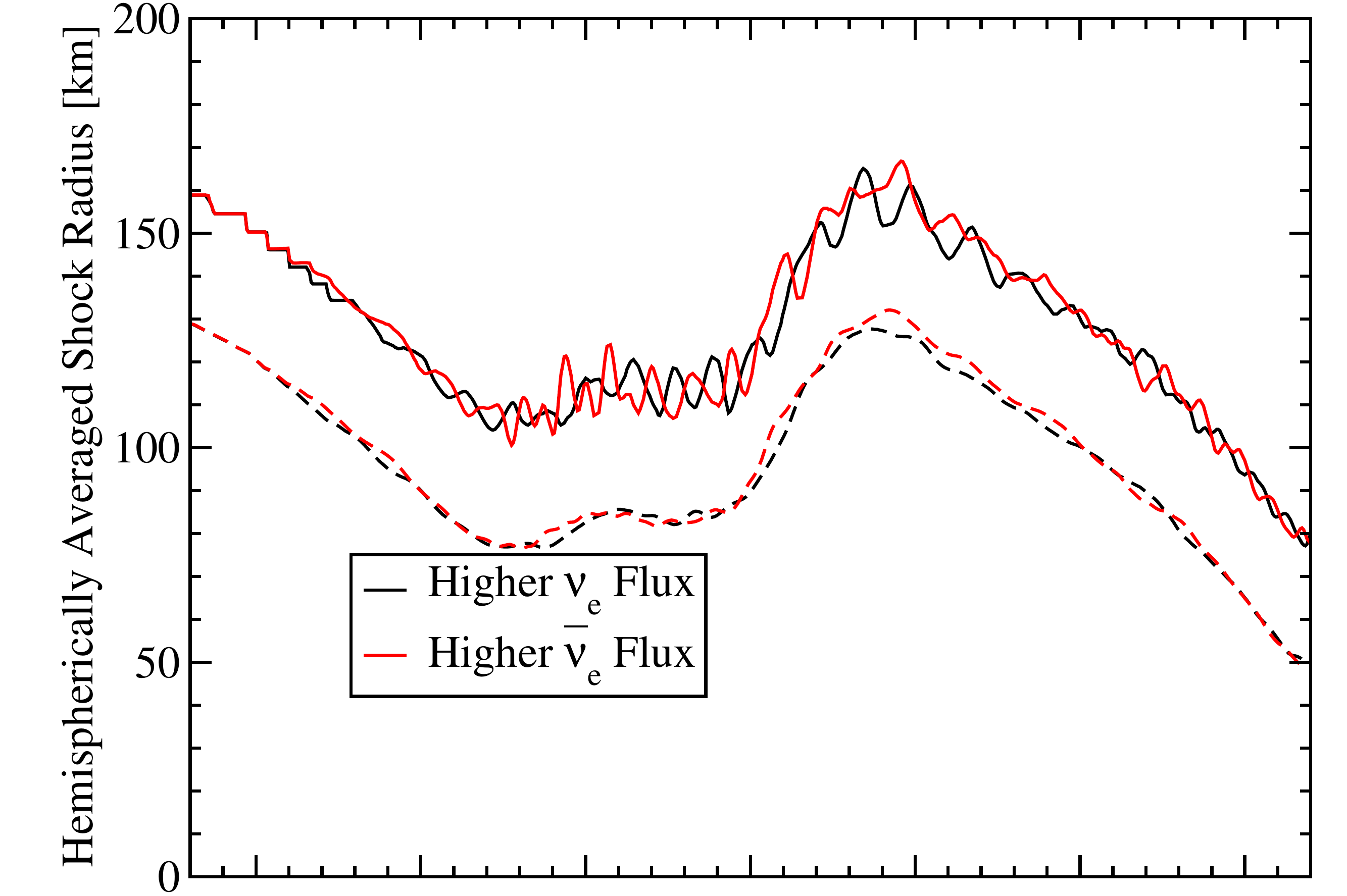}
  \includegraphics[width=0.33\textwidth]{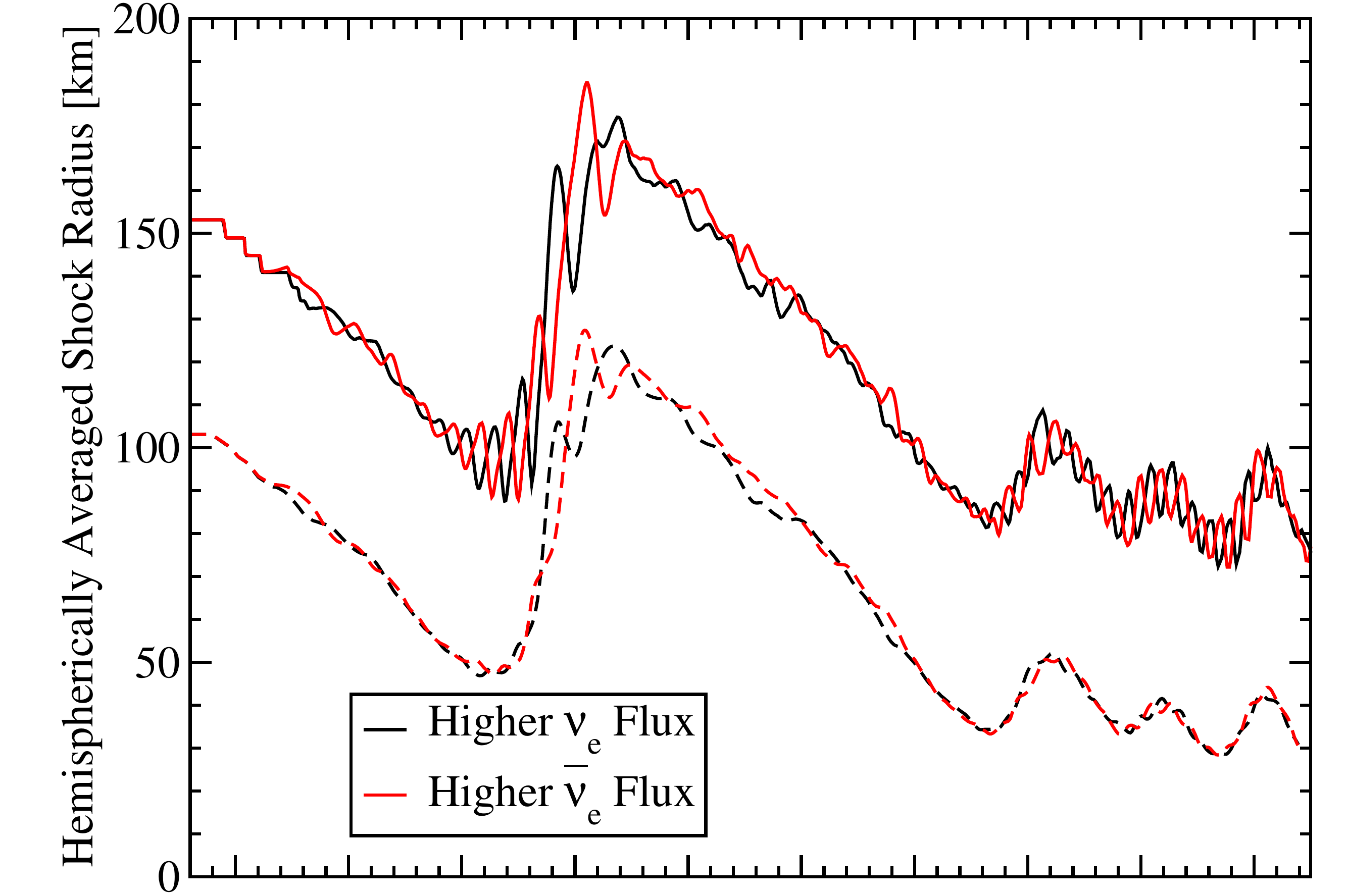}\\
  \includegraphics[width=0.33\textwidth]{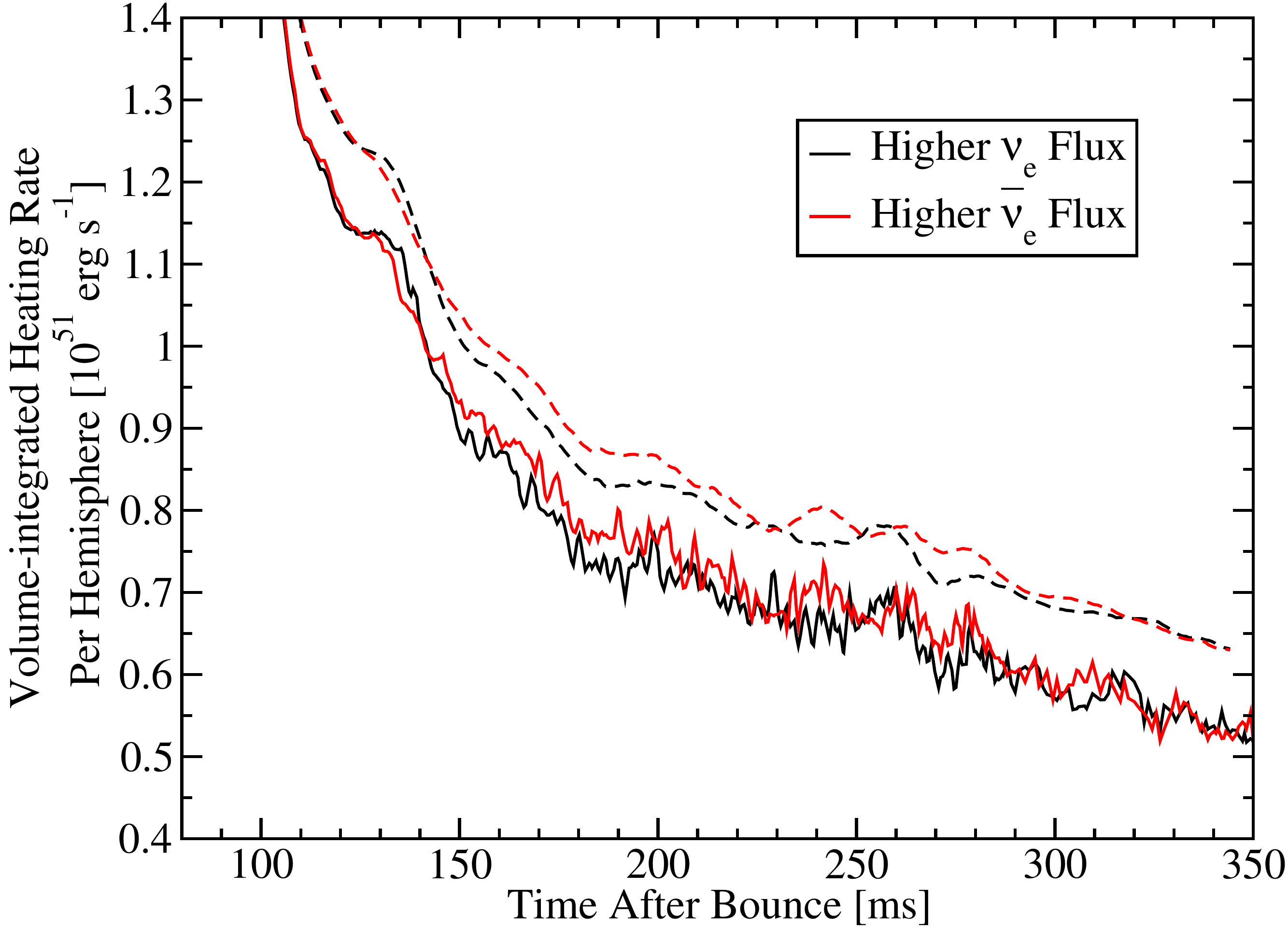}
  \includegraphics[width=0.33\textwidth]{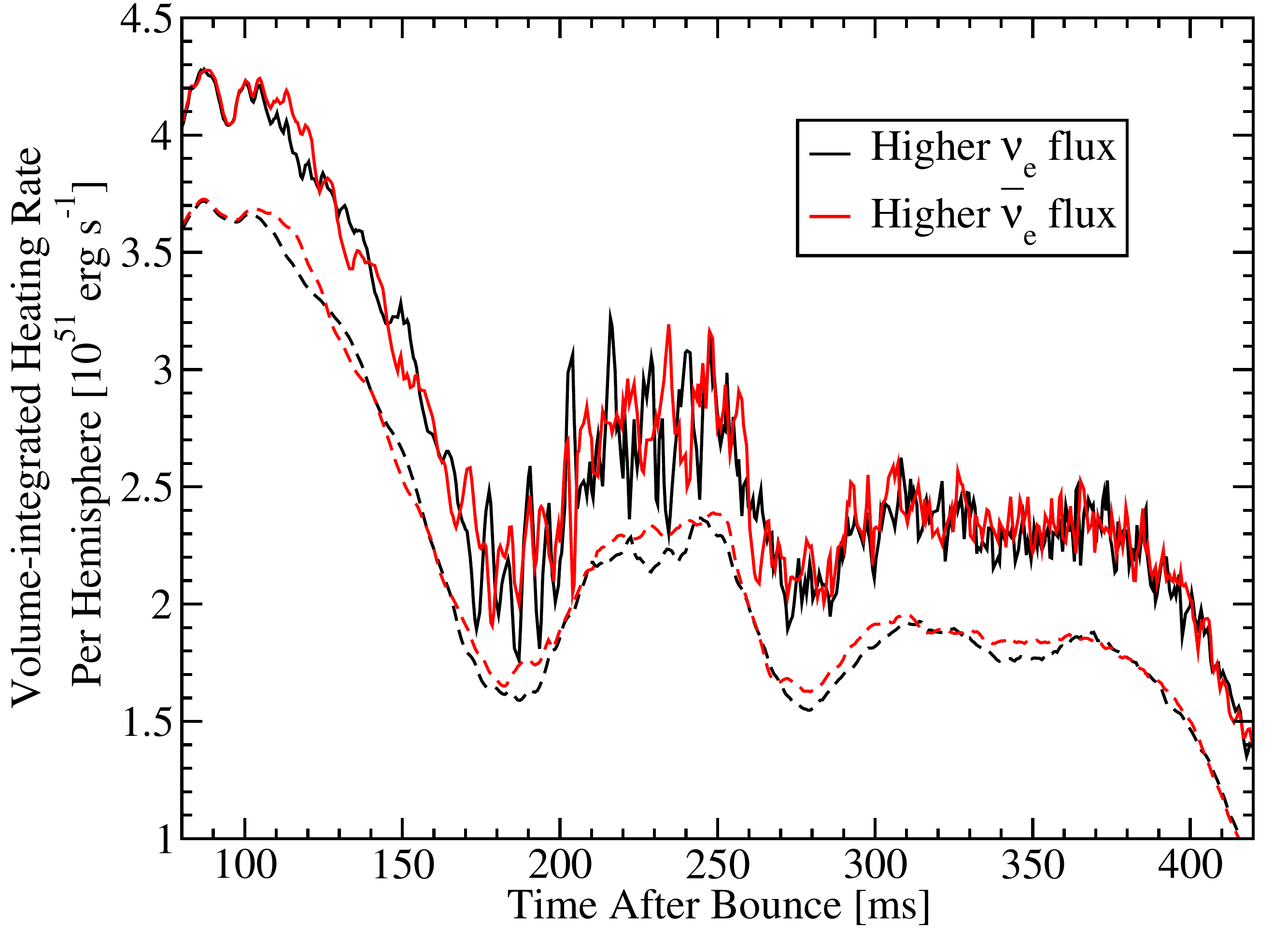}
  \includegraphics[width=0.33\textwidth]{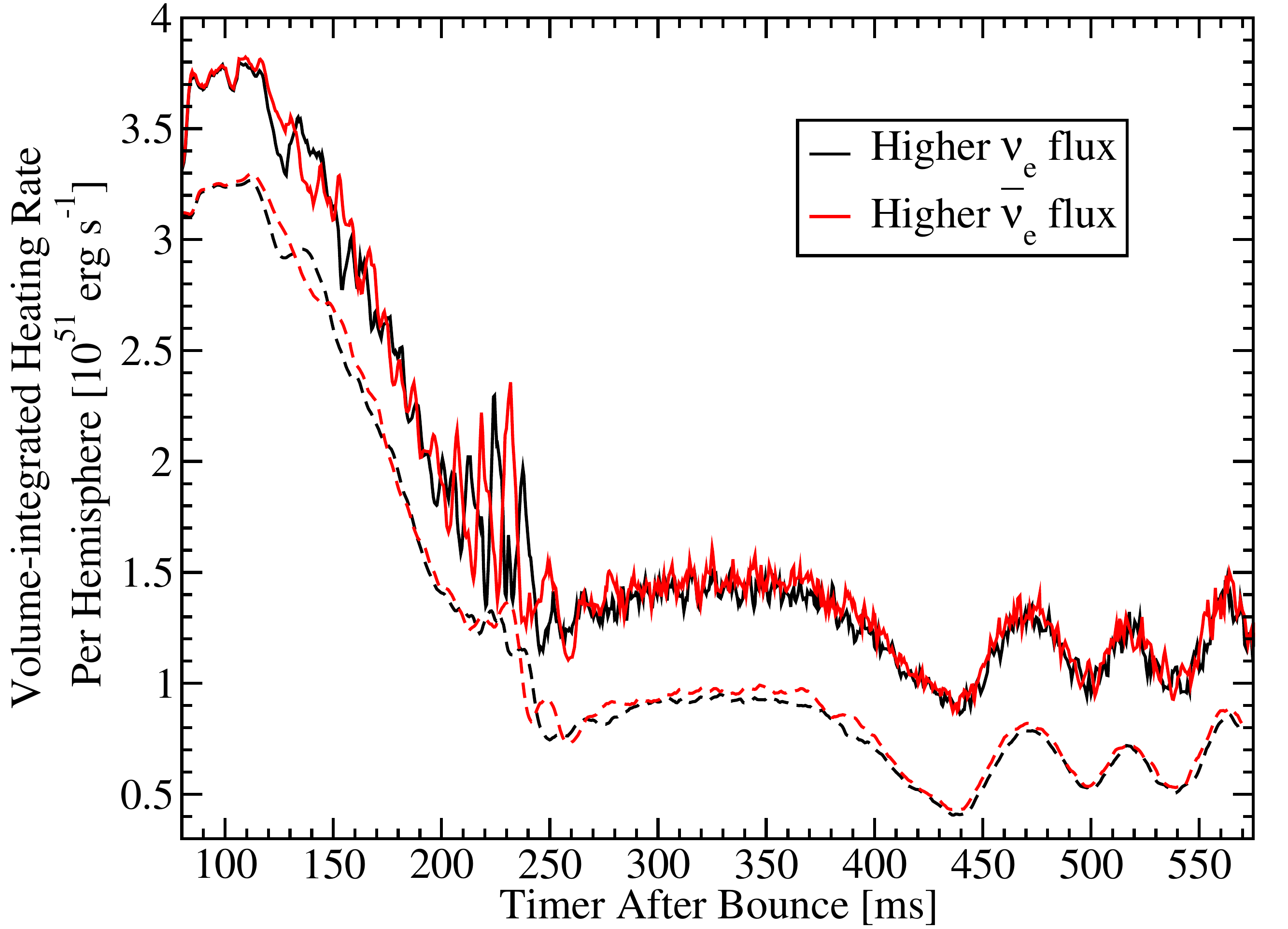}
  \caption{Time evolution of quantities in the postshock accretion layer
    of the 11.2\,$M_\odot$ ({\em left column}), 20\,$M_\odot$
    ({\em middle column}), and 27\,$M_\odot$
    ({\em right column}) models for the hemisphere of large
    lepton-number flux, i.e., large $\nu_e$ flux (black lines), and the
    opposite hemisphere of larger $\bar\nu_e$ flux (red lines).
    The dashed lines show running averages over 10\,ms intervals,
    shifted vertically for better visibility.
    The two hemispheres of the analysis follow the slowly
    drifting LESA dipole direction.
    {\em  Top:\/} Mass accretion rate, measured half way between the
    average radius of the stalled SN shock and the average gain radius.
    {\em Middle:\/} Average shock radius.
    {\em Bottom:\/} Volume-integrated neutrino-heating rate in the
    gain layer.
    The plots visualize important components of the
    crucial feedback loop consisting of asymmetric accretion rate,
    asymmetric lepton-number flux, asymmetric neutrino heating rate, and
    dipole deformation of the shock front as explained in the main text
    and Fig.~\ref{fig:cartoon}. In all models the mass accretion rate is
    systematically and persistently higher in the hemisphere with larger
    $\nu_e$ flux (black lines), whereas the shock radius and heating rate
    are greater in the opposite hemisphere.
    In the 20 and 27\,$M_\odot$ models with episodes of prominent SASI
    activity (marked in Fig.~\ref{fig:dipole11})
    the effect is most clearly visible outside of the SASI phases,
    when convective overturn is the dominant hydrodynamic instability
    in the neutrino-heated postshock layer.
  \label{fig:accretionasymmetry}}
\end{center}
\end{figure*}

The most plausible physical connection between the asymmetries deep in the
PNS and hydrodynamical properties of the envelope derives from asymmetric
mass-accretion flows. To study this hypothesis we consider the time evolution
of the mass accretion flow in our usual two hemispheres defined by maximal
and minimal lepton-number emission, shown in
Fig.~\ref{fig:accretionasymmetry} for all three considered progenitors
(top row). At a time when the dipole begins to
form in earnest, we notice a significant hemispheric asymmetry of the mass
accretion rate such that the hemisphere of larger lepton-number flux also has
the systematically larger mass accretion rate.

%------------------------------------------------------------------------------------------
%------------------------------------------------------------------------------------------

%------------------------------------------------------------------------------------------
\begin{figure*}
\begin{center}
  \includegraphics[width=.45\textwidth]{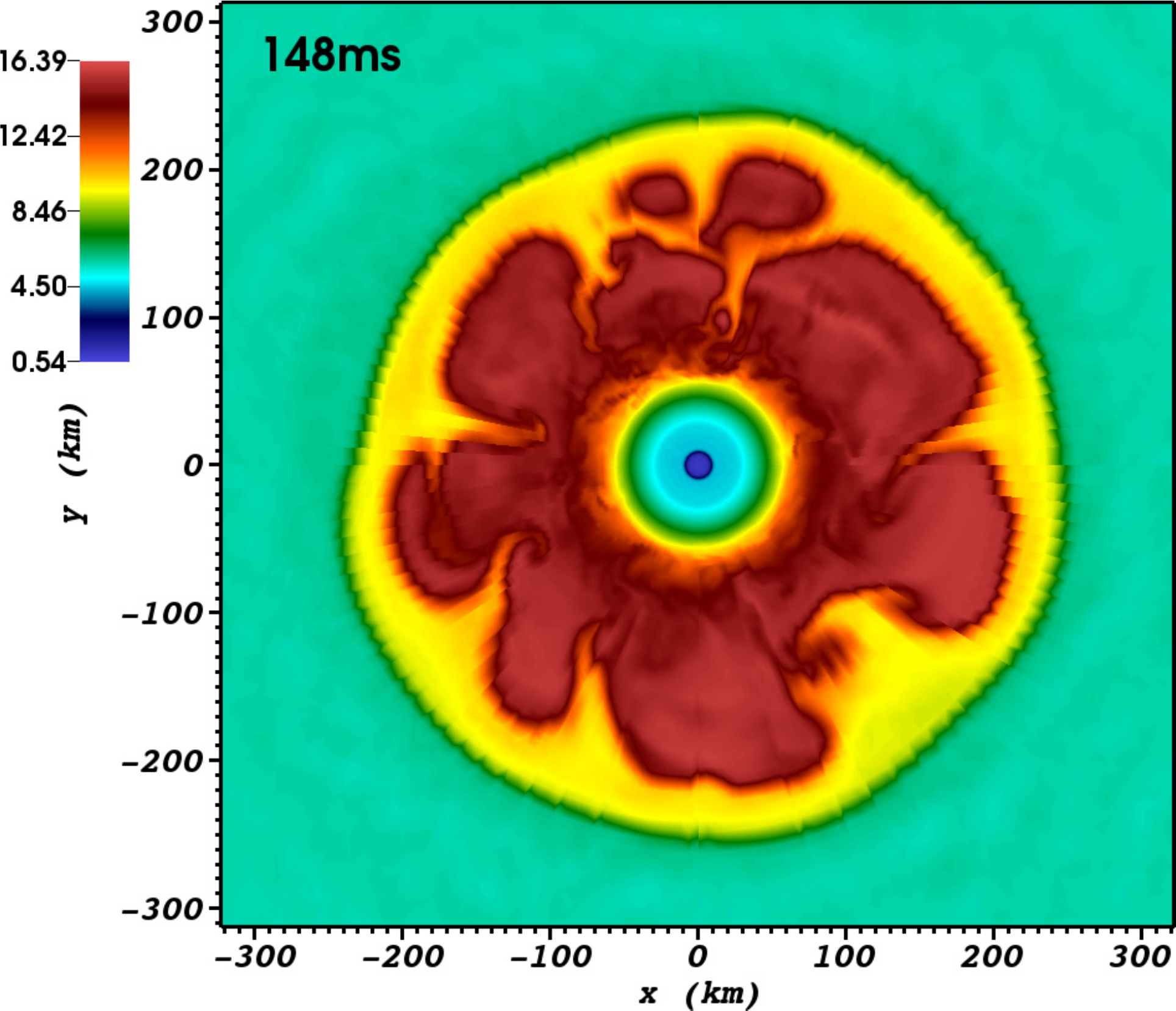}\hspace{8pt}
  \includegraphics[width=.45\textwidth]{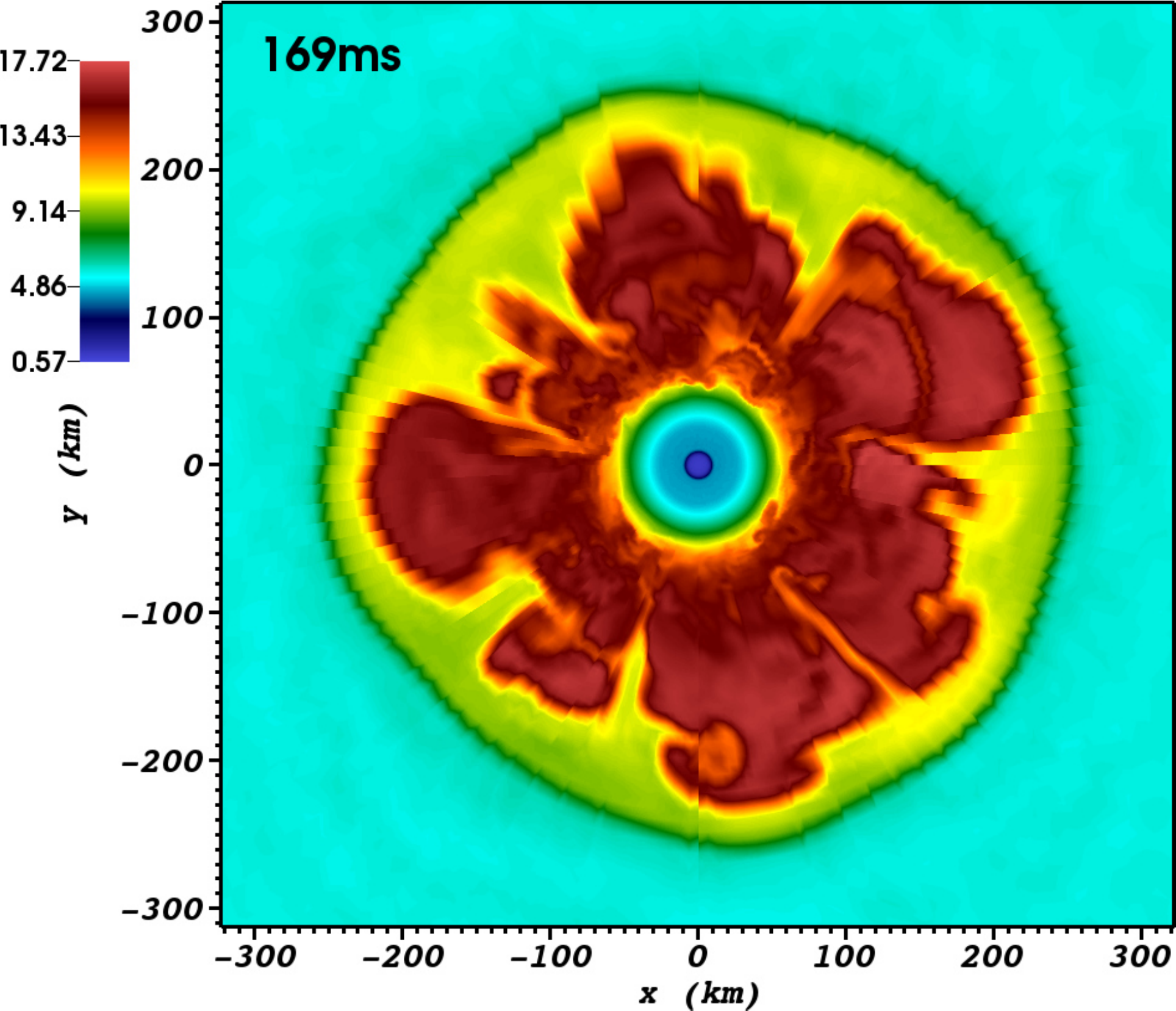}\\[3pt]
  \includegraphics[width=.45\textwidth]{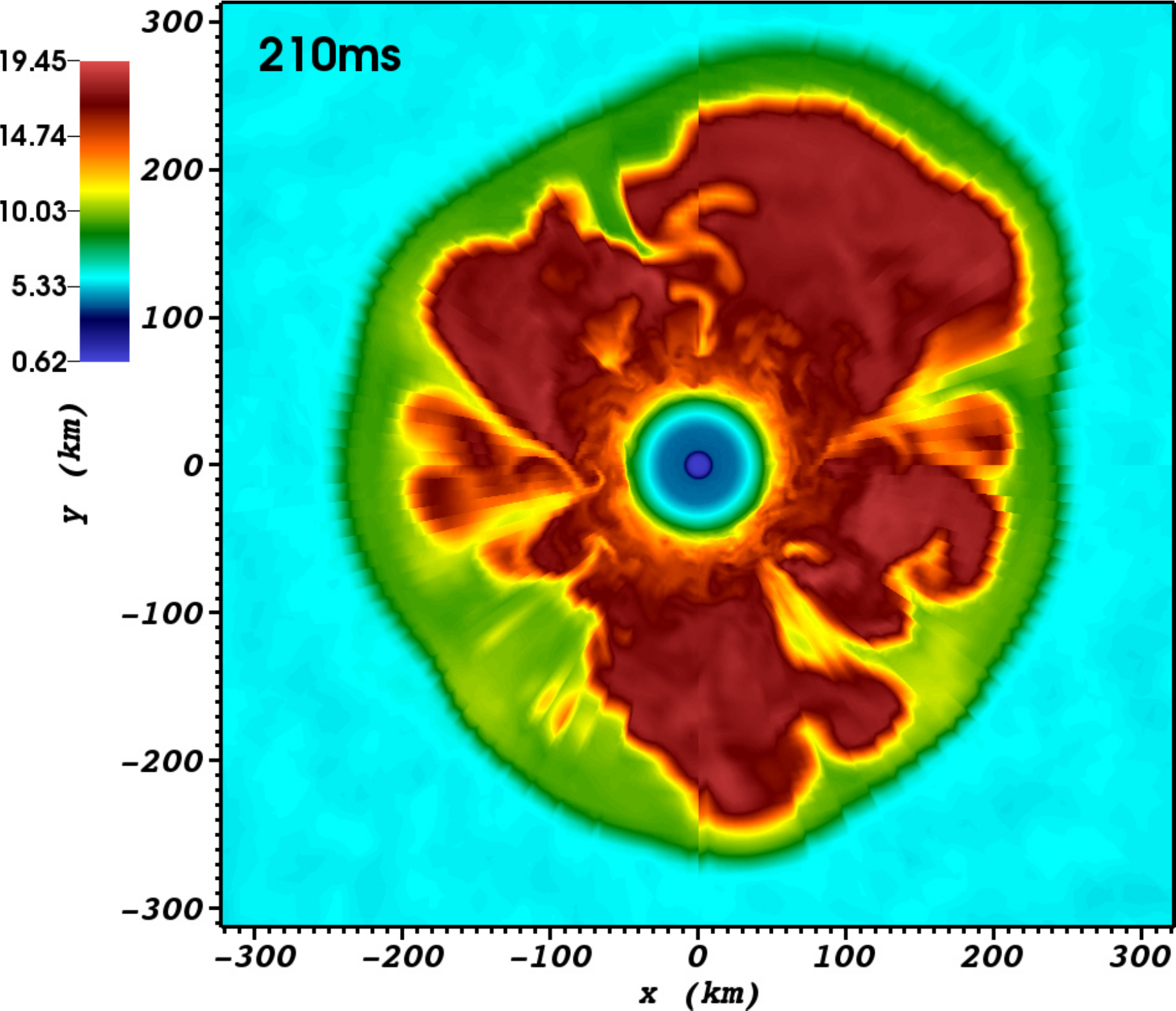}\hspace{8pt}
  \includegraphics[width=.45\textwidth]{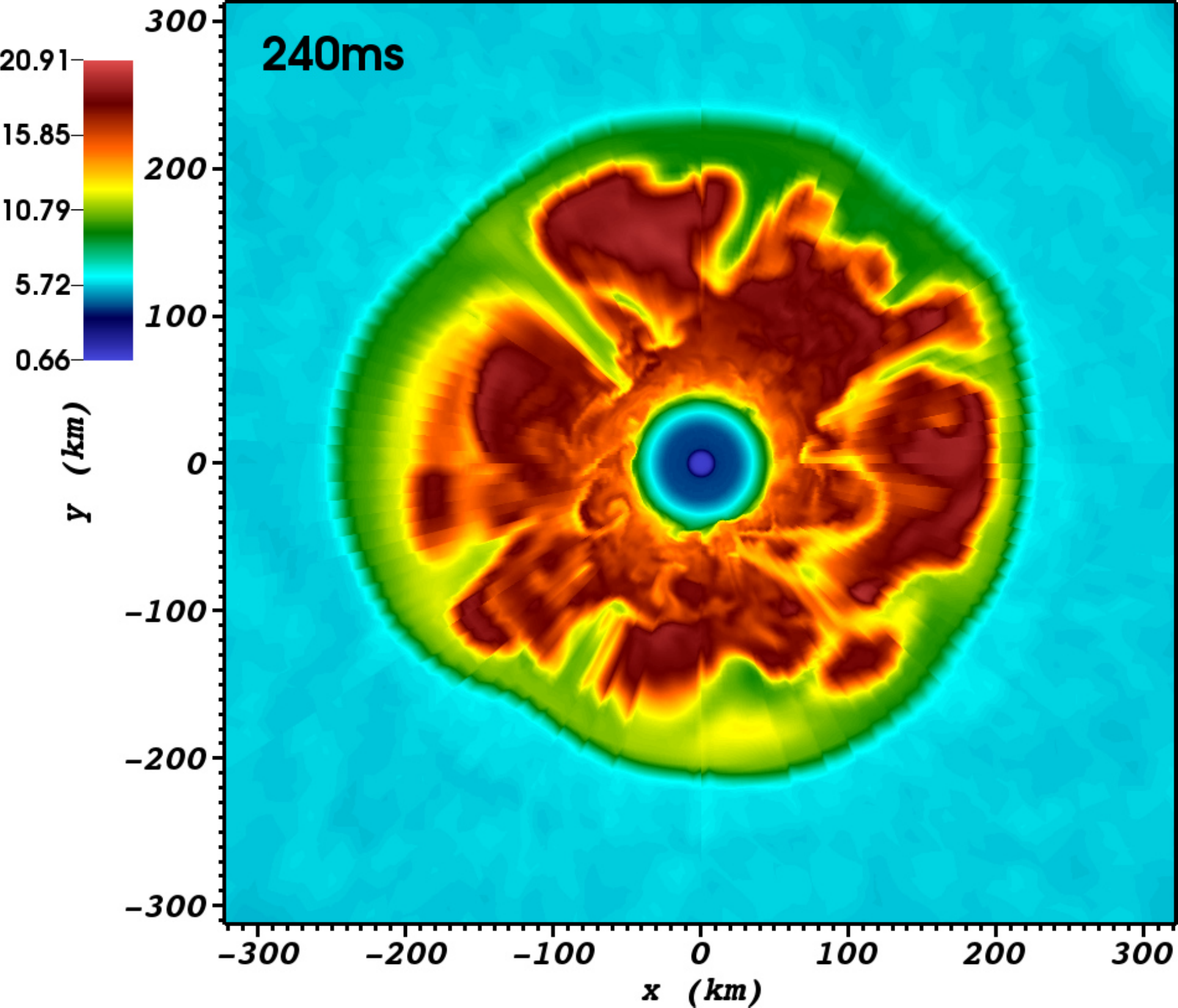}
  \caption{Entropy distribution in the postshock region of the
    11.2\,$M_\odot$ simulation at the indicated p.b.\ times. The cut plane
    is the same as in Figs.~\ref{fig:neutrinoasymmetry3} and \ref{fig:yecuts},
    i.e., it contains the dipole direction, where
    downward is the direction of maximum lepton-number flux.
    While a global asymmetry between the upper and lower
    hemispheres is hardly visible during the early stages
    when the lepton emission anisotropy just begins to develop (see
    Fig.~\ref{fig:fluxvariations}), stronger convection in the upper
    hemisphere during the later stages ({\em bottom panels}) can be inferred
    from the larger buoyant bubbles and larger average shock radius on this
    side.
\label{fig:entropycuts}}
  \end{center}
\end{figure*}
%------------------------------------------------------------------------------------------

This flow fluctuates strongly because of convective perturbations
and (in the 20 and 27\,$M_\odot$ models) SASI mass motions, but on
average exhibits a time-dependent anisotropy of 30--50\% for the
11.2\,$M_\odot$ case and of 10--25\% for the other two progenitors.
On average it
carries considerably more fresh lepton number to the central compact object
on one side than on the other. The lepton-rich flow partially deleptonizes
by neutrino emission before it spreads out below the PNS surface to settle
into the PNS mantle\footnote{Consistent with our discussion of the radial
evolution of the lepton-emission dipole in Sect.~\ref{sec:dipole-radius} that
was based on Fig.~\ref{fig:neutrinoasymmetry1}, we now independently confirm
that the lepton-number loss ($\Delta Y_e\approx 0.4$) associated with the
accretion-rate difference between the two hemispheres, $\Delta\dot M\lesssim
0.08\,M_\odot$\,s$^{-1}$ for the 11.2\,$M_\odot$ model
(Fig.~\ref{fig:accretionasymmetry}, left column), accounts for a
lepton-number flux of at most 4$\times 10^{55}$\,s$^{-1}$ and therefore can
explain at most 25\% of the lepton-emission dipole.}, but it is still more
lepton-rich than the deleptonized material that is already present in this
region.  Because it is specifically lighter (containing a higher number
fraction of electrons and protons instead of heavier neutrons), the
lepton-rich gas does not efficiently mix with the deleptonized plasma.
Instead, it accumulates on one side of the PNS and pushes the more
deleptonized fluid out of the way, towards the opposite hemisphere.

The full explanation for the $Y_e$ distribution is actually even more
complex. The accretion inflow of lepton number cannot explain the entire
$Y_e$ asymmetry in the PNS mantle. On the one hand, the accretion downflows
deleptonize nearly identically during their infall from different directions,
fairly independently of the local mass-flux density in the convective
downdrafts that carry the accretion flow towards the PNS. Some of the
greenish areas in the lower hemisphere of Fig.~\ref{fig:yecuts}, which have
no counterpart on the upper side, have possibly inherited their electons from
accretion flows. However, the red, orange and yellow bulges, which reach
outward from the dense, high-$Y_e$ core most prominently in the lower
hemisphere, cannot come from the same origin. They are located in the
convective shell of the PNS, which is interior to the neutrinospheres, and
they thus suggest an enhanced efficiency of the convective lepton-number
transport out from the inner core. The convective region inside of the PNS
can be recognized as a circular ring of short-wavelength color variations
between $\sim$12\,km and $\sim$25\,km in the right panel of
Fig.~\ref{fig:neutrinoasymmetry3}. This region is more pronounced in the
lower hemisphere, indicating stronger PNS convection effects in this
direction, by which lepton number is pulled up from the central, lepton-rich
high-density core. This dredge-up explains the presence of high-$Y_e$ patches
(red, orange, and yellow in Fig.~\ref{fig:yecuts}) extending outward from the
inner core region.

Convective activity inside the PNS is constrained to a shell that surrounds
the convectively stable core. The size of the convection cells is
roughly defined by the radial scale of the active layer. Therefore, a
volume-filling convective dipole mode is absent and the formation of a
low-mode dipolar asymmetry of the lepton transport is puzzling. This argument
supports a driving mechanism connected to the global accretion asymmetry. It
seems that convective transport of lepton (electron) number in the deeper PNS
mantle region might be amplified when accretion downflows impinge into the
PNS, presumably because shear flows, turbulence, and gravity waves, which are
instigated in the outer layers of the PNS mantle as a consequence of such
violent impacts, can act in a destabilizing way on convectively nearly
neutral or marginally stable statifications. Also the contraction of the PNS
might play an important role during the growth phase of the dipole mode.
Postshock overturn develops in parallel to a phase of strong PNS contraction.
Since the NS mantle layer settles to increasing densities, convection cannot
reach a steady state in which perturbations connected to the impact of
accretion downflows may be washed out on the typical timescales of
small-scale lepton transport. Instead, hemispheric asymmetries imprinted at
early times, when matter still resides at low densities with short restoring
timescales, may be carried to higher densities and thus may get frozen in for
much longer periods, becoming the initial seeds for a subsequent growth
towards the global dipolar asymmetry. A more detailed analysis of the
dynamical interaction of accretion stream impacts in the outer layers of the
PNS and convection in the contracting PNS mantle is deferred to future~work.

The low-mode dipolar asymmetry of the lepton-number distribution in the
PNS is therefore a consequence of the strong gravity, which on the
one hand tends to spread out flows and to smoothen structures along
equipotential surfaces and on the other hand defines an
environment where buoyancy effects play an extremely important role.

\subsection{Asymmetry of gain-layer convection and shock-wave radius}
\label{sec:shock-wave}

One may wonder if the asymmetric accretion flow has a visible correspondence
in the structure of the large-scale convection in the gain region. To
investigate this question we show in Fig.~\ref{fig:entropycuts} entropy
distributions of the 11.2\,$M_\odot$ model
in the same cut planes that were used in Fig.~\ref{fig:yecuts}
for the $Y_e$ distribution, i.e., the plane contains the dipole direction and
downward is the direction of maximum lepton-number flux. In contrast to the
clear hemispheric differences of the mass accretion rate in the postshock
layer (Fig.~\ref{fig:accretionasymmetry}, top left panel),
associated differences in the region of convective overturn are not
clearly visible in the entropy cuts or in the earlier visualizations of
Fig.~\ref{fig:3dconv11}. A closer inspection of the entropy cuts
(Fig.~\ref{fig:entropycuts}) reveals that in the lower two panels ($t=210$
and 240\,ms p.b.) the convective plumes are bigger and push the shock to a
slightly larger radius in the upper hemisphere. Moreover, the convective
downdrafts that carry the accretion flow to the PNS are more numerous (lower
left panel) and the flow close to the PNS is more vigorous (lower right
panel) on the opposite side. Because the convective mass motions are highly
turbulent and time variable, however, one should be cautious with conclusions
based on selected snapshots.

To be more quantitative, the average shock radii in both hemispheres are
shown in the middle panel (left column)
of Fig.~\ref{fig:accretionasymmetry}. Indeed, from
$\sim$150\,ms onwards a clear and persistent difference, aside from
fluctuations, of 10--20\,km (6--7\% of the average shock radius) is found.

The one-sided increase of the
average shock-wave radius is actually the physical cause for
the accretion-flow asymmetry, because the spherical infall from larger radial
distances is deflected and channelled preferentially toward the hemisphere of
smaller shock-wave radius. One should notice the anti-correlation between
mass-accretion flow and shock-wave radius in
Fig.~\ref{fig:accretionasymmetry}---the hemisphere of increased accretion
flow (black line) is the hemisphere of reduced shock-wave radius. It is
important in this context to remember that the preshock accretion flow is
spherically symmetric except for imposed, small-scale random perturbations.

\subsection{Asymmetry of neutrino-heating rates}
\label{sec:neutrino heating}

But how can this dipolar shock deformation be maintained in a persistent
and long-lasting
manner despite vigorous convection behind the shock? Here an interesting,
self-sustaining feedback mechanism comes into play, in which the
lepton-emission dipole asymmetry itself facilitates, supports, and stabilizes
the conditions for its existence. The key point is that the lepton-flux
asymmetry implies a neutrino-heating rate asymmetry in the gain region below
the shock. In particular, the heating rate is larger in the hemisphere of
smaller lepton-number flux (red lines in Fig.~\ref{fig:accretionasymmetry}),
which is the hemisphere of more vigorous gain-region convection and increased
shock-wave radius. In the bottom panels of Fig.~\ref{fig:accretionasymmetry}
we show volume-integrated heating rates in the gain layers of both
hemispheres as functions of time, and indeed there is a systematic offset
between the two hemispheres. Stronger neutrino heating leads to more powerful
convective buoyancy and this naturally pushes the radius of the stalled shock
farther out in one hemisphere compared to the opposite hemisphere, where the
neutrino-energy deposition is weaker. So in the hemisphere of small
lepton-number flux we have an increased heating rate, increased shock-wave
radius, reduced mass-accretion flow, and therefore reduced lepton-number
flux.

%------------------------------------------------------------------------------------------
\begin{figure}
\begin{center}
  \includegraphics[width=.85\columnwidth]{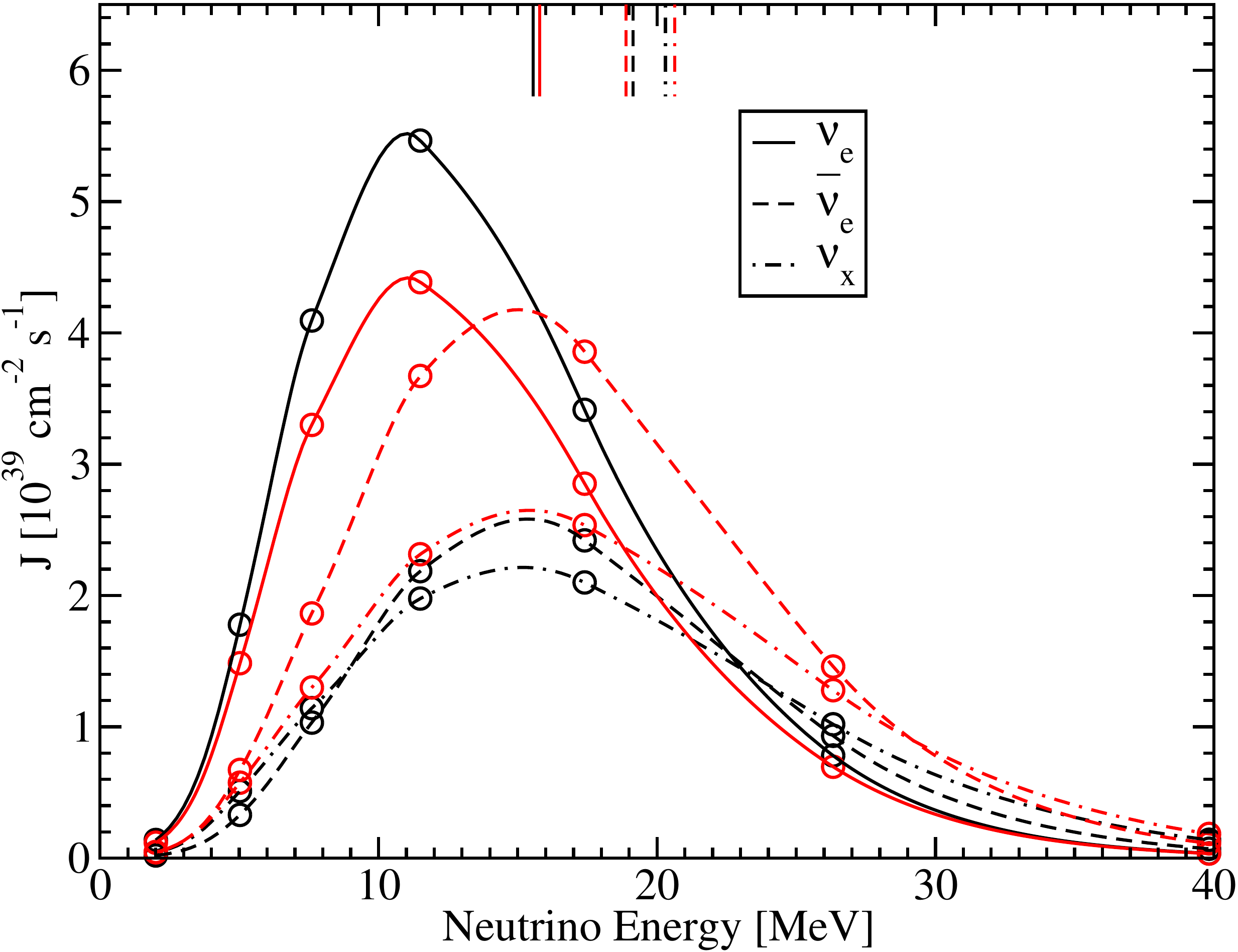}
  \caption{Energy spectra of $\nu_e$, $\bar\nu_e$, and $\nu_x$ for the
  $11.2\,M_\odot$ model at 210\,ms p.b.\
  (same time as in Fig.~\ref{fig:neutrinoasymmetry1}).
  The spectra are for rays in the direction of maximal (black)
  and minimal (red) lepton-number flux, evaluated at a distance of 400\,km in the
  comoving frame of the accretion flow, which is spherically symmetric at this radius.
  We provide the monochromatic energy moment
  $J=\Delta E\,c/(4\pi\,\Delta\epsilon_\nu)$ with $\Delta E$ being the energy
  density in neutrino energy bin $\Delta\epsilon_\nu$. The tick marks
  at the upper edge of the plot mark the rms energies.
  The neutrino spectral shape is very similar in these opposite directions for
  each species, whereas the differences in overall normalization reflect the
  dipolar flux asymmetries.
\label{fig:neutrinoasymmetry2}}
\end{center}
\end{figure}
%------------------------------------------------------------------------------------------

In order to further discuss the heating asymmetry we now consider the
characteristic spectral properties of the radiated neutrinos for our
11.2\,$M_\odot$ model. The $\bar\nu_e$
are emitted by the PNS with significantly higher average energies than
$\nu_e$. Typically, $\langle\epsilon_{\bar\nu_e}\rangle$ exceeds
$\langle\epsilon_{\nu_e}\rangle$ by 3--3.5\,MeV, averaged over all emission
directions as illustrated by Fig.~\ref{fig:neutrinoemission}. These spectral
differences also exist separately in both hemispheres, and they manifest
themselves also in the rms energies as shown in
Fig.~\ref{fig:neutrinoasymmetry2}, where we display the emitted energy
spectra for our previous 210\,ms snapshot of the $11.2\,M_\odot$ model.
Comparing the radiated energy spectra of $\nu_e$, $\bar\nu_e$, and $\nu_x$ on
two selected radial rays close to the directions of maximal (black) and
minimal lepton-number flux (red), the spectra differ primarily in the
normalizations, whereas the spectral shapes are very similar
(Fig.~\ref{fig:neutrinoasymmetry2}). Correspondingly, the rms energies,
indicated by tick marks at the upper edge of the plot, are 15.6, 19.1 and
20.3\,MeV for $\nu_e$, $\bar\nu_e$ and $\nu_x$, respectively, in the
direction of maximum lepton-number flux, and very similar values of 15.8,
18.9 and 20.6\,MeV are found in the opposite direction. The normalized
moments of the energy spectra for each neutrino type are nearly identical in
both hemispheres.

The increased heating rate in the hemisphere of
small lepton-number flux (more similar $\nu_e$ and $\bar\nu_e$ number fluxes)
can now be understood as follows.
For the dominant processes of $\nu_e$ absorption on free neutrons and
$\bar\nu_e$ absorption on free protons the heating rate per nucleon can be
approximated by (cf.\ \citealt{Janka_2001})
\begin{equation}
\dot q \propto \frac{\sigma_0}{r^2}\left ( L_{\nu_e}
                                   \left\langle\epsilon_{\nu_e}^2\right\rangle
                                   Y_n
                                   + L_{\bar\nu_e}
                                   \left\langle\epsilon_{\bar\nu_e}^2\right\rangle
                                   Y_p
                                   \right ) \,,
\label{eq:heatingrate}
\end{equation}
where $\sigma_0$ is the normalizing cross section, $r^{-2}$ describes the
radial flux dilution far away from the neutrinosphere, $Y_n$ and $Y_p$ are
the number fractions of free neutrons and protons, respectively, $L_\nu$
stands for the neutrino luminosities, and $\langle \epsilon_\nu^2 \rangle$
denotes the squared rms energy of the energy flux. Since $\bar\nu_e$ are
radiated with higher rms energies, the neutrino heating will be stronger in
the hemisphere where the $\bar\nu_e$ number emission is relatively enhanced,
despite the nearly isotropic luminosity sum of $\nu_e$ plus $\bar\nu_e$. The
effect is amplified by the fact that in this hemisphere, the $\bar\nu_e$
energy flux even exceeds that of $\nu_e$ (cf.\ Fig.~\ref{fig:skymaps11},
lower left panel).

Quantitatively, the numerical difference between the heating rates in the two
hemispheres seen in Fig.~\ref{fig:accretionasymmetry} (bottom left panel) is
roughly 5\%. This finding can easily be verified by estimating the ratio of
the heating rates between the directions of low and high lepton-number flux,
$\dot q_\mathrm{low}/\dot q_\mathrm{high}$. Using Eq.~(\ref{eq:heatingrate}),
assuming $Y_p$ and $Y_n$ to vary little with direction, and adopting a ratio
of the squared rms energies of $\langle \epsilon_{\bar\nu_e}^2\rangle/
\langle \epsilon_{\nu_e}^2\rangle \approx 1.46$ in both hemispheres
(Fig.~\ref{fig:neutrinoasymmetry2}) as well as amplitudes of 10--15\% for the
dipolar asymmetry of the $\nu_e$ and $\bar\nu_e$ energy fluxes
(Figs.~\ref{fig:fluxvariations} and \ref{fig:skymaps11}), we obtain $\dot
q_\mathrm{low}/\dot q_\mathrm{high}\sim 1.04$--1.06. The numerical results in
the bottom panels of Fig.~\ref{fig:accretionasymmetry} are based on an
integration over the volumes of the gain layer in both hemispheres
while these hemispheres were moved with the slowly wandering 
direction of the LESA dipole.

%, taking
%  into account only the volumes of rising plumes but excluding the regions of
%  supersonic accretion downflows, because material in the convective downdrafts
%  ends up in the cooling layer. Energy deposited by neutrino heating in the
%  downflows is reemitted by neutrino radiation in the cooling layer and
%  therefore has no direct effect on the shock behavior.

\subsection{Asymmetries in models with SASI activity}
\label{sec:modelswithsasi}

As expected from the fact that the neutrino emission exhibits all the
characteristic LESA features also in our 20 and 27\,$M_\odot$ models,
the hemispheric asymmetries described above mostly for the 11.2\,$M_\odot$
case are also found in the other two progenitors
(Fig.~\ref{fig:accretionasymmetry}). However, the SASI mass motions in
these cases lead to short-timescale neutrino-emission modulations
superimposed on the dipolar lepton emission asymmetry (see
\citealt{Tamborra_2014}), and the large-amplitude SASI-induced variations
of the postshock accretion flow
can mask the global, hemispheric differences of the mass accretion rate,
shock radius, and neutrino-heating rate associated with
the LESA phenomenon during the phases of strong SASI activity.

This problem for the analysis is more conspicuous for the 27\,$M_\odot$
simulation (Fig.~\ref{fig:accretionasymmetry}, right column), where
the higher mass accretion rate, smaller shock radius and neutrino
heating rate in the hemisphere of excess $\nu_e$ flux are clearly
visible only during the SASI-quiet episodes, i.e. before 170\,ms and
between 260\,ms and 420\,ms. In constrast, the familiar hemispheric
asymmetries of these quantities can be seen at essentially all times
in the 20\,$M_\odot$ model, despite the violent SASI in the time interval
from 170\,ms to 305\,ms.

This difference between both models is connected to the different
relative orientations of the SASI plane and the LESA dipole vector in
these models during the first SASI episodes (cf.~Fig.~\ref{fig:lesasasisky}).
While in the 27\,$M_\odot$ model the LESA dipole lies in the plane of the
SASI spiralling, it is nearly orthogonal to the SASI plane in the
20\,$M_\odot$ case. In the former model the mass-accretion, gain-layer
convection, and heating asymmetries described in
Sects.~\ref{sec:mass-accretion}--\ref{sec:neutrino heating}
are therefore strongly perturbed by the violent mass flows associated
with the SASI, whereas in the 20\,$M_\odot$ model these hemispheric
asymmetries seem to be less affected by the SASI flows because they
proceed in the direction perpendicular to the SASI plane.
It is important to note that usually also during SASI-active phases
convective overturn as a consequence of neutrino heating is still present.

Therefore SASI mass motions in the postshock layer interfere with the
LESA phenomenon in different ways, depending on the orientation of
the LESA direction relative to the plane of SASI sloshing and spiralling.
If the lepton-emission dipole happens to coincide with the SASI plane
the violent modulations of the postshock accretion flow by the SASI
seem to be able to prevent further growth of the LESA dipole
(cf.~Sect.~\ref{sec:dipole-time}) and to enforce a gradual drift
of the dipole direction, see the right panels of Figs.~\ref{fig:dipole11}
and \ref{fig:lesasasisky} for the first SASI phase (between 170 and
260\,ms p.b.) of our 27\,$M_\odot$ model. If, in contrast, the LESA
vector is incidentally perpendicular to the SASI plane, the growth of
the LESA amplitude is not impeded and the LESA direction may describe
quasi-periodic wobbling around a mean orientation, see the SASI-active
episode of our 20\,$M_\odot$ simulation from 170\,ms to 305\,ms in
the middle panels of Fig.~\ref{fig:dipole11} and the left panels of
Fig.~\ref{fig:lesasasisky}.
The LESA-SASI interference is therefore complex and any behavior
intermediate between these extrema might be possible.

%\newpage

%------------------------------------------------------------------------------------------
\begin{figure*}
\begin{center}
  \includegraphics[width=0.735\textwidth]{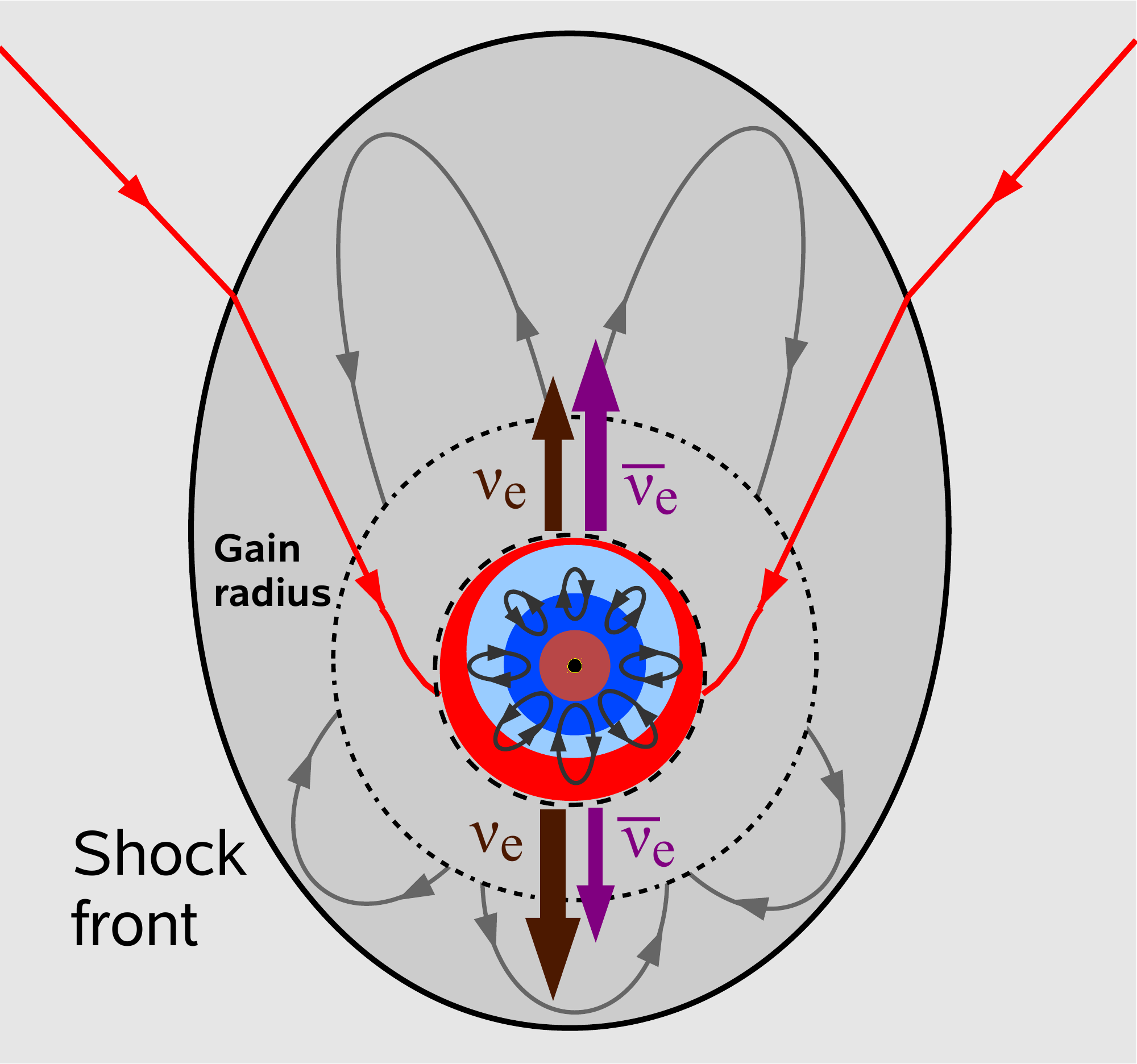}
\caption{Schematic visualization of the physics components that provide
  the feedback loop for the self-sustained lepton-emission asymmetry. The outer
  thick, solid black line indicates the accretion shock, whose dipolar
  deformation is strongly exaggerated. The dotted circular line marks
  the gain radius and the inner dashed circular line the
  neutrinospheres close to the surface of the PNS.  Inside the PNS,
  the bright-red and inner dark-red
  circular regions indicate the spherical density distribution around
  the mass center (small, black dot), whereas the displaced, blue
  circular shapes indicate schematically the deformation of the $Y_e$
  distribution (see Fig.~\ref{fig:yecuts}). The black elliptical loops
  interior to the neutrinospheres visualize convection inside the PNS, whereas
  the light gray loops visualize convective overturn between gain radius
  and shock. PNS convection is stronger in the lower hemisphere
  (cf.\ Fig.~\ref{fig:neutrinoasymmetry1}), whereas gain-region convection
  is more
  powerful on the opposite side. The red lines mark accretion-stream
  lines, which are deflected by the deformed shock front. The brown
  and magenta arrows show the hemispheric asymmetry of the $\nu_e$ and
  $\bar\nu_e$ energy fluxes. Note that the sum of the fluxes is nearly
  isotropic, showing only a percent-level dipole variation,
  whereas the hemispheric differences of the
  $\nu_e$ and $\bar\nu_e$ number and energy fluxes can reach 20--30\%
  of their average values (cf.\ Figs.~\ref{fig:fluxvariations} and
  \ref{fig:skymaps11}). While the convective overturn in the
  neutrino-heating layer fluctuates strongly in time, the
  asymmetry of the lepton-number distribution in the PNS mantle layer
  and the corresponding anisotropic lepton-number emission as well as
  the deformation of the accretion shock can be stable for hundreds of
  milliseconds. \label{fig:cartoon}}
\end{center}
\end{figure*}
%------------------------------------------------------------------------------------------

%%%%%%%%%%%%%%%%%%%%%%%%%%%%%%%%%%%%%%%%%%%%%%%%%%%%%%%%%%%%%%%%%%%%%%%%%%%%%%%%%
\section{Overall Picture of the LESA Phenomenon}
\label{sec:explanation}
%%%%%%%%%%%%%%%%%%%%%%%%%%%%%%%%%%%%%%%%%%%%%%%%%%%%%%%%%%%%%%%%%%%%%%%%%%%%%%%%%

\subsection{Two interlocking cycles}

From our discussion so far a picture of the LESA phenomenon emerges that
involves a machinery consisting of two major interacting parts. One consists
of asymmetric PNS convection and concomitant asymmetric lepton-number
emission. The other consists of asymmetric mass-accretion regulated by
asymmetric neutrino emission through asymmetric neutrino heating in the gain
layer.

We re-capitulate and summarize the cog-wheels of this machinery in the sketch
of Fig.~\ref{fig:cartoon}. It is oriented in the same way as our previous
cut-planes, i.e., the lepton-number emission maximum is in the downward
direction. In our line plots, black curves correspond to properties in the
downward direction or lower hemisphere, red curves to the upward direction
(or hemisphere), which is the
hemisphere of minimal lepton-number flux emission.

In the inner parts of Fig.~\ref{fig:cartoon}, the mass center is marked by a
black dot, surrounded by concentric red circular regions which show the
density stratification inside the newly formed NS. The long-dashed line
indicates the location of the neutrinosphere(s). Blue circles represent
levels of constant electron fraction ($Y_e$). The upward displacement of the
light-blue region visualizes schematically a dipolar asymmetry of the
deleptonization in the NS mantle region enclosed by the neutrinosphere: While
in the top hemisphere the layer below the neutrinosphere has deleptonized
more strongly due to the preceding and ongoing emission of electron neutrinos
(the light-blue region there is bigger), the bottom hemisphere contains a
larger fraction of electrons (indicated by the larger red area).

This dipolar asymmetry of the $Y_e$ distribution in the PNS mantle is a
consequence of a pronounced hemispheric difference in the mass flow towards
the compact object. The latter accretes matter at a significantly higher rate
on one side and thus receives a greater inflow of fresh lepton number in this
hemisphere (bottom in Fig.~\ref{fig:cartoon}). Even more important, however,
is an indirect effect connected with the mass accretion. As the accretion
streams impinge onto the PNS they instigate turbulence and gravity waves,
which enhance convection in the deeper mantle layers of the PNS, dredging up
electrons from the central, lepton-rich dense core (visualized in
Fig.~\ref{fig:cartoon} by the larger convective loops in the lower hemisphere).
Lepton number carried away by the outgoing neutrino fluxes is therefore
replenished by electrons delivered by accretion downdrafts as well as those
pulled outward in convective flows that reach deeper into the PNS core. Since
the underlying processes are nonisotropic and more efficient in one
hemisphere, the deleptonized mantle layer of the PNS exhibits a long-lasting
dipolar asymmetry of the $Y_e$ distribution.

The anisotropic accretion flow towards the PNS is indicated by the two red
accretion-stream lines. Such an accretion asymmetry in the convectively
stirred postshock layer should be understood in a time-averaged sense,
emerging effectively from a strongly fluctuating accretion-flow pattern. The
convective activity in the region between gain radius (short-dashed line) and
shock is symbolized by the up- and downgoing loops for the convective
overturn motions of rising and sinking gas. The accretion asymmetry is caused
by a dipolar shock deformation, which---again in a time-averaged sense---is
associated with a larger radius of the accretion shock (thick, black outer
ellipsoidal line) in the upper hemisphere. This leads to a deflection of the
radial accretion flow when the collapsing matter crosses the shock front,
because the tangential velocity component is conserved whereas the
perpendicular velocity component is reduced by the flow deceleration
according to the shock-jump conditions. The deflection of the postshock flow
feeds, on average, the lower PNS hemisphere with a higher mass accretion rate.

In this picture, the dipole asymmetry of the neutrino lepton-number flux is a
self-sustained, self-stabilizing phenomenon, because the different spectral
properties of $\nu_e$ and $\bar\nu_e$ ensure more efficient neutrino heating
on the side of the lower mass accretion rate and thus lower radiated
lepton-number flux. Stronger heating in this hemisphere supports stronger
convection and a greater shock radius, thus maintaining the shock deformation
that is responsible for the deflection of the accretion flow and the enhanced
mass accretion rate reaching the opposite side of the PNS. Neutrinos
therefore play a crucial role in this nonradial instability, which leads to a
long-lasting, stable asphericity of the postbounce accretion situation.

%\newpage
\subsection{Memory effect in the PNS}

The LESA effect is not ``switched off'' by the appearance of strong SASI
activity, a surprising finding because one might imagine that violent
sloshing and spiral modes could interrupt the feedback loop which is the
driving cause for LESA. However, LESA has substantial inertia built into its
machinery, where the $Y_e$ distribution and related PNS convection asymmetry
play the role of a flywheel that keeps going even if the driving engine has
been temporarily disabled. Once SASI has subsided, the continuing asymmetric
lepton emission from the PNS region quickly restarts the engine and puts the
feedback loop back into operation. This picture is not necessarily
contradicted by the observation that the onset of SASI can considerably shift
the LESA dipole direction as seen, for example, after 170\,ms p.b.\ in the
$27\,M_\odot$ model (bottom right panel in Fig.~\ref{fig:dipole11}).
In fact, if our picture of an outer feedback loop coupled to asymmetric
mass accretion is correct, one would expect that massive, large-scale
perturbations of the postshock accretion flow like those connected
to the onset and presence of violent
SASI episodes can lead to a drift of the direction of the
lepton-emission dipole. Similarly, it does not appear astonishing that
vigorous, strongly time-dependent convective overturn activity in the
neutrino-heating layer creates stochastic variations and fluctuations
that can be strong enough to induce a gradual, slow movement of the
LESA dipole on secular timescales.

How long can the PNS memory effect last? The life time of the $Y_e$ asymmetry
in the PNS depends on two competing effects, on the one hand the inflow of
fresh electron number in the asymmetric accretion flow and caused by
convective transport out of the high-density PNS core, and on the other hand
the loss of lepton number by the anisotropic lepton fluxes, which strive for
destroying the hemispheric $Y_e$ asymmetry. The temporal decay of this
asymmetry in the PNS mantle can therefore be approximately described by the
following differential equation:
\begin{equation}
\frac{1}{2\,m_\mathrm{u}}\,
\frac{{\mathrm{d}}\left ( \Delta Y_e M_\mathrm{shell}
\right )}{{\mathrm{d}}t} = \frac{1}{m_\mathrm{u}}
\left [ \Delta\dot{M}\,\Delta Y_e + \left ( \frac{\delta (MY_e)}{\delta t}
\right )_\mathrm{mix}\right ] - \Delta N \,,
\label{eq:dipoledecay}
\end{equation}
where $\Delta Y_e$ is the difference of the electron fraction in the two
hemispheres within a shell of mass $M_\mathrm{shell}$, $\Delta \dot M$ is the
mass-accretion rate difference (which we assume to carry an excess lepton
fraction of $\Delta Y_e$ into the PNS mantle, $[\delta (MY_e)/\delta
t]_\mathrm{mix}$ is the rate of electron-number change
associated with enhanced convective
mixing, $\Delta N$ the neutrino-lepton flux difference between both
hemispheres, and $m_\mathrm{u}$ the baryon mass.
Equation~(\ref{eq:dipoledecay}) yields a rough estimate of the decay
timescale:
\begin{equation}
t_\mathrm{decay} \sim
\frac{0.5\,\Delta Y_e M_\mathrm{shell}}{m_\mathrm{u}\Delta N}
\,.
\label{eq:decaytime}
\end{equation}
In this expression we have ignored the positive contributions on the rhs of
Eq.~(\ref{eq:dipoledecay}) because the first term turns out to be subdominant
when numbers from the 11.2\,$M_\odot$
simulation ($\Delta Y_e\sim 0.05$, $\Delta \dot M\sim
0.05\,M_\odot$\,s$^{-1}$, cf.\ Figs.~\ref{fig:yecuts}
and~\ref{fig:accretionasymmetry}) are used, and the second term is difficult
to calculate without digging into the details of the dynamic interaction
between the PNS convection and the impact of accretion flows on the PNS. In
any case, the inflow of fresh lepton number associated with the positive
source term can only stretch the decay timescale so that our estimate
provides a firm lower limit. With numbers deduced from our analysis
of the 11.2\,$M_\odot$ model,
$M_\mathrm{shell}\sim 0.4\,M_\odot$, $\Delta
Y_e\sim 0.05$ (Fig.~\ref{fig:yecuts}), and $\Delta N\sim 12\times
10^{55}$\,s$^{-1}$ (Fig.~\ref{fig:neutrinoasymmetry1}), we obtain
$t_\mathrm{decay} \sim 100$\,ms. Therefore, once the lepton-emission dipole
has developed to its full strength, it will continue to exist for at least
100\,ms even if the asymmetries of accretion and PNS convection disappear.

%%%%%%%%%%%%%%%%%%%%%%%%%%%%%%%%%%%%%%%%%%%%%%%%%%%%%%%%%%%%%%%%%%%%%%%%%%%%%%%%%
\section{Summary and outlook}
\label{sec:conclusions}
%%%%%%%%%%%%%%%%%%%%%%%%%%%%%%%%%%%%%%%%%%%%%%%%%%%%%%%%%%%%%%%%%%%%%%%%%%%%%%%%%

In this paper we have described a new type of nonradial deformation mode that
we discovered in our 3D stellar core-collapse simulations using
energy-dependent, three-flavor neutrino transport, applied to progenitor
stars of 11.2, 20, and 27\,$M_\odot$. During the first 100--150\,ms of
postbounce evolution, a long-lasting, only slowly evolving
dipolar neutrino-emission asymmetry
establishes itself. It persists throughout the  postbounce accretion phase of
the stalled SN shock for hundreds of ms, even in those models that show
violent SASI activity for parts of their postbounce evolution. This
multi-dimensional phenomenon has not been identified unambiguously in any
previous 2D simulation. We call the new phenomenon LESA for ``Lepton-number
Emission Self-sustained Asymmetry.'' LESA emerges from an instability, i.e.,
the spherically symmetric state is not stable and the asymmetry grows from
any perturbation, no matter how small. In contrast to convection or
SASI, the nature of LESA is not simply hydrodynamical, but rather a
neutrino-hydrodynamical instability and as such the first of its kind
identified in the SN context.

The dipole mode manifests itself most conspicuously in the lepton-number flux
($\nu_e$ minus $\bar\nu_e$), whose dipole amplitude can reach 100\% of its
$4\pi$ directional average, i.e., in one direction the lepton-number flux can
exceed twice the average, in the opposite direction it can be even somewhat
negative ($\bar\nu_e$ number flux exceeds that of $\nu_e$). While dipole
amplitudes of the individual $\nu_e$ and $\bar\nu_e$ number and energy fluxes
of 10--20\% can be observed in our 3D calculations of the 11.2\,$M_\odot$
progenitor (and somewhat smaller ones in the 27\,$M_\odot$ model),
heavy-lepton neutrinos as well as the sum of $\nu_e$ and $\bar\nu_e$ exhibit
only percent-level dipolar luminosity asymmetries.

The neutrino-emission dipole originates from a hemispheric asymmetry of the
electron distribution in the mantle layer of the PNS interior to the
neutrinospheres, although the density stratification is perfectly spherical
and concentric in these regions of extreme gravitational field
strength. In the hemisphere of higher $\nu_e$ flux, convection in the
deeper layers of the PNS mantle is enhanced compared to the opposite side
and dredges up electrons more efficiently from the dense, lepton-rich central
core. These electrons are mixed outward towards the neutrinospheres and lead
to less deleptonized conditions in one hemisphere, supporting the higher
$\nu_e$ flux. The convective activity seems to be instigated and fostered by
shear flows, turbulent motions, and gravity waves caused by the violent
impacts of accretion streams in the neutrinospheric region. This
connection is suggested by the observed correlation between PNS-convection
asymmetry and a considerable asymmetry of the mass-accretion flow, which is
stronger in the hemisphere of enhanced PNS convection. The accretion
asymmetry also contributes to the lepton-number emission dipole, because the
accretion streams carry electron-rich matter and thus supply the PNS with
fresh lepton number, but this effect is subdominant.

The global accretion asymmetry is maintained by anisotropic neutrino heating
in the gain layer behind the stalled SN shock, because $\bar\nu_e$ leave the
neutrinosphere with higher mean energy than $\nu_e$. Therefore, neutrino
heating is stronger on the side of lower lepton-number flux, despite the
nearly isotropic energy flux of $\nu_e$ plus $\bar\nu_e$. Stronger neutrino
heating enhances convective overturn in the postshock layer, pushes the shock
to a larger stagnation radius and thus produces a dipolar deformation of the
shock surface. This shock deformation in turn deflects the accretion flow
falling through the shock and, in the time-averaged sense, amplifies the
accretion flux to the hemisphere of the PNS facing away from the greater
shock radius (Fig.~\ref{fig:cartoon}).

Anisotropic neutrino heating therefore establishes a feedback mechanism
between the neutrino-emission asymmetry on one side and shock deformation and
accretion asymmetry on the other. It thus mediates a complex, mutual
dependence between lepton-number transport by neutrino fluxes and convection
inside the PNS on the one hand and anisotropic convective overturn in the
gain layer on the other. This feedback, which involves neutrinos as crucial
players, allows the global dipolar asymmetry to become a self-sustained
phenomenon, which exists in stable conditions over many dynamical
timescales despite the presence of vigorous and highly time-dependent
convective overturn in the postshock region and even through phases of
violent SASI activity. Stochastic fluctuations of this convective overturn
or of the convection in the PNS mantle are probably responsible for
initiating the development of the hemispheric asymmetry. The convective SN
core seems to be generically unstable against such a dipolar mode of
asymmetry. LESA and SASI are independent phenomena, but the latter
can influence the former in a complex manner, depending on the relative
orientations of the associated vector directions.

LESA could have important implications for a variety of physical processes in
the SN core, most importantly nucleosynthesis
in the neutrino-heated ejecta, and potentially NS~kicks and
neutrino-flavor conversion.

Concerning nucleosynthesis, we recall that charged-current reactions of
$\nu_e$ and $\bar\nu_e$ with neutrons and protons do not only heat the SN
blast wave but also set the n/p ratio in the neutrino-driven outflow and thus
determine the nucleosynthesis conditions in the innermost SN ejecta. Since
$\nu_e$ absorption converts neutrons to protons while $\bar\nu_e$ captured on
protons create neutrons, the exposure to a higher flux of $\bar\nu_e$ tends
to neutronize the expelled matter. The lepton-number emission asymmetry
could persist until the explosion sets in or even beyond, fuelled by
continued anisotropic PNS convection and/or asymmetric accretion beyond the
onset of the explosion. In this case a considerable hemispheric asymmetry of
the electron fraction in the ejecta could be expected with possibly favorable
conditions for neutron-rich material in the direction where the $\bar\nu_e$
flux has its emission maximum. We speculate that the O-Ne-Mg core explosion
of \citet{Wanajo_2011}, where convective overturn but not SASI played a
role, may be a case where we have encountered consequences of the dipolar
lepton-emission asymmetry in a 2D model. The moderate hemispheric differences
of $Y_e$ in the early neutrino-driven wind (with variations in the range
$0.40\lesssim Y_e \lesssim 0.54$) in this simulation might give an impression
of the corresponding effects that could be obtained in future 3D explosions.

Asymmetric neutrino emission imparts a recoil on the nascent NS. We
assume a dipolar emission anisotropy of the total neutrino-energy loss rate
(the summed contributions of $\nu$ and $\bar\nu$ of all flavors),
\begin{equation}
\frac{{\mathrm{d}}\dot E_\mathrm{tot}}{{\mathrm{d}}\Omega} =
\frac{\dot E_\mathrm{tot}}{4\pi}
\left ( 1 + \alpha\,\cos\vartheta \right ) \,,
\label{eq:lumdipole}
\end{equation}
where $\alpha$ is the dipole amplitude and $\vartheta$ the zenith angle
relative to the dipole direction. In this case the NS acceleration is
\begin{equation}
a_\mathrm{NS} = \frac{1}{M_\mathrm{NS}c}\int_{4\pi}\mathrm{d}\Omega\,
\frac{{\mathrm{d}}\dot E_\mathrm{tot}}{{\mathrm{d}}\Omega} \,\cos\vartheta
= \frac{\alpha}{3\,c}\,\frac{\dot E_\mathrm{tot}}{M_\mathrm{NS}} \,.
\label{eq:nsacceleration1}
\end{equation}
With $\dot E_{53}\equiv\dot E_\mathrm{tot}/(10^{53}\,\mathrm{erg\,s}^{-1})$
and $M_{1.5} \equiv M_\mathrm{NS}/(1.5\,M_\odot$) we obtain
\begin{equation}
a_\mathrm{NS} \approx 37\,\,\frac{\alpha}{0.01}\,
\frac{\dot E_{53}}{M_{1.5}}\,\,\mathrm{km\,s}^{-2}\,.
\label{eq:nsacceleration2}
\end{equation}
If $\alpha\sim 0.01$--0.02 and the duration of the emission asymmetry lasts
only some hundred milliseconds, corresponding to the duration of the
accretion phase, the recoil velocity will not exceed several
10\,km\,s$^{-1}$, depending on the time-integrated neutrino-energy release
$E_\mathrm{tot}$. But the NS kick velocity could reach 100--200\,km\,s$^{-1}$
for a canonical value of $3\times 10^{53}\,$erg for the NS gravitational
binding energy, if the neutrino-emission dipole continues to exist for the
whole period of PNS neutrino cooling. With the luminosity maximum coinciding
with the $\bar\nu_e$ emission peak in our 3D simulations, the NS acceleration
will point in the direction of the strongest $\nu_e$ emission. Even in the
optimistic (and highly speculative) case that the neutrino-emission dipole
survives for seconds, however, the estimated recoil velocity is dwarfed by
those that can typically be expected from the ``gravitational tug-boat
mechanism'' associated with the anisotropic ejection of matter in 3D
simulations of SN explosions \citep{Wongwathanarat_2010,Wongwathanarat_2013}.

Another small but amusing mechanical consequence of LESA is an angular
momentum transfer, i.e., a spin-up of the nascent NS. Weak interactions
violate parity maximally, implying that a relativistic $\nu_e$ has negative
helicity and carries the spin angular momentum $-\hbar/2$ relative to its
direction of motion, whereas a $\bar\nu_e$ has positive helicity and carries
$+\hbar/2$. We denote the lepton-number flux dipole amplitude with $A_{\rm
Dipole}$, i.e., the quantity plotted in Fig.~\ref{fig:dipole11}. The angular
momentum transfer rate then has the magnitude
\begin{equation}
\dot J=\frac{\hbar}{2}\,\frac{A_{\rm Dipole}}{3}\,,
\end{equation}
because our normalization of the dipole amplitude implies that it is three
times the total lepton-number flux projected on the dipole direction. With a
typical value $A_{\rm Dipole}=3\times10^{56}\,{\rm s}^{-1}$ and recalling
that $\hbar=1.054\times10^{-27}\,{\rm cm}^2\,{\rm g}\,{\rm s}^{-1}$ we find a
typical angular-momentum transfer rate of $\dot J=5\times10^{28}\,{\rm
cm}^2\,{\rm g}\,{\rm s}^{-2}$. We recall that the moment of inertia of a
homogeneous sphere with mass $M$ and radius $R$ is $I=2 M R^2/5$ and its
angular momentum is $J=I\,2\pi f$ with $f$ the rotation frequency. Using as
benchmark values $M=1.5\,M_\odot$ and $R=15\,{\rm km}$, a typical PNS spin-up
rate is of the order of $\dot f\sim 3\times10^{-18}\,{\rm Hz}\,{\rm s}^{-1}$,
a very small value indeed. A much larger effect, however, is non-radial neutrino
emission that can transfer orbital angular momentum and cause a spin-down of
a rotating NS as first shown by \citet{Mikaelian_1977} and
\citet{Epstein_1978}.

Our entire study has ignored neutrino flavor conversion. The LESA phenomenon
depends on a subtle hemispheric asymmetry of neutrino heating rates.
Certainly this effect would be modified if the $\nu_e$ and $\bar\nu_e$
fluxes would partially swap flavor with $\nu_x$ and $\bar\nu_x$ which have
different spectral properties and different number fluxes. Moreover,
flavor conversion also modifies the neutrino influence on the n/p ratio in
the context of nucleosynthesis as first pointed out by \citet{Qian_1993}
and \citet{Qian_1995}.

The thorniest problem in the context of neutrino flavor conversion
is the role of neutrino-neutrino refraction, which causes many
complications because of the feedback of flavor conversion on itself
\citep{Duan_2010}. The justification
for ignoring flavor conversion in the dense region below the stalled shock
wave is the so-called multi-angle matter effect, which tends to suppress
self-induced flavor conversion
\citep{Esteban_2008,Sarikas_2012,Raffelt_2013,Chakraborty_2011,Saviano_2012,Chakraborty_2014}.
In particular, \citet{Dasgupta_2012} have
studied the onset of self-induced flavor conversion in a somewhat simplified
3D model. These authors conclude that flavor conversion always begins outside
the shock wave.

It is conceivable, however, that these conclusions must be modified in the
LESA context, notably in those directions where the lepton number flux is
small. The asymmetry between the $\nu_e$ and $\bar\nu_e$ number
fluxes, sometimes denoted with the parameter $\epsilon$,
is a crucial ingredient for
the stability of the neutrino flux in flavor space~\citep{Esteban_2007}.
Moreover, the LESA
phenomenon also modifies the electron-density profile which defines the
matter effect for neutrino flavor oscillations. Therefore, it remains to
be verified that flavor conversion is indeed irrelevant for the LESA
phenomenon.

The observations reported in this paper raise many interesting questions. It
is obvious that the LESA phenomenon needs much more work and analysis to
develop a full understanding, especially concerning how asymmetric PNS
convection is stimulated by asymmetric mass accretion and how SASI
and LESA interact. Many of our
explanations remain tentative and require further confirmation.

In particular, further studies are necessary to reveal how generic the
lepton-number emission dipole is. How exactly does its amplification work,
what is the underlying mechanism? What is the role of the PNS contraction in
this context, and, if it is important, how sensitive is the phenomenon to the
nuclear equation of state and the neutrino opacities? How big is the
saturation amplitude of LESA and what does it depend on? Could its amplitude
be larger than in our present 3D simulations and could its dipolar
neutrino-heating asymmetry affect the onset of neutrino-driven SN explosions?
How long does LESA last? Does LESA require the inner engine of
hemispherically asymmetric PNS convection as a necessary ingredient,
or is the role of PNS convection
only that of an amplifier of the lepton-number flux asymmetry and
that of a stabilizing factor which enables long-term memory? A feedback loop
seems possible that is solely based on the outer engine, in which the
asymmetric accretion and lepton-number emission on the one hand is
intertwined with the dipolar neutrino-heating asymmetry and shock deformation
on the other. Answers to these questions are indispensable to reach firm
conclusions on the importance of LESA for NS kicks, the explosion mechanism
and asymmetries, and SN nucleosynthesis.

Linear growth studies would help to develop deeper insights. It would be
especially useful to construct toy models that capture the essential parts of
the mechanism but reduce its complexity and thus allow a better control of
the interplay of its different components. Such studies would be particularly
useful because numerical models are always prone to artifacts. For example,
it is unclear whether our findings are affected by approximations involved in
our treatment, e.g., the ray-by-ray-plus transport, which does not include
nonradial neutrino fluxes, the use of a monopole description of the
gravitational potential, or the spherical core of 10\,km radius, which fixes
the PNS to its location at the grid center. It is therefore essential that
other groups investigate their neutrino-hydrodynamics simulations for hints
of effects similar to our LESA phenomenon.

\bigskip\noindent
{\bf Note added.} After our work had been circulated in preprint
form, the authors of \citet{Couch_2013a} informed us that
they also found LESA features in their 3D models. On the other
hand, \citet{Dolence_2014} do not find any evidence for LESA
in their 2D simulations (see their Fig.~11, where they plot the
evolution of the dipole-to-monopole ratio for the lepton-number flux).
Performing a similar analysis for our own 2D models, we do find a
strong LESA effect, commensurate with our 3D findings. However,
in 2D we see an oscillatory, high-amplitude lepton-flux asymmetry,
which can change its north-south orientation on timescales of order
10\,ms. This behavior differs from typical features of LESA in 3D,
like its persistence and directional stability, and asymmetric
accretion and heating are difficult to diagnose in such a
strongly time-variable situation. While a more
detailed future study of our 2D models may shed more light on this
question, we conclude that the appearance or not of LESA is not
directly related to the dimensionality of the simulation.

One difference between our works is that we treat neutrino
transport in the ray-by-ray-plus approximation, an approach also
used by \citet{Couch_2013a}, whereas \citet{Dolence_2014} use a
multi-dimensional
flux-limited diffusion scheme, leading them to speculate that LESA
is an artifact of the ray-by-ray technique. We note, however, that
there are numerous other differences, including neutrino interaction
rates, and that our 2D models have rather different physical
characteristics at comparable epochs, in particular differ strongly
in the evolution of the shock radius and actually explode.
Understanding the differences in these results requires a systematic
study of the influence of all physics inputs and their numerical
representation. The ray-by-ray approximation ignores lateral
radiative neutrino transport, although it includes, of course,
advective transport of neutrinos trapped in the stellar medium
(see \citealt{buras_06a}). It seems rather implausible that lateral
radiative diffusion, caused by relatively weak transverse
gradients in the PNS, could suppress a global dipole asymmetry that
would otherwise occur. Similarly, it is difficult to imagine that the
accretion and lepton-emission asymmetries on the largest possible
(hemispheric) scale could be wiped out by angular smoothing of the
neutrino heating in the gain layer on much smaller scales
(cf.\ \citealt{Sumiyoshi_2014}).
Such findings would be intriguing in their own
right. In any case, a satisfactory answer to this question depends
on a true physical understanding of the LESA phenomenon and on
observing it in simulations where the neutrino treatment does not
have the shortcomings of the ray-by-ray-plus approximation.

\section*{acknowledgements}

We are grateful to Ewald M\"uller for discussions.  This research was
supported by the Deutsche Forschungsgemeinschaft through the Transregional
Collaborative Research Center SFB/TR 7 ``Gravitational Wave Astronomy'' and
the Cluster of Excellence EXC 153 ``Origin and Structure of the Universe''
(http://www.universe-cluster.de) and by the EU through ERC-AdG No.\
341157-COCO2CASA. I.T.~acknowledges partial support from
the Netherlands Organization for Scientific Research (NWO).
The results described in this paper could
only be achieved with the assistance of high performance computing resources
(Tier-0) provided by PRACE on CURIE TN (GENCI@CEA, France) and SuperMUC
(GCS@LRZ, Germany). We also thank the Rechenzentrum Garching for computing
time on the IBM iDataPlex system \emph{hydra}.

%\newpage

%%%%%%%%%%%%%%%%%%%%%%%%%%%%%%%%%%%%%%%%%%%%%%%%%%%%%%%%%%%%%%%%%%%%%%%%%%%%%%%%%
\bibliographystyle{apj}

\begin{thebibliography}{47}
%%%%%%%%%%%%%%%%%%%%%%%%%%%%%%%%%%%%%%%%%%%%%%%%%%%%%%%%%%%%%%%%%%%%%%%%%%%%%%%%%

\expandafter\ifx\csname natexlab\endcsname\relax\def\natexlab#1{#1}\fi

\bibitem[{{Bethe}(1990)}]{bethe_90} {Bethe}, H.~A.\ 1990, Rev.~Mod.~Phys., 62,
    801

\bibitem[Blondin, Mezzacappa, \& DeMarino (2003)]{Blondin_2003}
  Blondin, J.~M., Mezzacappa, A., \& DeMarino, C.\ 2003, \apj, 584, 971

\bibitem[Blondin \& Mezzacappa (2006)]{Blondin_2006}
  Blondin, J.~M. \& Mezzacappa, A.\ 2006, \apj, 642, 401

\bibitem[Blondin \& Mezzacappa (2007)]{Blondin_2007}
  Blondin, J.~M. \& Mezzacappa, A.\ 2007, Nature, 445, 58

\bibitem[Blondin \& Shaw (2007)]{Blondin_2007a}
  Blondin, J.~M. \& Shaw, S.\ 2007, \apj, 656, 366

\bibitem[Buras et~al.\ (2003)]{Buras_2003}
  Buras, R., Janka, H.-Th., Keil, M., Raffelt, G.~G., \& Rampp, M.\ 2003
  \apj, 587, 320

\bibitem[{{Buras} {et~al.}(2006{\natexlab{a}}){Buras}, {Janka}, {Rampp}, \&
  {Kifonidis}}]{buras_06b}
{Buras}, R., {Janka}, H.-T., {Rampp}, M., \& {Kifonidis}, K.\
  2006{\natexlab{a}}, \aap, 457, 281

\bibitem[{{Buras} {et~al.}(2006{\natexlab{b}}){Buras}, {Rampp}, {Janka}, \&
  {Kifonidis}}]{buras_06a}
{Buras}, R., {Rampp}, M., {Janka}, H.-T., \& {Kifonidis}, K.\
  2006{\natexlab{b}}, \aap, 447, 1049

\bibitem[Burrows \& Fryxell (1992)]{Burrows_1992}
  Burrows, A. \& Fryxell, B.~A.\ 1992, Science, 258, 430

\bibitem[Burrows, Hayes, \& Fryxell (1995)]{Burrows_1995}
  Burrows, A., Hayes, J., \& Fryxell, B.~A.\ 1995, \apj, 450, 830

\bibitem[Burrows \& Lattimer (1988)]{Burrows_1988}
  Burrows, A. \& Lattimer, J.~M.\ 1988, Phys.\ Rep., 163, 51

\bibitem[{{Chakraborty} {et~al.}(2011)}]{Chakraborty_2011}
  {Chakraborty}, S., {Fischer}, T., {Mirizzi}, A., {Saviano}, N., \&
  {Tom{\`a}s}, R.\ 2011, \prl, 107, 151101

\bibitem[{{Chakraborty} {et~al.}(2014)}]{Chakraborty_2014}
  {Chakraborty}, S., {Mirizzi}, A., {Saviano}, N., \& {de Sousa
    Seixas}, D. 2014, \prd, 89, 093001

\bibitem[{{Colella} \& {Woodward}(1984)}]{colella_84} {Colella}, P., \&
    {Woodward}, P.~R.\ 1984, J.~Comp.~Phys., 54, 174

\bibitem[Couch (2013)]{Couch_2013}
  Couch, S.~M.\ 2013, \apj, 775, 35

\bibitem[Couch \& O'Connor (2014)]{Couch_2013a}
  Couch, S.~M. \& O'Connor, E.~P.\ 2014, \apj, 785, 123

\bibitem[Dessart et~al.\ (2006)]{Dessart_2006}
  Dessart, L., Burrows, A., Livne, E., \& Ott, C.~D.\ 2006, \apj,
  645, 534

\bibitem[Dolence et~al.\ (2013)]{Dolence_2013}
  Dolence, J.~C., Burrows, A., Murphy, J.~W., \& Nordhaus, J.\ 2013,
  \apj, 765, 110

\bibitem[Dolence, Burrows, \& Zhang (2014)]{Dolence_2014}
  Dolence, J.~C., Burrows, A., \& Zhang, W.\ 2014, arXiv:1403.6115

\bibitem[{{Dasgupta}, {O'Connor} \& {Ott}(2012)}]{Dasgupta_2012}
 {Dasgupta}, B., {O'Connor}, E.~P., \& {Ott}, C.~D.\ 2012, \prd, 85, 065008

\bibitem[{{Duan}, {Fuller} \& {Qian}(2010)}]{Duan_2010}
  {Duan}, H., {Fuller}, G.~M., \& {Qian}, Y.-Z.\ 2010,
  {Ann.\ Rev.\ Nucl.\ Part.\ Sci.}, 60, 569

%\bibitem[{{Duan} {et~al.}(2011)}]{Duan_2011}
%  Duan, H., Friedland, A., McLaughlin, G.~C., \& Surman, R. 2011,
%  J.\ Phys.\ G., 38, 035201

\bibitem[{{Epstein}(1978)}]{Epstein_1978}
  Epstein, R.\ 1978, \apj, 219, L39

\bibitem[Epstein (1979)]{Epstein_1979}
  Epstein, R.\ 1979, \mnras, 188, 305

\bibitem[{{Esteban-Pretel} {et~al.}(2007)}]{Esteban_2007}
  Esteban-Pretel, A., Pastor, S., Tom\`as, R., Raffelt, G.~G., \&
  Sigl, G.\ 2007 \prd, 76, 125018

\bibitem[{{Esteban-Pretel} {et~al.}(2008)}]{Esteban_2008}
  {Esteban-Pretel}, A., {et~al.} 2008, \prd, 78, 085012

\bibitem[Fern{\'a}ndez (2010)]{Fernandez_2010}
  Fern{\'a}ndez, R.\ 2010, \apj, 725, 1563

\bibitem[Foglizzo (2002)]{Foglizzo_2002}
  Foglizzo, T.\ 2002, \aap, 392, 353

\bibitem[Foglizzo et~al.\ (2007)]{Foglizzo_2007}
  Foglizzo, T., Galletti, P., Scheck, L., \& Janka, H.-Th.\ 2008,
  \apj, 654, 1006

\bibitem[Foglizzo, Scheck, \& Janka (2006)]{Foglizzo_2006}
  Foglizzo, T., Scheck, L., \& Janka, H.-Th.\ 2006, \apj, 652, 1436

\bibitem[Foglizzo et~al.\ (2012)]{Foglizzo_2012}
  Foglizzo, T., Masset, F., Guilet, J., \& Durand, G.\ 2012, \prl, 108, 051103

\bibitem[Fryer \& Warren (2002)]{Fryer_2002}
  Fryer, C.~L. \& Warren, M.~S., 2002, \apjl, 574, L65

\bibitem[Fryer \& Warren (2004)]{Fryer_2004}
  Fryer, C.~L. \& Warren, M.~S., 2004, \apj, 601, 391

\bibitem[{{Fryxell} {et~al.}(1989){Fryxell}, {M\"uller}, \&
  {Arnett}}]{fryxell_89}
{Fryxell}, B.~A., {M\"uller}, E., \& {Arnett}, D.\ 1989, Max-Planck-Institut
  f\"ur Astrophysik, Preprint Nr.~449

\bibitem[Guilet \& Fern{\'a}ndez (2013)]{Guilet_2013}
  Guilet, J. \& Fern{\'a}ndez, R.\ 2013, arXiv:1310.2616

\bibitem[Guilet \& Foglizzo (2012)]{Guilet_2012}
  Guilet, J. \& Foglizzo, T.\ 2012, \mnras, 421, 546

\bibitem[Hanke et~al.\ (2012)]{Hanke_2012}
  Hanke, F., Marek, A., M\"uller, B., \& Janka, H.-Th.\ 2012, \apj, 755, 138

\bibitem[Hanke et~al.\ (2013)]{Hanke_2013}
  Hanke, F., M\"uller, B., Wongwathanarat, A., Marek, A., \& Janka, H.-Th.\
  2013, \apj, 770, 66

\bibitem[Herant, Benz, \& Colgate (1992)]{Herant_1992}
  Herant, M., Benz, W., \& Colgate, S.\ 1992, \apj, 395, 642

\bibitem[Herant et~al.\ (1994)]{Herant_1994}
  Herant, M., Benz, W., Hix, W.~R., Fryer, C.L., \& Colgate, S.~A.\ 1994,
  \apj, 435, 339

\bibitem[Iwakami et~al.\ (2008)]{Iwakami_2008}
  Iwakami, W., Kotake, K., Ohnishi, N., Yamada, S., \& Sawada, K.\ 2008,
  \apj, 678, 1207

\bibitem[Iwakami et~al.\ (2009)]{Iwakami_2009}
  Iwakami, W., Ohnishi, N., Kotake, K., Yamada, S., \& Sawada, K.\ 2009,
  \apss, 322, 43

\bibitem[Janka (2001)]{Janka_2001}
  Janka, H.-Th.\ 2001, \aap, 368, 527

\bibitem[Janka et~al.\ (2003)]{Janka_2003}
  Janka, H.-Th., Buras, R., Kifonidis, K., Plewa, T., \& Rampp, M.\ 2003,
  in: From Twilight to Highlight: The Physics of Supernovae,
  Eds.\ W.~Hillebrandt \& B.~Leibundgut, Springer, Berlin, p.~39

\bibitem[Janka et~al.\ (2004)]{Janka_2004}
  Janka, H.-Th., Buras, R., Kifonidis, K., Rampp, M., \& Plewa, T.\ 2004,
  in: Stellar Collapse, Ed.\ C.~L.\ Fryer, Kluwer, Dordrecht, p.~65

\bibitem[Janka et~al.\ (2012)]{Janka_2012}
  Janka, H.-Th., Hanke, F., H\"udepohl, L., Marek, A., M\"uller, B. \&
  Obergaulinger, M.\ 2012, PTEP 2012, 01A309

\bibitem[Janka \& M\"uller (1993)]{Janka_1993}
  Janka, H.-Th. \& M\"uller, E.\ 1993, in: Frontiers of Neutrino Astrophysics,
  Eds. Y.~Suzuki \& K.~Nakamura, p.~203

\bibitem[Janka \& M\"uller (1995)]{Janka_1995}
  Janka, H.-Th. \& M\"uller, E.\ 1995, \apjl, 448, L109

\bibitem[Janka \& M\"uller (1996)]{Janka_1996}
  Janka, H.-Th. \& M\"uller, E.\ 1996, \aap, 306, L167

\bibitem[Keil, Janka, \& M\"uller (1996)]{Keil_1996}
  Keil, W., Janka, H.-Th., \& M\"uller, E.\ 1996, \apjl, 473, L111

\bibitem[Keil, Raffelt, \& Janka (2003)]{Keil_2003}
  Keil, M., Raffelt, G.~G., \& Janka, H.-Th.\ 2003, \apj, 590, 971

\bibitem[{{Lattimer} \& {Swesty}(1991)}]{lattimer_91} {Lattimer}, J.~M., \&
    {Swesty}, F.~D.\ 1991, Nucl.~Phys.~A, 535, 331

\bibitem[Lund et~al.\ (2010)]{Lund_2010}
   Lund, T., Marek, A., Lunardini, C., Janka, H-Th., Raffelt, G.\ 2010,
   \prd, 82, 063007

\bibitem[{{Marek} \& {Janka}(2009)}]{marek_09} {Marek}, A. \& {Janka}, H.\
    2009, \apj, 694, 664

\bibitem[Marek, Janka, \& M\"uller (2009)]{Marek_2009}
  Marek, A., Janka, H-Th., \& M\"uller, E.\ 2009, \aap, 496, 475

\bibitem[{{Marek} {et~al.}(2006){Marek}, {Dimmelmeier}, {Janka},
    {M{\"u}ller}, \& {Buras}}]{marek_06}
  {Marek}, A., {Dimmelmeier}, H., {Janka}, H.-T., {M{\"u}ller}, E., \& {Buras},
  R.\ 2006, \aap, 445, 273

%\bibitem[{{Mart{\'\i}nez-Pinedo} {et~al.}(2011)}]{Martinez_2011}
%  Mart{\'\i}nez-Pinedo, G., Ziebarth, B., Fischer, T., \& Langanke,
%  K.\ 2011, Eur.\ Phys.~J.~A, 47, 98

\bibitem[Mezzacappa et al.\ (1998)]{Mezzacappa_1998}
  Mezzacappa, A., Calder, A.~C., Bruenn, S.~W., Blondin, J.~M.,
  Guidry, M.~W., Strayer, M.~R., \& Umar, A.~S.\ 1998, \apj, 495, 911

\bibitem[{{Mikaelian}(1977)}]{Mikaelian_1977}
  Mikaelian, K.~O.\ 1977, \apj, 214, L23

\bibitem[Miller, Wilson, \& Mayle (1993)]{Miller_1993}
  Miller, D.~S., Wilson, J.~R., \& Mayle, R.~W.\ 1993, \apj, 415, 278

\bibitem[{{M{\"u}ller} {et~al.}(2012{\natexlab{b}}){M{\"u}ller}, {Janka}, \&
  {Marek}}]{mueller_12}
  {M{\"u}ller}, B., {Janka}, H.-T., \& {Marek}, A.\ 2012{\natexlab{a}}, \apj,
  756, 84

\bibitem[{{M{\"u}ller} {et~al.}(2012{\natexlab{a}}){M{\"u}ller}, {Janka}, \&
  {Heger}}]{mueller_12b}
  {M{\"u}ller}, B., {Janka}, H.-T., \& {Heger}, A.\ 2012{\natexlab{b}}, \apj,
  761, 72

\bibitem[M\"uller, Janka, \& Marek (2013)]{Mueller_2013}
  M\"uller, B., Janka, H.-Th., \& Marek, A.\ 2013, \apj, 766, 43

\bibitem[M\"uller \& Janka (1994)]{Mueller_1994}
  M\"uller, E. \& Janka, H.-Th.\ 1994, in: Reviews in Modern Astronomy, 7,
  Ed.\ G.~Klare, Astronomische Gesellschaft, Hamburg, p.~103

\bibitem[M\"uller \& Janka (1997)]{Mueller_1997}
  M\"uller, E. \& Janka, H.-Th.\ 1997, \aap, 317, 140

\bibitem[M\"uller, Janka, \& Wongwathanarat (2012)]{Mueller_2012}
  M\"uller, E., Janka, H.-Th., \& Wongwathanarat, A.\ 2012, \aap, 537, A63

\bibitem[Murphy \& Burrows (2008)]{Murphy_2008}
  Murphy, J.~W. \& Burrows, A.\ 2008, \apj, 688, 1159

\bibitem[Murphy, Ott, \& Burrows (2009)]{Murphy_2009}
  Murphy, J.~W., Ott, C.~D., \& Burrows, A.\ 2009, \apj, 707, 1173

\bibitem[Nordhaus et~al.\ (2010a)]{Nordhaus_2010}
  Nordhaus, J., Burrows, A., Almgren, A., \& Bell, J.\ 2010, \apj, 720, 694

\bibitem[Nordhaus et~al.\ (2010b)]{Nordhaus_2010b}
  Nordhaus, J., Brandt, T.~D., Burrows, A., Livne, E., \& Ott, C.~D.\ 2010,
  \prd, 82, 103016

\bibitem[Nordhaus et~al.\ (2012)]{Nordhaus_2012}
  Nordhaus, J., Brandt, T.~D., Burrows, A., \& Almgren, A.\ 2012,
  \mnras, 423, 1805

%\bibitem[{{Qian} \& {Fuller}(1995)}]{Qian_1995}
%  Qian, Y.-Z. \& Fuller, G.~M.\ 1995, \prd, 52, 656

\bibitem[Qian \& Fuller (1995)]{Qian_1995}
  Qian, Y.-Z. \& Fuller, G.~M.\ 1995, \prd, 52, 656

\bibitem[{{Qian} {et~al.}(1993)}]{Qian_1993}
  Qian, Y.-Z., Fuller, G.~M., Mathews, G.~J., Mayle, R.~W.,
  Wilson, J.~R., \& Woosley, S.~E.\ 1993, \prl, 71, 1965

\bibitem[Raffelt (2001)]{Raffelt_2001}
  Raffelt, G.~G.\ 2001, \apj, 561, 890

\bibitem[{{Raffelt}, {Sarikas} \& {de Sousa Seixas}(2013)}]{Raffelt_2013}
  {Raffelt}, G.~G., {Sarikas}, S., \& {de Sousa Seixas}, D.\ 2013,
  \prl, 111,  091101

\bibitem[{{Rampp} \& {Janka}(2002)}]{rampp_02} {Rampp}, M., \& {Janka}, H.-T.\
    2002, \aap, 396, 361

\bibitem[Rantsiou et~al.\ (2011)]{Rantsiou_2011}
  Rantsiou, E., Burrows, A., Nordhaus, J., \& Almgren, A.\ 2010, \apj, 732, 57

\bibitem[{{Sarikas} {et~al.}(2012)}]{Sarikas_2012}
  {Sarikas}, S., {Raffelt}, G.~G., {H{\"u}depohl}, L., \& {Janka}, H.-T.\ 2012,
  \prl, 108, 061101

\bibitem[{{Saviano} {et~al.}(2012)}]{Saviano_2012}
  {Saviano}, N., {Chakraborty}, S., {Fischer}, T., \& {Mirizzi}, A.\ 2012,
  \prd, 85, 113002

\bibitem[Scheck et~al.\ (2004)]{Scheck_2004}
  Scheck, L., Plewa, T., Janka, H.-Th., Kifonidis, K., \& M\"uller, E.\ 2004,
  \prl, 92, 011103

\bibitem[Scheck et~al.\ (2006)]{Scheck_2006}
  Scheck, L., Kifonidis, K., Janka, H.-Th., \& M\"uller, E.\ 2006,
  \aap, 457, 963

\bibitem[Scheck et~al.\ (2008)]{Scheck_2008}
  Scheck, L., Janka, H.-Th., Foglizzo, T., \& Kifonidis, K.\ 2008,
  \aap, 477, 931

\bibitem[Sumiyoshi et~al.\ (2014)]{Sumiyoshi_2014}
  Sumiyoshi, K., Takiwaki, T., Matsufuru, H., \& Yamada, S.\ 2014,
  arXiv:1403.4476

\bibitem[Tamborra et~al.\ (2013)]{Tamborra_2013}
  Tamborra, I., Hanke, F., M\"uller, B., Janka, H.-Th., \& Raffelt, G.\ 2013,
  \prl, 111, 121104

\bibitem[Tamborra et~al.\ (2014)]{Tamborra_2014}
  Tamborra, I., Raffelt, G., Hanke, F., Janka, H.-Th., \& M\"uller, B.\ 2014,
  arXiv:1406.0006

\bibitem[Takiwaki, Kotake, \& Suwa (2014)]{Takiwaki_2013}
  Takiwaki, T., Kotake, K., \& Suwa, Y.\ 2014, \apj, 786, 83

\bibitem[Wanajo, Janka, \& M\"uller (2011)]{Wanajo_2011}
  Wanajo, S., Janka, H.-Th., \& M\"uller, B.\ 2011, \apjl, 726, L15

\bibitem[Wongwathanarat, Janka, \& M\"uller (2010)]{Wongwathanarat_2010}
  Wongwathanarat, A., Janka, H.-Th., \& M\"uller, E.\ 2010, \apjl, 725, L106

\bibitem[Wongwathanarat, Janka, \& M\"uller (2013)]{Wongwathanarat_2013}
  Wongwathanarat, A., Janka, H.-Th., \& M\"uller, E.\ 2013, \aap, 552, A126

\bibitem[Woosley \& Heger (2007)]{Woosley_2007}
  Woosley, S.~E. \& Heger, A.\ 2007, PhR, 442, 269

\bibitem[{{Woosley} {et~al.}(2002){Woosley}, {Heger}, \&
    {Weaver}}]{woosley_02}
    {Woosley}, S.~E., {Heger}, A., \& {Weaver}, T.~A.\ 2002,
    Rev.~Mod.~Phys.,
    74, 1015

\bibitem[Yamasaki \& Foglizzo (2008)]{Yamasaki_2008}
  Yamasaki, T. \& Foglizzo, T.\ 2008, \apj, 679, 607


\end{thebibliography}

\end{document}